\pdfoutput=1
\documentclass[3p]{elsarticle}

\usepackage{natbib}
\usepackage[]{lineno}
\usepackage{multirow}
\usepackage{amsmath}
\usepackage{stfloats}
\usepackage[export]{adjustbox}
\usepackage{amssymb}
\usepackage{gensymb}
\usepackage{comment}
\usepackage{xr-hyper}
\usepackage{hyperref}
\usepackage{graphicx}
\usepackage[labelfont=bf]{caption}
\usepackage{xcolor}
\usepackage{multicol}
\usepackage{float}
\usepackage{subcaption}
\usepackage[T1]{fontenc}
\graphicspath{{Figures}}
\setcitestyle{round,numbers,sort&compress}
\journal{Science Advances}

\begin{document} 
\includegraphics[scale=1]{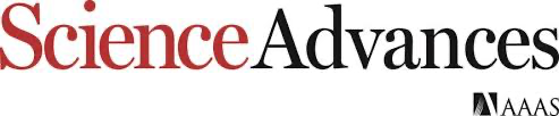}
\section*{FRONT MATTER} 

\section*{Title}
\begin{itemize}
    \item Revealing hidden defects through stored energy measurements of radiation damage
    \item Stored energy measurements of radiation damage
\end{itemize}

\section*{Authors}

Charles A. Hirst$^{1*}$, Fredric Granberg$^2$, Boopathy Kombaiah$^3$, Penghui Cao$^4$, Scott Middlemas$^3$, R. Scott Kemp$^1$, Ju Li$^{1,5}$, Kai Nordlund$^2$, Michael P. Short$^{1*}$.

\section*{Affiliations}
\begin{itemize}
    \item $^1$ Department of Nuclear Science and Engineering, Massachusetts Institute of Technology, Cambridge, MA 02139, USA.
    \item $^2$ Department of Physics, University of Helsinki, P.O. Box 43, FIN-00014, Finland.
    \item $^3$ Materials and Fuels Complex, Idaho National Laboratory, Idaho, ID 83415, USA.
    \item $^4$ Department of Mechanical and Aerospace Engineering, University of California, Irvine, CA 92697, USA.
    \item $^5$ Department of Materials Science and Engineering, Massachusetts Institute of Technology, Cambridge, MA 02139, USA.
    \item $^{*}$ Corresponding author(s) emails: \textit{cahirst@mit.edu; hereiam@mit.edu}
\end{itemize}

\section*{Abstract}

\noindent With full knowledge of a material's atomistic structure, it is possible to predict any macroscopic property of interest. In practice, this is hindered by limitations of the chosen characterisation techniques. For example, electron microscopy is unable to detect the smallest and most numerous defects in irradiated materials. Instead of spatial characterisation, we propose to detect and quantify defects through their excess energy. Differential scanning calorimetry (DSC) of irradiated Ti measures defect densities 5 times greater than those determined using transmission electron microscopy (TEM). Our experiments also reveal two energetically-distinct processes where the established annealing model predicts one. Molecular dynamics (MD) simulations discover the defects responsible and inform a new mechanism for the recovery of irradiation-induced defects. The combination of annealing experiments and simulations can reveal defects hidden to other characterisation techniques, and has the potential to uncover new mechanisms behind the evolution of defects in materials.

\section*{Teaser}

\noindent The combination of annealing experiments \& simulations can uncover new defect evolution mechanisms in materials.

\section*{MAIN TEXT}

\section*{INTRODUCTION}
\noindent At the most fundamental level, a material's properties are determined by its structure. Thus with full knowledge of the structure it is possible to predict a material's behaviour. In practice this is limited by the (in)ability of a given characterisation technique to resolve the full structure, especially at the atomic-level. This general problem is exemplified by the study of irradiation-induced defects in materials. 

Irradiation alters materials through the creation of defects~\cite{Wirth2007}. To predict how the properties will change it is critical to characterise the type, size, and number density of these defects. Despite many techniques being used -- transmission electron microscopy (TEM), positron annihilation spectroscopy (PAS), and resistivity measurements -- each of these methods has limitations that restrict their ability to fully characterise the defects in irradiated materials. Simulations and experiments show that the majority of defect clusters are below $\sim$10s of point defects in size and are thus below the resolution limit for TEM~\cite{Yi2015,Zhou2007}. Consequently, TEM often underestimates the defect density by an order of magnitude~\cite{Meslin2010,Reza2020}. PAS can detect individual vacancies but is not sensitive to interstitials~\cite{Eldrup1997} which prevents the characterisation of a significant fraction of defects. Resistivity measurements have been extensively used but require knowledge of the resistivity contributions from each defect type to interpret the microstructure~\cite{Horak1975}. This is complex and is computationally intractable to be simulated for systems larger than $\sim$100s of atoms~\cite{Imbalzano2015}. Instead of determining a material's structure through spatial characterisation or property measurement it may be more effective to probe another dimension: the energy space.

By definition, defects are imperfections within a crystal and therefore all defects have an associated excess energy. In addition, the evolution of defects is limited by an energy barrier. Thus for every defect reaction there is a characteristic activation energy and energy of transformation. Determination of these parameters, through kinetic methods such as Kissinger analysis~\cite{Kissinger1957}, allows the defects involved to be deduced. While energetic transitions in more complex material systems may overlap, the deconvolution of simultaneously evolving microstructural features can be aided by correlative techniques.

That energy can be stored in a material, in the form of irradiation-induced defects, was first postulated by Eugene Wigner during the Manhattan Project~\cite{Wigner1946}. Since then, there have been many studies investigating Wigner energy in ceramic materials, including graphite~\cite{Iwata1985,Telling2003}, Si~\cite{Roorda1991}, and SiC~\cite{Snead2019}. Metals have been less well studied, with most analyses focused on defect annealing after cryogenic irradiation. These include experiments on Cu~\cite{Losehand1969}, Al~\cite{Isebeck1966}, Be~\cite{Nicoud1968}, and Mg~\cite{Delaplace1968}. However, evaluating defect populations through annealing experiments need not be limited to cryogenic temperatures. This concept is applicable to defects at all temperatures.

Exploring defects through their energetic dimensions allows the direct comparison between experimental annealing and molecular dynamics (MD) simulations of defect evolution~\cite{Caturla2000}. This combination of techniques can reveal defects that are hidden to other characterisation techniques and thus has the potential to uncover new mechanisms behind the evolution of defects. Experimental and simulated annealing is not limited to characterising defects arising from irradiation, but can be used to investigate damage resulting from other environmental factors experienced in the processing and operation of materials. In addition, this approach can be applied to study defects across the whole range of materials systems: from structural to optical to electronic materials.

\section*{RESULTS}
\noindent Here we demonstrate the excess energy idea by conducting Differential Scanning Calorimetry (DSC) experiments to anneal neutron-irradiated Ti and determine the stored energy corresponding to radiation damage recovery. Multiple energy release stages were observed between 300-600$\degree$C which contrasts with the established recovery model. TEM provides some insight into the defects responsible but can't fully account for the stored energy released. Experiments are correlated to MD simulations of radiation damage annealing to investigate the defects involved in the recovery mechanism. Comparisons between the experimental and simulated results support the use of stored energy to explore defects which may not be measurable by other characterisation techniques.

\section*{DSC annealing experiments}
\noindent Figure \ref{DSC} shows the difference in specific power, between the first heat of the (defective) sample and the mean of subsequent heats 2 to 5 (annealed sample), for irradiated and unirradiated samples. Significantly, the irradiated samples exhibit exothermic peaks whereas the unirradiated samples do not. It is also notable that there are two distinct peaks observed. The irradiated samples exhibit exothermic energy releases between 380 to 470$\degree$C (region of interest, ROI 1) and 500 to 590$\degree$C (ROI 2). The presence of multiple peaks indicates that two separate annealing processes occur between 300 and 600$\degree$C. This contrasts with the established recovery model~\cite{Schilling1978}, which describes one process occurring during stage V, and this implies that the process of recovery is more complex that previously thought.

Integrating the signal within each ROI yields values for the stored energy (J/g). ROI 1 corresponds to a release of 0.11$\pm$0.04~J/g in the irradiated samples, compared to 0.00$\pm$0.02~J/g in the unirradiated samples. ROI 2 corresponds to 0.26$\pm$0.06~J/g and 0.03$\pm$0.03~J/g for irradiated and unirradiated samples, respectively. These measured values of stored energy are the correct order of magnitude for radiation damage in metals~\cite{Snead2019} and show statistical significance between the irradiated and unirradiated samples. In order to explore the defect reactions behind each of the peaks, TEM was conducted to visualise the evolution of extended defects.

\section*{TEM characterisation}
\begin{figure}[!ht]
\centering
\includegraphics[scale=0.7]{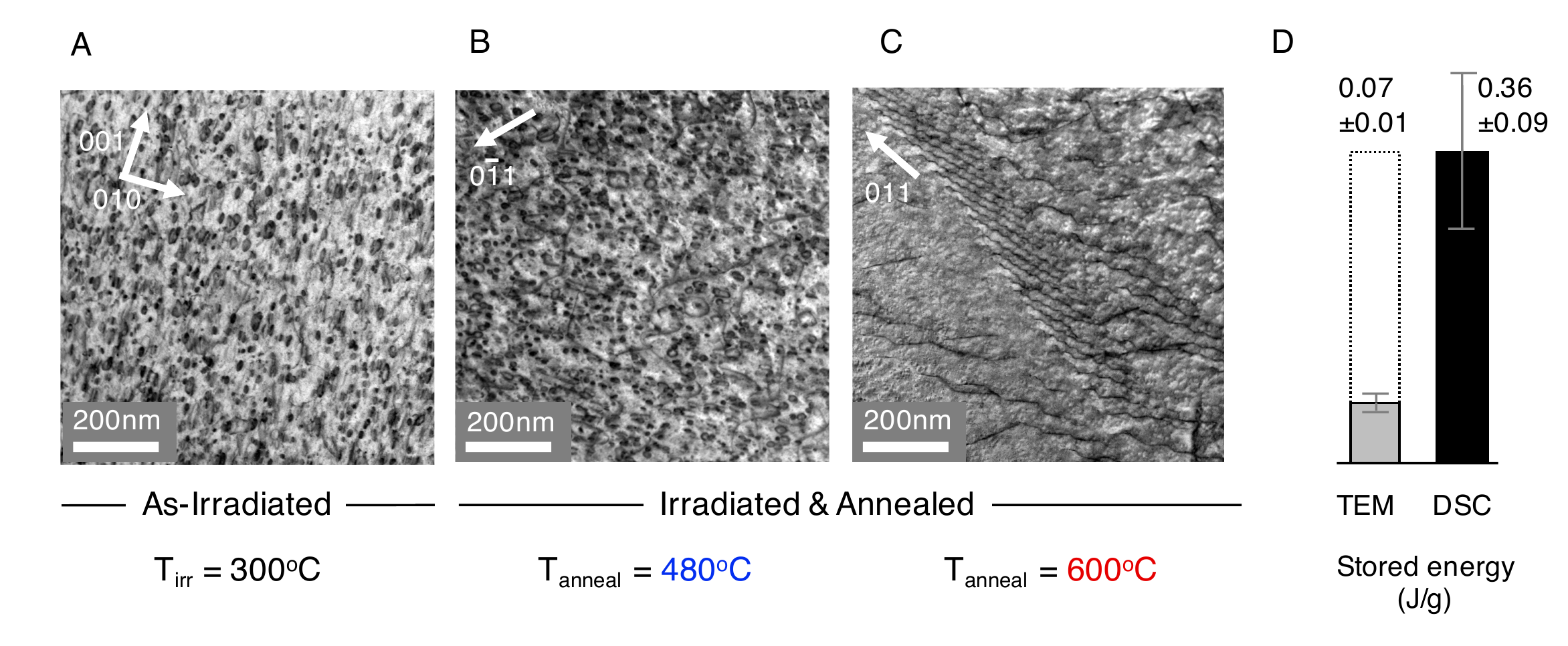}
\caption{\textbf{TEM provides insight into the annealing mechanism but can't fully account for the energy released.} (\textbf{A}) The as-irradiated microstructure shows a high density of $<$\textbf{a}$>$-type dislocation loops. STEM image taken by tilting the TEM specimen to two-beam diffraction condition of $g = 011$ along the zone axis of $[100]$. (\textbf{B}) Following annealing at 480$\degree$C the microstructure has changed very little. STEM image taken with the two-beam diffraction condition of $g = 0\overline{1}1$ along the $[311]$ zone axis. (\textbf{C}) After heating to 600$\degree$C the microstructure has significantly recovered. STEM image taken with $g = 01\bar{1}$ along the $[111]$ zone axis. (\textbf{D}) Calculating the stored energy contribution from the dislocation loops shows that the TEM-visible defects represent only a fraction of the energy measured in the DSC between 300--600$\degree$C. The error bars show the standard error of the stored energy integrated between 380--590$\degree$C for each sample (DSC) and the standard deviation of the energy calculation. Full details of the calculation are shown in section S\ref{energy_calc}.} \label{TEM} \vspace{25pt}
\end{figure}

\noindent Figure \ref{TEM} shows micrographs of the irradiated and annealed samples. In the as-irradiated sample there is a high density of $<$\textbf{a}$>$-type dislocation loops. This qualitatively matches that reported previously for Ti irradiated to a 3$\times$ greater fluence (1.7$\times$10$^{25}$~m$^{-2}$) at a similar temperature (316$\degree$C)~\cite{Griffiths1983}. The mean dislocation loop diameter in our samples is 19~nm and the number density is (4.0$\pm$0.7)$\times$10$^{21}$~m$^{-3}$. Following annealing to 480$\degree$C the microstructure is remarkably similar with the mean dislocation loop diameter still 19~nm and the number density unchanged at (4.0$\pm$0.8)$\times$10$^{21}$~m$^{-3}$. After heating to 600$\degree$C the dislocation loops have disappeared and the microstructure has fully recovered, showing features that are characteristic of annealed metals.

Calculating the energy per dislocation loop (details in section S\ref{energy_calc}) allows the stored energy contribution from TEM-visible defects to be compared to the DSC measurements. Figure \ref{TEM} D. shows that TEM-visible dislocation loops in the as-irradiated sample contribute a stored energy density of 0.07$\pm$0.01~J/g compared to our DSC measurements of 0.36$\pm$0.09~J/g (released between 300--600$\degree$C). This is significant as it demonstrates that TEM-visible defects cannot fully account for the stored energy release, and implies that there are defects being annealed which are not detected by the TEM. This finding is consistent with the well-known fact that TEM cannot resolve the full spectrum of defects~\cite{Zhou2007,Jenkins1994} in the material and supports the use of DSC measurements to quantify the magnitude of this discrepancy. In order to investigate the annealing mechanism and explore the nature of the `hidden' defects, we conduct MD simulations.

\section*{MD annealing simulations}
\begin{figure}[!ht]
\centering
\includegraphics[scale=0.9]{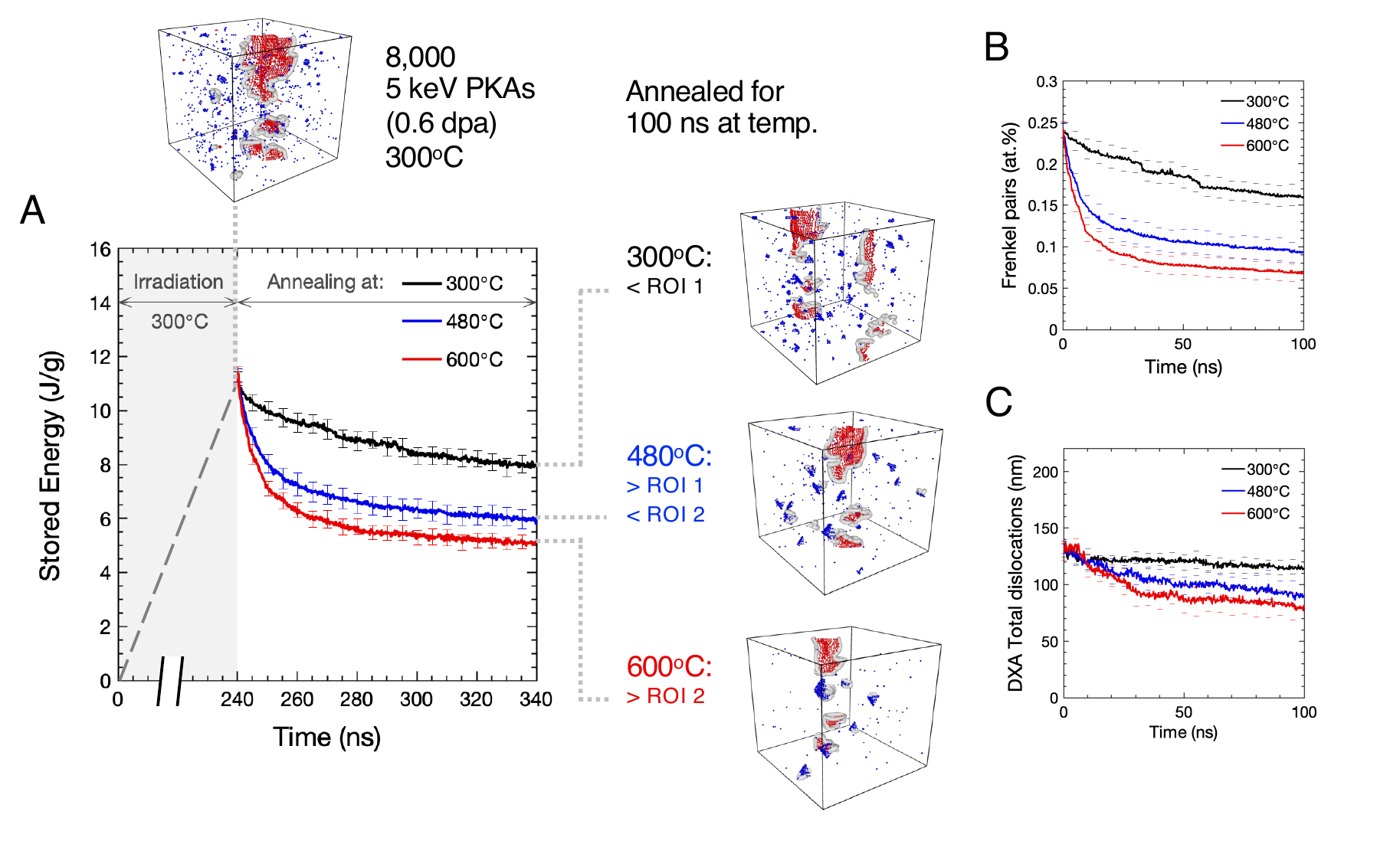}
\caption{\textbf{Simulated annealing of radiation damage shows that the stored energy recovery has a significant contribution from point defects.} Simulations of 8,000 primary knock-on atom (PKA) cascades generate defected microstructures. These are then annealed for 100~ns at 300$\degree$C, 480$\degree$C, or 600$\degree$C to determine the energy release (\textbf{A}) and the defects remaining. Wigner-Seitz (WS) and dislocation extraction algorithm (DXA) analyses show the defects as a function of time, with (\textbf{B}) a significant decrease in the point defect concentration and (\textbf{C}) a minimal decrease in the total dislocation length. All data shown is the mean of ten independent simulations and the errors bars are $\pm$ the standard error. \label{MD}}
\end{figure}

\noindent Figure \ref{MD} shows the evolution of irradiation-induced defects during annealing. Initially the microstructure consists of isolated vacancies, small vacancy clusters, and $<$\textbf{a}$>$-type interstitial dislocation loops. A representative atomic configuration is shown with dislocations represented as grey lines, interstitials as red spheres, and vacancies as blue spheres. Simulation cells were annealed for 100~ns at either 300$\degree$C, 480$\degree$C, or 600$\degree$C, and the corresponding stored energy is plotted as a function of time, in figure \ref{MD} A. For all temperatures, initially the stored energy decreases rapidly, then the rate diminishes until it effectively plateaus towards 100~ns. Annealing at 300$\degree$C exhibits step-wise behaviour with periods of gradual recovery between larger drops in stored energy. 

Figures \ref{MD} B. and C. show the corresponding Wigner-Seitz (WS) and dislocation extraction algorithm (DXA) analyses as measures of the point defect and dislocation populations, respectively. It can be seen that the point defect evolution closely matches the stored energy behaviour, while the total dislocation line length is either constant (300$\degree$C) or decreases slightly (480$\degree$C and 600$\degree$C). 

\begin{figure}[!ht]
\centering
\includegraphics[scale=1]{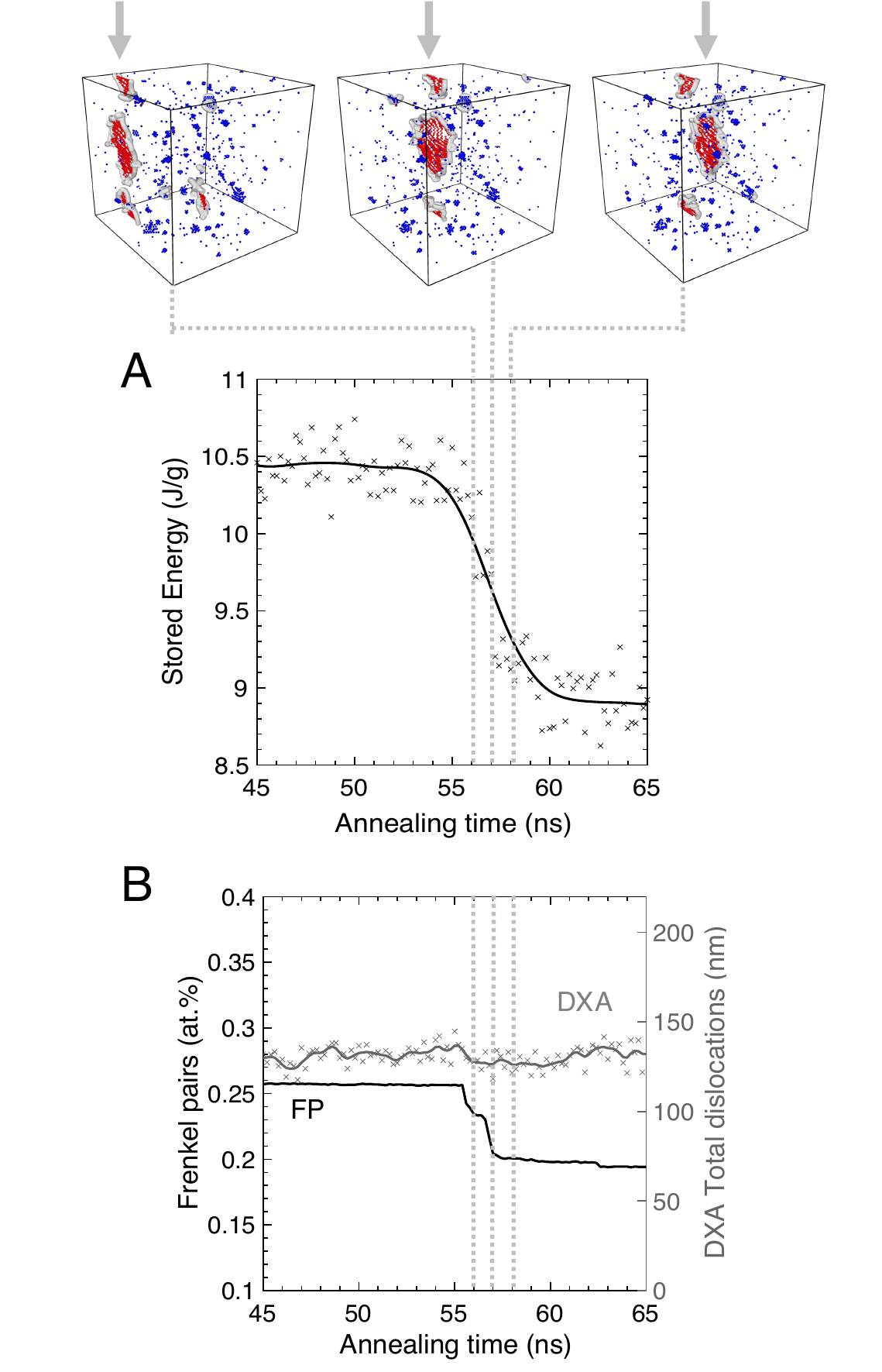}
\vspace{10pt}
\caption{\textbf{Stored energy recovery occurs via dislocation loops gliding and annihilating point defects.} (\textbf{A}) shows the stored energy of the system during relaxation at 300$\degree$C. Atomic configurations (above) highlight the associated migration of a dislocation loop. (\textbf{B}) WS and DXA analyses show that during this process the number of point defects decreases while the total line length of dislocations does not change. \label{Glide}}
\end{figure}

Figure \ref{Glide} investigates the mechanism behind the stored energy evolution in detail. During the significant drops in stored energy, dislocation loops glide considerable distances ($>$20~nm) and annihilate vacancies during this process. Figure \ref{Glide} B. shows that the motion of dislocation loops is associated with a significant decrease in point defect concentration while the total dislocation line length remains constant. This process is consistent with our experimental results that show that ROI 1 involves an exothermic defect reaction but yields microstructures that appear similar when evaluated using TEM. The annihilation of vacancies by the interstitial-type dislocation loop should ultimately lead to a decrease in the dislocation loop size. The discrepancy between dislocation line length and point defect concentration may result from a delay in the reorganisation of the dislocation loop.

The glide of loops and recombination of vacancies is responsible for the significant recovery seen initially at all temperatures and also periodically in the 300$\degree$C anneal. After the dislocation loops glide through the supercell, their migration is reduced. As the simulation temperature is constant throughout, it can be postulated that the driving force for glide comes from the stress fields of defects interacting and when the small defect clusters have been annihilated there is no longer a significant driving force for migration. The exhaustion of point defects may be an artefact of the limited simulation cell size and in a more realistic microstructure the dislocation loops may glide until sinking at a grain boundary or interacting with another dislocation.

\section*{DISCUSSION}
\noindent The evolution of irradiation-induced defects is more complex than previously thought. This may be due to previous studies being limited by the characterisation techniques employed. Exploring defects through their energetic dimensions can yield insight into their formation and evolution, as demonstrated by our work.

\section*{Combined annealing experiments and simulations infer defect evolution}
\noindent DSC measurements have been used to quantify the stored energy release from radiation damage annealing in metals. The observed temperature range of recovery matches that observed for hardness recovery of fast neutron-irradiated Ti~\cite{Higashiguchi1976}. Notably, our experiments show two distinct peaks corresponding to two separate processes where the established recovery model predicts only one~\cite{Schilling1978}. The temperature range of ROI 1 matches a prior PAS annealing study of neutron-irradiated Ti~\cite{Hasegawa1982} and the ROI 2 temperature range corresponds to recovery of cold-worked Ti~\cite{Hajizadeh2013,Prabha2013}. This suggests that the two stages involve vacancies and dislocations, respectively.

TEM characterisation supports these findings. The high density of dislocation loops observed in the as-irradiated sample remains after annealing to 480$\degree$C. Following annealing to 600$\degree$C the dislocation loops recover. Comparing the DSC and TEM results, by converting the TEM-measured defect density to a stored energy density, shows that TEM-visible defects make up only a fraction of the energy released. This indicates that there are defects involved in the annealing process that are below the resolution of the TEM.

MD simulations of primary knock-on atom (PKA) cascades generate microstructures that are qualitatively similar to the established recovery model~\cite{Schilling1978} and prior TEM results. Interstitial dislocation loops and smaller vacancy clusters match that expected for temperatures above stage III recovery and $<$\textbf{a}$>$-type interstitial dislocation loops have been observed previously in the TEM~\cite{Griffiths1983}. While the simulations do not contain vacancy dislocation loops or network dislocations, this may be explained by the high effective dose rate. The elevated dose rate results in greater recombination of defects and thus less growth of vacancy clusters into dislocation loops and less coalescence of interstitial dislocation loops into network dislocations. Additionally, since the simulations are only ($\sim$20~nm)$^{3}$ in volume, with periodic boundary conditions, extended dislocations are unlikely to form. While the presence of existing dislocations and grain boundaries would influence the evolution of radiation damage, not including them in our simulations is motivated by the difference in length scales between damage production and existing microstructure. The areal density of defect clusters in our experimental samples is much larger (2.5$\times$10$^{14}$~m$^{-2}$) than that of network dislocations (2.4$\times$10$^{13}$~m$^{-2}$) and is also significantly larger (by many orders of magnitude) than that of grain boundaries. Thus the shortest defect-defect distance is between point defect clusters and the high density of dislocation loops. As a result this will be the most significant interaction both in terms of the reaction rate and also in terms of the stored energy density.

Analysing the defect annealing simulations shows a strong correlation between the stored energy and the Frenkel pair concentration within the system. Investigating the mechanism behind the stored energy recovery reveals that dislocation loops glide through a field of point defects annihilating them. Previous \textit{in-situ} TEM heating experiments~\cite{Topping2018} of proton-irradiated Zr observed gliding of $<$\textbf{a}$>$-loops between 300-425$\degree$C. The considerable decrease in system energy driven by point-defect induced migration of dislocations has also been observed in simulations by Derlet and Dudarev~\cite{Derlet2020}. The observed mechanism is similar to the effect of dislocation channelling in a highly damaged metal. In that process dislocations become mobile and sweep straight regions of material free of smaller dislocations, creating a defect-free ``channel''~\cite{Mastel1963,Rubia2000}. However, the current mechanism is clearly distinct from this since the dislocations are much smaller, the damage level at which the effect occurs is lower, and the defects cleared away are point defects and small defect clusters.

\section*{A new mechanism for elevated-temperature irradiation damage recovery}
\begin{figure}[!hb]
\centering
\includegraphics[scale=0.75]{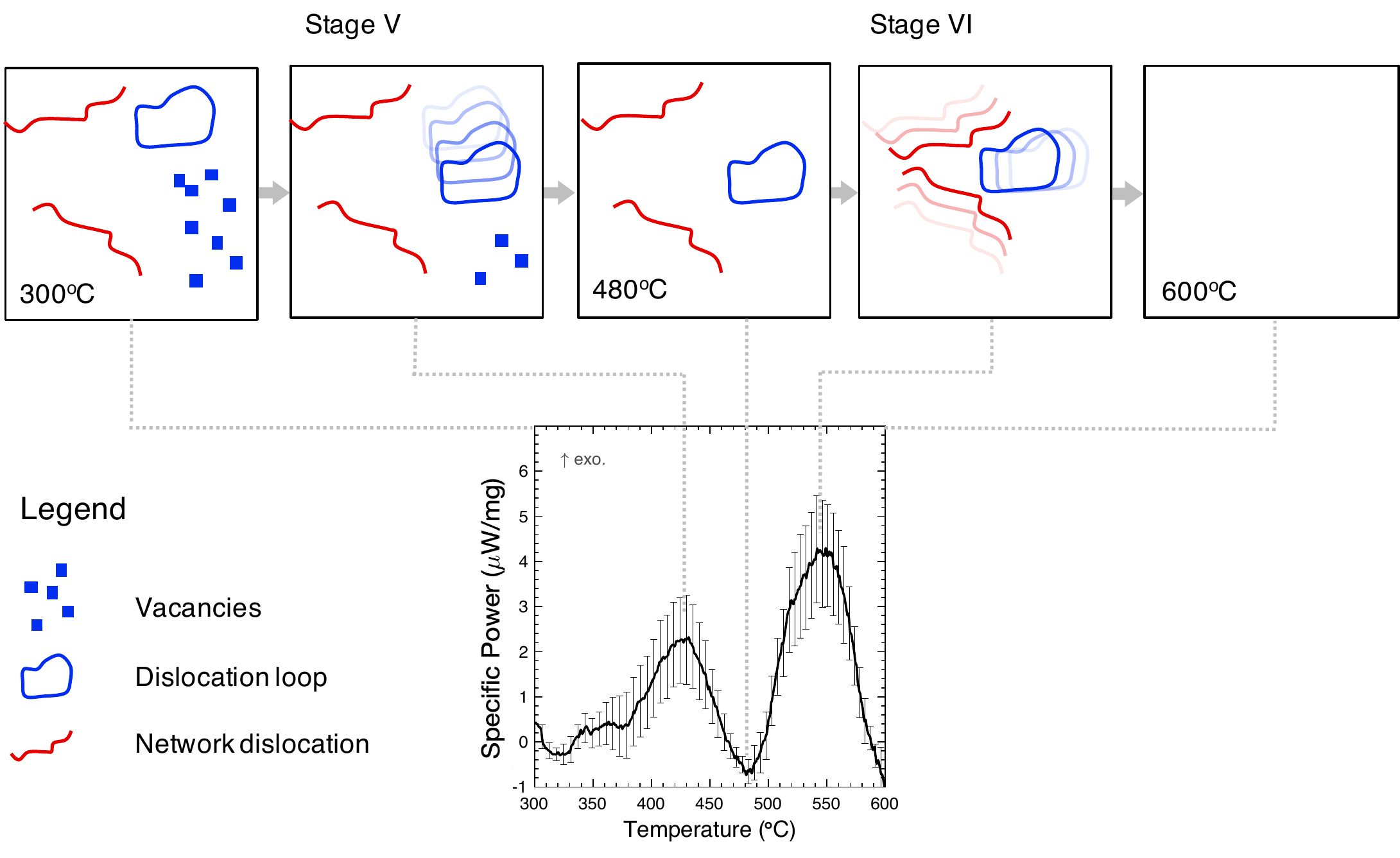}
\caption{\textbf{Recovery of neutron-irradiated Ti consists of two stages: dislocation loop glide followed by dislocation recovery.} DSC, TEM, and MD annealing results suggest that between 300-600$\degree$C there are multiple recovery stages. Stage V: glide of dislocation loops through a field of point defects and small clusters, annihilating them. Stage VI: recovery of dislocation loops and network dislocations. \label{Mechanism}}
\end{figure}

\noindent Interpreting all our results leads to the following proposed mechanism for recovery, seen in figure \ref{Mechanism}. Initially the irradiated microstructure consists of isolated vacancies and small vacancy clusters, dislocation loops, and network dislocations that form due to the impingement of dislocation loops. Heating between 300-480$\degree$C leads to Stage V: dislocation loops become mobile and glide through the system sweeping up vacancies. Heating between 480-600$\degree$C leads to Stage VI: dislocation loops and network dislocations become mobile and annihilate leading to an annealed microstructure.

This mechanism contrasts with the established recovery model~\cite{Schilling1978} in a number of ways. Firstly, we observe two distinct processes at temperatures corresponding to (so-called) Stage V where the model predicts only one. Given that multiple substages exist for earlier recovery stages~\cite{Fu2004}, this finding is not completely unexpected. Stages V and VI have also previously been reported for neutron-irradiated W with the authors attributing the annealing of vacancies and the recovery of complex defects such as dislocation loops, respectively~\cite{Bykov1972}. This prior work agrees well with our postulated mechanism. Secondly, the glide of dislocation loops also does not feature in the recovery model which assumes the evaporation of point vacancies from sessile vacancy defect clusters and their annihilation at larger interstitial clusters. Additionally, the coalescence of interstitial loops into network dislocations is not captured by the model. This may be due to the irradiation conditions (cryogenic, electron) and characterisation techniques (resistivity) used for many of the prior studies. Electron irradiation creates isolated Frenkel pairs within the material. Upon heating through stages I to IV, many defects will have recombined and the remaining dislocation loops may not be large enough to coalesce and form network dislocations. Finally, resistivity is less sensitive to network dislocations than to small defect clusters and the resistivity values may well have recovered close to the pre-irradiation value. Our work highlights the importance of understanding the conditions in which previous mechanisms have been discovered and the limitations of their extrapolation to different scenarios. As a result, our proposed mechanism may be more applicable to practical investigations of radiation damage at reactor-relevant temperatures.

To conclusively determine this mechanism, additional characterisation is being conducted. X-ray diffraction will be used to determine the defect densities before and after annealing at 480$\degree$C and 600$\degree$C and thus deduce the change in defect populations. This will be supported by PAS measurements to validate the change in vacancy concentration with temperature. Additionally, \textit{in-situ} TEM heating will be used to observe the evolution of larger defects directly. Further DSC experiments, at different heating rates, will be conducted to determine the activation energy for each of the annealing peaks. These can be correlated to nudged elastic band simulations to determine the activation energy of loop migration with and without point defects. Simulations with larger supercell sizes may enable the study of loop coalescence into extended defects, exploring dislocation recovery, however computational cost may be a limiting factor. It should also be noted that the MD annealing simulations are likely only representative of the first annealing peak observed in the DSC (ROI 1), as there are no network dislocations or grain boundaries that would enable the sinking of dislocation loops as predicted in our mechanism for the second annealing peak (ROI 2). While there is a difference in composition between our experimental samples, which are commercially pure Ti, and our simulations, which are completely pure Ti, solute atoms may be trapped at point vacancies and their clusters rather than at dislocation lines. Atom probe tomography is being conducted to confirm the location of solutes and thus determine their effect on our proposed recovery mechanism.

\section*{Microstructural understanding is only as good as our characterisation techniques}
\noindent Our DSC experiments show that two energetically-distinct processes occur in place of stage V recovery and consequently the annealing mechanism for irradiation-induced defects is more complex than previously thought. This is supported by Blewitt et al.'s paper which shows a discrepancy between resistivity and yield stress recovery of neutron-irradiated Cu~\cite{Blewitt1961}. This implies that different populations of defects present in the material are responsible for the resistivity and yield stress and demonstrates the perils of using certain characterisation techniques to investigate the microstructure of a material. Dennett et al. recently demonstrate that ``\textit{the evolution in elastic properties during swelling is found to depend significantly on the entire size spectrum of defects, from the nano- to meso-scales, some of which are not resolvable in imaging.}''~\cite{Dennett2021} Limitations on the sensitivity of characterisation techniques, such as electron microscopy, restrict analysis to a subset of the defects present and may result in the development of inaccurate models. As a result, using these techniques to correlate the structure of a defected material to its behaviour will be unsuccessful~\cite{Meslin2010,Reza2020}.

Instead of spatial characterisation, defects can be identified and quantified through their excess energy. Fundamentally, all defects in a material contribute to its energetic structure and therefore all defects have the ability to be detected through changes in their population. Significantly, annealing experiments can be directly compared to MD simulations to gain insight into the defect reactions occurring. This also has the potential to experimentally validate atomistic simulations, thereby answering the key question that exists for all simulated observations. In conclusion, exploring microstructure through the lens of stored energy can be applied to the whole spectrum of material systems, can reveal defects unable to be detected by other characterisation techniques, and has the potential to uncover new mechanisms behind the evolution of defects.

\section*{MATERIALS AND METHODS}
\section*{Materials}
\noindent Samples were sectioned from a 1/2" CP-2 titanium nut. The composition is given in table \ref{Composition}. The nut was subject to 73 days of irradiation in the Advanced Cladding Irradiation (ACI) facility of the MIT reactor. The conditions in the ACI loop simulate a pressurised water reactor (PWR) with controlled coolant chemistry and temperature of 300~$\pm$~2~$\degree$C. The nut was irradiated at a fast neutron flux of 1.0$\times$10$^{14}$~cm$^{-2}$~s$^{-1}$ ($>$~0.1~MeV) to a total fluence of 6.3$\times$10$^{20}$~cm$^{-2}$. This corresponds to a dose of 0.76~dpa which was calculated using the NRT formula~\cite{Norgett1975} and total damage energy production cross section (ENDF/B-VIII.0 MT=444) with E$_d$~=~30~eV~\cite{Was2007}.

After irradiation, the nut was sectioned on a low-speed saw into approximately (4~mm)$^{3}$ samples, with the average mass 67~mg, for DSC analysis. While mechanical deformation induces cold work to the sectioned faces, this does not contribute significantly to the stored energy due to the surface area to volume ratio of the samples. In addition, this contribution will be identical for unirradiated and irradiated samples and can thus be accounted for.

Unirradiated samples were sectioned from an identical nut, a subset of these were annealed in a vacuum furnace at 300$\degree$C for 16~hours or at 400$\degree$C for 168~hours, to replicate the time spent at 300$\degree$C in the reactor. Unirradiated samples show no measurable difference in stored energy with prior annealing, indeed they show no measurable release of stored energy between 50-600$\degree$C.

\section*{DSC experiments}
\begin{figure}[!ht]
\centering
\includegraphics[scale=0.6]{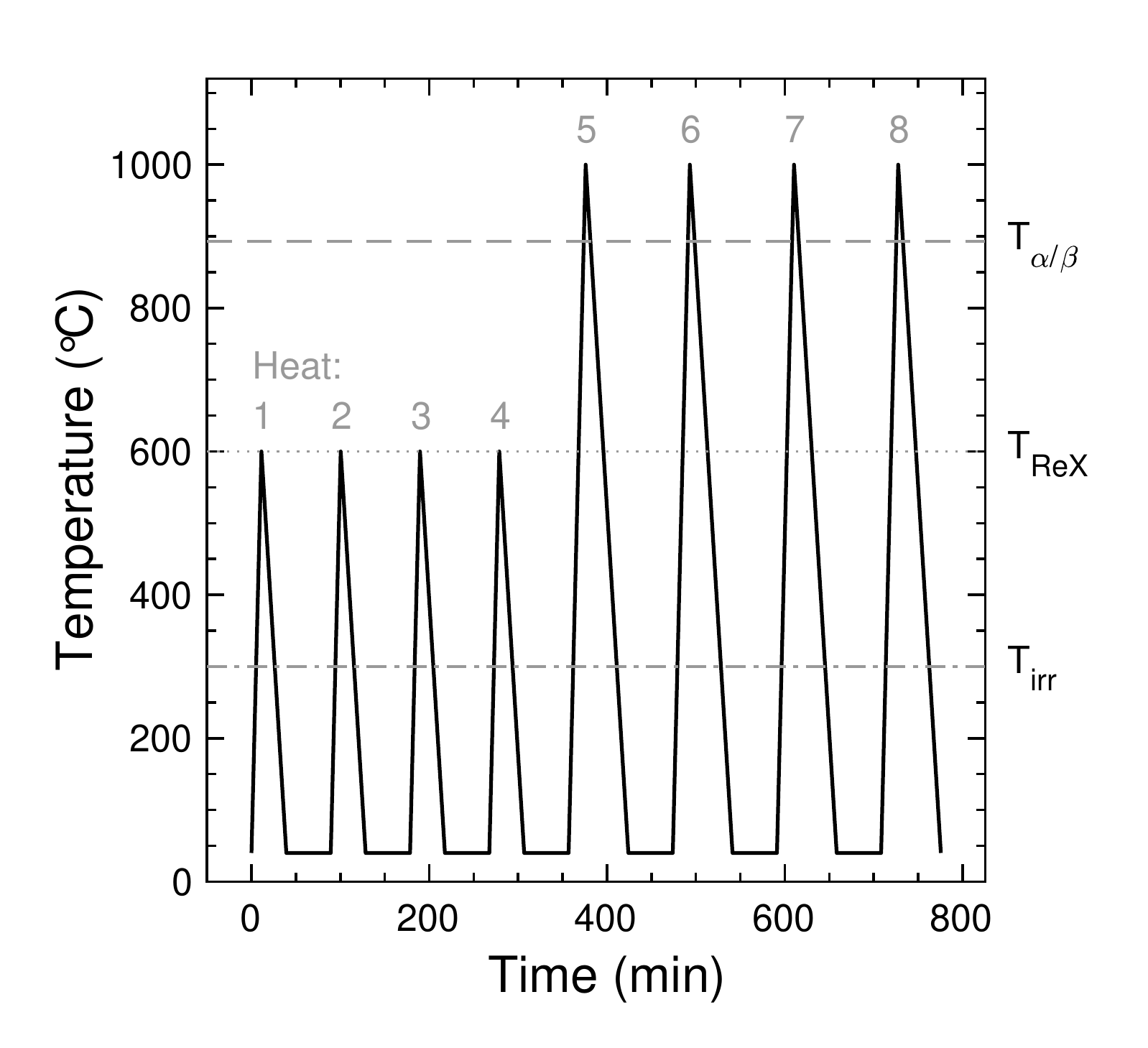}
\caption{\textbf{DSC Heating profile.} Samples were heated initially (1 to 4) to anneal out radiation damage and generate an annealed baseline, then were heated (5 to 8) through the $\alpha/\beta$ phase transition to measure the enthalpy of transformation. T$_{ReX}$ = Recrystallisation temperature. \label{heatprofile}}
\end{figure}

\noindent Samples were annealed using a Netzsch 404 F3 DSC with a type P sensor for increased sensitivity. Crucibles were 0.19~mL Pt/Rh to maximise the sample size and Y$_{2}$O$_{3}$ spray was used to ensure samples did not adhere to the crucibles. Samples were heated at 50~K/min in a UHP Ar atmosphere, according to the heating profile shown in Figure \ref{heatprofile}. Samples were heated to 600$\degree$C four times, first to anneal out the radiation damage (heat 1) and then to generate an annealed baseline (heats 2 to 5) to compare to the first heating run. Samples were then heated to 1000$\degree$C four times to undergo the $\alpha/\beta$ phase transition (heats 5 to 8) and measure the enthalpy of transformation. The $\alpha/\beta$ enthalpy of transformation is $-$87$\pm$4~J/g~\cite{Cezairliyan1977}. The measured enthalpy was used to validate the instrument sensitivity calibration, which was conducted after the experiments (using a sapphire standard and the Netzsch C$_{p}$ software package). For more details on the calibration procedure see section S\ref{sensitivity}.

DSC data analysis involved fitting a cubic baseline, to the areas outside the regions of interest (ROIs), and subtracting this to evaluate the specific power at the correct scale ({\textmu}W/mg). The effect of the crucible was then corrected for. Heats 2 to 5 (annealed sample) were averaged and subtracted from heat 1 (defected sample) to determine the stored energy released on the first heat. Nine irradiated and nine unirradiated samples were then averaged to increase the signal to noise ratio. Error bars show $\pm$ the summation in quadrature of the standard errors arising from averaging the crucible corrections, heats 2 to 5, and the different samples. The stored energy was evaluated by integrating the signal within each ROI (1: 380-470$\degree$C, 2: 500-590$\degree$C). The uncertainty on the integrals was calculated as the summation in quadrature of standard errors arising from averaging the integral from each sample and each correction run. For more details on the analysis procedure see section S\ref{dsc_analysis}.

\section*{TEM characterisation}
\noindent Samples were annealed to different temperatures in the DSC prior to preparation for TEM analysis. Four samples were selected: one as-irradiated (T = 300$\degree$C), one annealed to 480$\degree$C (ROI 1 $<$ T $<$ ROI 2), one annealed to 600$\degree$C (ROI 2 $<$ T), and one unirradiated. Each sample then had one TEM lamella prepared using a Tescan Lyra 3 focused ion beam microscope. The sample thickness, and thus defect density, was determined using energy-filtered TEM log ratio method~\cite{Malis1988}. The mean free path for inelastic scattering of 200~keV electrons in Ti = 106~nm with an uncertainty of 19\%~\cite{Shinotsuka2015}. For measuring the dislocation loop diameter from the TEM micrographs, \textit{ImageJ} software was used to determine the Feret diameter.

In order to correlate the TEM-determined defect densities to the DSC measurements, the energy per dislocation loop was calculated from elasticity theory~\cite{Hirth1982,Liu2020}. The energy per length was determined and then multiplied by the dislocation loop size and density to obtain the stored energy density (J/g). This was compared to the stored energy from DSC integrated over both ROIs (380-590$\degree$C) with error bars calculated similarly to above. The full details of the calculation are included in section S\ref{energy_calc}.

\section*{MD simulations}
\section*{Displacement cascades}
\noindent To generate Ti microstructures that were representative of neutron irradiation, MD simulations of consecutive collision cascades were performed using the PARCAS code~\cite{Nordlund1998}. An adaptive timestep was used to accurately follow the trajectories of the energetic particles~\cite{Nordlund1995}. The interatomic potential by Ackland et al., `A92'~\cite{Ackland1992} with close up repulsion by G. Ackland~\cite{A92-pot} was used and ten independent simulations were conducted to increase the statistics of the results. Simulation cells of 492,800 atoms were subject to repeated 5~keV PKAs at 300$\degree$C to achieve doses up to 0.6~dpa (8,000 PKAs with a threshold displacement energy of 30~eV). Electronic stopping was active on all atoms with kinetic energy of 5~eV or more. After each cascade was initiated, the simulation cells were held at 300$\degree$C for 30~ps to allow for unstable defect configurations to relax. This was done in a two step manner, first with border cooling not to affect the cascade region and then when the cell had equilibrated, a thermostat and barostat on the whole simulation cell to reach a zero overall pressure were applied. The box was randomly shifted after each cascade to obtain a homogeneous irradiation.

\section*{Defect annealing}
\noindent Following the PKA cascades the simulation cells were annealed using the LAMMPS code~\cite{Plimpton1995}. The simulation cell of 492,800 atoms corresponds to $\sim$(20~nm)$^{3}$ in volume and the periodic boundary conditions create an infinite single crystal. The simulations therefore correspond to a system without grain boundaries. Using the NVT ensemble, the defected supercells were relaxed with a 2~fs timestep for 5$\times 10^{7}$ timesteps, resulting in a total duration of 100~ns. Supercells were relaxed at 300$\degree$C, 480$\degree$C and 600$\degree$C, which respectively correspond to below ROI 1, between ROI 1 and 2, and above ROI 2 in figure \ref{DSC}. Cell configurations were output every 0.2~ns and minimised using the conjugate gradient algorithm to relax unstable defect configurations before calculation of the stored energy. The total stored energy was calculated by comparing the potential energy of the defected supercell to that of a pristine crystal. Ovito~\cite{Stukowski2010a} was used to visualise the system with dislocation extraction algorithm (DXA)~\cite{Stukowski2012} analysis used to identify dislocations and Wigner-Seitz (WS) analysis used to detect point defects~\cite{Nordlund1998}.

\bibliographystyle{ieeetr}
\bibliography{main.bib}

\section*{Acknowledgements} 
\noindent The authors would like to thank colleagues from the MIT Nuclear Reactor Laboratory for providing the sample, in particular Dr. Edward Lamere for radiation protection guidance and Dr. Guiqiu Zheng for assisting with the sectioning procedure. The authors acknowledge Dr. Cheng Sun for assisting with sample shipping and their preparation for TEM. CAH thanks Dr. Felice Frankel and Dr. Mary O'Reilly for assistance preparing the figures and Dr. Weiyue Zhou for helpful discussions. This research made use of the resources of the High Performance Computing Center at Idaho National Laboratory, which is supported by the Office of Nuclear Energy of the U.S. Department of Energy and the Nuclear Science User Facilities under Contract No. DE-AC07-05ID14517. FG and KN acknowledges that this work has partially been carried out within the framework of the EUROfusion Consortium. The views and opinions expressed herein do not necessarily reflect those of the European Commission. Computer time granted by the IT Center for Science -- CSC -- Finland and the Finnish Grid and Cloud Infrastructure persistent identifier urn:nbn:fi:research-infras-2016072533 are gratefully acknowledged.

\section*{Funding}
\noindent National Science Foundation Faculty Early Career Development Program Grant DMR-1654548 - MPS \\
Idaho National Laboratory Nuclear University Consortium Laboratory Directed Research and Development Grant No. 19A39-070 - SM, MPS \\
Euratom Research and Training programme 2014--2018 and 2019--2020 Grant Agreement No. 633053 - FG, KN

\section*{Competing interests}
\noindent Authors declare that they have no competing interests.

\section*{Author Contributions}
\noindent Conceptualisation: CAH, PC, RSK, JL, KN, MPS \\
\noindent Methodology: CAH, FG, PC, SM, MPS \\
\noindent Investigation: CAH, FG, BK \\
\noindent Visualisation: CAH, MPS \\
\noindent Supervision: MPS  \\
\noindent Writing - original draft: CAH, FG, BK \\
\noindent Writing - editing and review: all authors.

\section*{Data and materials availability}
\noindent All data needed to evaluate the conclusions in the paper are present in the paper and/or the Supplementary Materials. All data can be accessed at repository: \url{https://doi.org/10.5281/zenodo.6485226}.

\newpage
\section*{FIGURES}

\begin{figure}[!ht]
\centering
\includegraphics[scale=0.8]{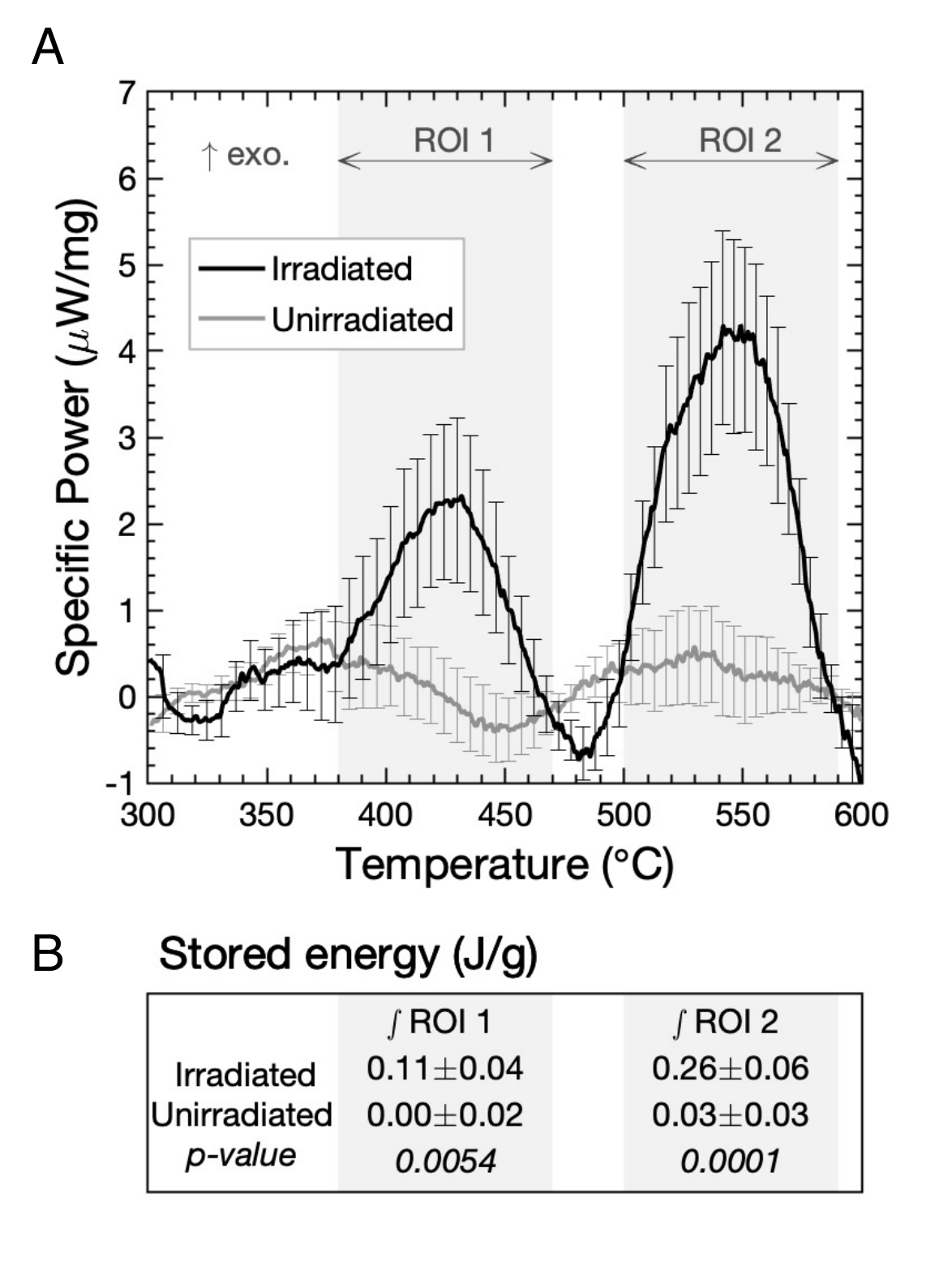}
\vspace{10pt}
\caption{\textbf{Irradiated samples release stored energy during annealing.} (\textbf{A}) Curves show the specific power difference between the first heat of the (defected) sample and the mean of subsequent heats 2 to 5 (annealed sample). Each dataset is the mean of 9 samples and the error bars show $\pm$ the summation in quadrature of the standard errors arising from averaging multiple corrections, heats, and samples. (\textbf{B}) Integrating the stored energy within each region of interest (ROI) shows that irradiated samples yield statistically significant results. Uncertainties are calculated as the summation in quadrature of the standard errors arising from averaging the integrals of sample and correction runs. The full data analysis procedure is described in section S\ref{dsc_analysis}.
\label{DSC}}
\end{figure}

\clearpage

\clearpage

\clearpage

\clearpage
\section*{FIGURES - MATERIALS AND METHODS}
\begin{table}[h]
\centering
\begin{tabular}{|c|c|c|c|c|c|c|c|c|c|c|}
\hline
& \textbf{Ti} & \textbf{O} & \textbf{Fe} & \textbf{N} & \textbf{C} & \textbf{Al} & \textbf{Ni} & \textbf{Mn} & \textbf{Si} & \textbf{Sn}\\ \hline
wt.\%  & bal. & 0.117 & 0.081 & 0.029 & 0.028 & 0.013 & 0.013 & 0.011 & 0.0073 & 0.0070 \\ \hline
\end{tabular}
\caption{\textbf{Composition of CP-2 Ti hex nut.} [O, N] determined using inert gas fusion, [C] using combustion infrared detection, and all other elements using direct current plasma emission spectroscopy.}
\label{Composition}
\end{table}

\clearpage

\newpage
\begin{center}
    \includegraphics[scale=1]{logo_SciAdv.png}

    \vspace{20pt}

    \Large{Supplementary Materials for}

    \vspace{10pt}

    \large{\textbf{Revealing hidden defects through stored energy measurements of radiation damage}}

    \vspace{10pt}

    \noindent \small{Charles A. Hirst$^{*}$, Fredric Granberg, Boopathy Kombaiah, Penghui Cao, Scott Middlemas, R. Scott Kemp, Ju Li, Kai Nordlund, Michael P. Short$^{*}$}
    
    \vspace{10pt}
    
    $^{*}$Corresponding author(s) emails: cahirst@mit.edu; hereiam@mit.edu
    
    \vspace{10pt}

\end{center}

\normalsize
\noindent \textbf{This PDF file includes:}

\vspace{5pt}

\noindent Supplementary text\\
\noindent Figs. S1 to S16 \\
Tables S1 to S2 \\
References (\textit{35 - 77})

\tableofcontents

\renewcommand{\thefigure}{S\arabic{figure}}
\setcounter{figure}{0}
\renewcommand{\thetable}{S\arabic{table}}
\setcounter{table}{0}

\clearpage
\section*{Supplementary Text}
\section{Background}
\subsection{Limitations of existing characterisation techniques}
\subsubsection{Transmission electron microscopy (TEM)}
\noindent TEM is commonly used to image radiation damage by scattering electrons off the locally distorted lattice around defects~\cite{Jenkins1994}. However, the resolution limit of conventional TEM is typically between 1-2~nm~\cite{Zhou2007}. This corresponds to a cluster of $\sim$55 defects for a dislocation loop with diameter of 1~nm in BCC or FCC, or $\sim$350 vacancies for a 1~nm diameter void in BCC~\cite{Caturla2000a}. The resolution limit thus prevents TEM from characterising point defects and small defect clusters. In addition, not all defects larger than the resolution limit will be visible, due to the $\mathbf{g \cdot b}$ visibility criterion~\cite{Wirth2015}. Simulations and experiments show that the size distribution of irradiation-induced defect clusters follows a power law~\cite{Yi2015}. This means that the majority of clusters in the material are below $\sim$10s of defects in size and thus the majority of clusters may be below the visibility limit for TEM.

With recent developments in aberration-corrected scanning TEM (STEM), it is possible to image radiation damage at high resolution. Liu et al. compare STEM to TEM of Kr-irradiated SiC and show that TEM is unable to detect 2/3 of the defects that are visible in STEM~\cite{Liu2017}. Jiang et al. also observe small defect clusters in electron-irradiated SiC~\cite{Jiang2016}. However, for both of these studies, the ability to image the clusters is attributable to the high migration energy and low mobility of carbon interstitials in SiC. This would not be achievable in metals where the interstitial migration energy is over an order of magnitude lower. The thickness requirements for S/TEM samples result in free surfaces which act as sinks for mobile defects. Therefore, in metals, up to the entire sample may be considered to be a denuded zone and point defects may no longer be present. The mobility of defects scales inversely with cluster size~\cite{Osetsky2000}, thus the defects that most likely to occur are both unable to be imaged in TEM and also least likely to be present in the sample.

Many studies have shown that the limitations of TEM prevent it from characterising the full population of defects in a sample. Meslin et al. report a defect number density that is an order of magnitude lower for TEM than for small angle neutron scattering (SANS) or positron annihilation spectroscopy (PAS) of neutron-irradiated Fe~\cite{Meslin2010}. Reza et al. observe a similar scale discrepancy between TEM and transient grating spectroscopy (TGS) of self-ion irradiated W~\cite{Reza2020}. Song et al. also report an order of magnitude between defect number densities determined by TEM and lattice strain measurements of Fe$^{3+}$ irradiated FeCr~\cite{Song2020}.

Vastly underestimating the defect density prevents TEM from accurately determining the corresponding change in properties. Wei{\ss} et al. show a factor of 2 between measured and calculated (from TEM) change in hardness for neutron irradiated EUROFER~\cite{Weiss2012}. Reza et al. report the same discrepancy between TGS-measured and TEM-determined thermal diffusivity for self-ion irradiated W~\cite{Reza2020}. Notably, when Reza et al. account for small defects through the addition of results from molecular dynamics (MD) simulations, the combination of TEM and MD matches TGS measurements. This result confirms the theory that point defects play a significant role in the thermal diffusivity of a material and further reinforces the need to accurately characterise small defects in order to evaluate irradiation-induced changes in properties. It also highlights the inability of TEM to validate radiation damage simulations as the true number of Frenkel pairs cannot be determined by electron microscopy.

\subsubsection{Resistivity measurements}
\noindent Interpretation of resistivity measurements in metals requires solving Matthiessen's rule, an equation which details the electron scattering contribution from various defects as a function of their size and density~\cite{Kittel2004}. A consequence of this is the sublinear relationship between resistivity and cluster size~\cite{Martin1972}, which reduces the sensitivity of these measurements to larger defects at lower concentrations. As such, resistivity measurements are well suited for characterising isolated point defects at low temperatures (such as those resulting from cryogenic electron irradiation) but less well suited to characterisation of larger defects (such as dislocation loops) following neutron irradiation at reactor-relevant temperatures.

\subsection{Stages of radiation damage recovery}
\noindent The evolution of defects in metals is strongly linked to temperature. The difference in migration energy between interstitials and vacancies leads to characteristic stages of radiation damage recovery as a function of temperature~\cite{Schilling1978}.
For the case of a metal irradiated close to absolute zero, initially all defects are `frozen' in the lattice. As the temperature increases, the following defect reactions occur:

\begin{itemize}
    \item Stage I - following recombination of close Frenkel pairs, interstitials become mobile. They migrate and either annihilate with vacancies or cluster with other interstitials.
    \item Stage II - increasing interstitial mobility leads to the growth of interstitial clusters into small dislocation loops.
    \item Stage III - vacancies become mobile and can annihilate at interstitial clusters. Vacancy clustering also occurs, resulting in a microstructure containing small vacancy clusters and larger interstitial loops.
    \item Stage IV - both interstitial loops and vacancy clusters grow with increasing temperature.
    \item Stage V - vacancy clusters become thermally unstable, emitting single vacancies. These migrate to interstitial loops and annihilate.
\end{itemize}

These stages of recovery occur at approximately the same fixed fraction of the melting point for metals of a given crystal structure. For the FCC metals Al, Cu, Au, Ni, stage III occurs at 220, 250, 290 and 340~K respectively, this corresponds to $\sim$0.21~T$_{m}$. Stage V recovery occurs at 420, 550, 650 and 760~K respectively, which corresponds to $\sim$0.45~T$_{m}$~\cite{Hiroshi2015}. Prior resistivity studies have explored this recovery mechanism in detail, revealing further substages~\cite{Takaki1983}. Stage I contains contributions from recombination of close, correlated and uncorrelated Frenkel pairs. Stage II exhibits a substage which may be related to the migration of small interstitial clusters. These studies show that great insight into a material's structure can be obtained through detailed annealing experiments.

The above recovery model was developed from cryogenic irradiations, often using electrons. The picture becomes more complex for neutron irradiation due to the intracascade formation of defect clusters at temperatures below stage I. In addition, the presence of solutes can result in trapping of defects, resulting in enhanced stages II and IV~\cite{Schilling1978}. The effect of pre-existing defects is also not considered, with the assumption that dissociated point vacancies migrate to interstitial clusters and annihilate rather than recombine at dislocation lines or grain boundaries.

\subsection{Wigner energy in metals}
\noindent The concept of Wigner energy has more commonly been applied to ceramics than metals. Radiation damage in ceramics can lead to the formation of amorphous regions~\cite{Nordlund2018}. Due to the large number of displaced atoms within these regions, ceramics can store energy values of the order 10$^3$~J/g~\cite{Snead2019}. The large values of stored energy result in easily detectable annealing peaks in differential scanning calorimetry (DSC) experiments. Radiation damage in metals, on the other hand, takes the form of Frenkel pairs and their clusters~\cite{Nordlund2018}. The smaller number of defected atoms in metals leads to stored energy values typically less than 1~J/g~\cite{Snead2019}. As a result, investigating defect annealing using DSC is more challenging for metals than in ceramics.

While most defect annealing studies have focused on cryogenic irradiations, experiments following room temperature irradiation have been conducted. These include Cu~\cite{Blewitt1961,Pedchenko1971}, Mo~\cite{Pedchenko1971,Kinchin1958}, W~\cite{Kinchin1958}, and Ni and Fe-based alloys~\cite{Toktogulova2010}. However, very few studies have been performed after irradiation at higher, reactor-relevant, temperatures~\cite{Lee2007,Ennaceur2018}.

One of the advantages of stored energy over resistivity is its ability to be simulated for systems up to $\sim$10$^{6}$ atoms. Insight into the precise mechanism of energy release can be gained through simulated defect annealing. MD~\cite{Caturla2000} and kinetic Monte Carlo (kMC)~\cite{Beland2013} simulations are often used to evaluate defect evolution as a function of temperature. Changes in the population of defects and the resulting potential energy difference can be readily determined through these methods. This knowledge can be used to aid interpretation of the defect reactions measured in DSC experiments.

\subsection{Prior TEM characterisation of irradiated Ti}
\noindent TEM has previously been used to study radiation damage in Ti following mixed-spectrum neutron~\cite{Brimhall1971,Jostsons1980}, fast neutron~\cite{Griffiths1983,Griffiths1987}, and dual-ion irradiation~\cite{Jones1980}.
Irradiation between 300-400$\degree$C forms a high density of $<$\textbf{a}$>$-type dislocation loops. The majority of these dislocation loops are vacancy-type and have a diameter smaller than that of the interstitial-type loops~\cite{Jostsons1980}.
This microstructure is consistent with that corresponding to Stage III recovery in metals. In some cases, it is reported that the interstitial loops have grown large enough to intersect and form network dislocations~\cite{Griffiths1983}.
After irradiation at 430$\degree$C, the ratio of vacancy to interstitial loops decreases and at higher irradiation temperatures the predominant form of damage is a low density of $<$\textbf{a}$>$-type network dislocations~\cite{Griffiths1983}.
Following irradiation at 630$\degree$C, there are no dislocations observed with damage in the form of small defect clusters~\cite{Jones1980}.
However, in this study this microstructure may have been influenced by the implantation of helium that would stabilise the formation of small defect clusters. The absence of dislocations at these temperatures is consistent with cold work recovery of Ti, which indicates that dislocations become mobile above 550$\degree$C~\cite{Contieri2010,Hajizadeh2013}.

\section{DSC methods}
\subsection{Data analysis process} \label{dsc_analysis}
\begin{figure}[!h]
\centering
\includegraphics[scale=1.1]{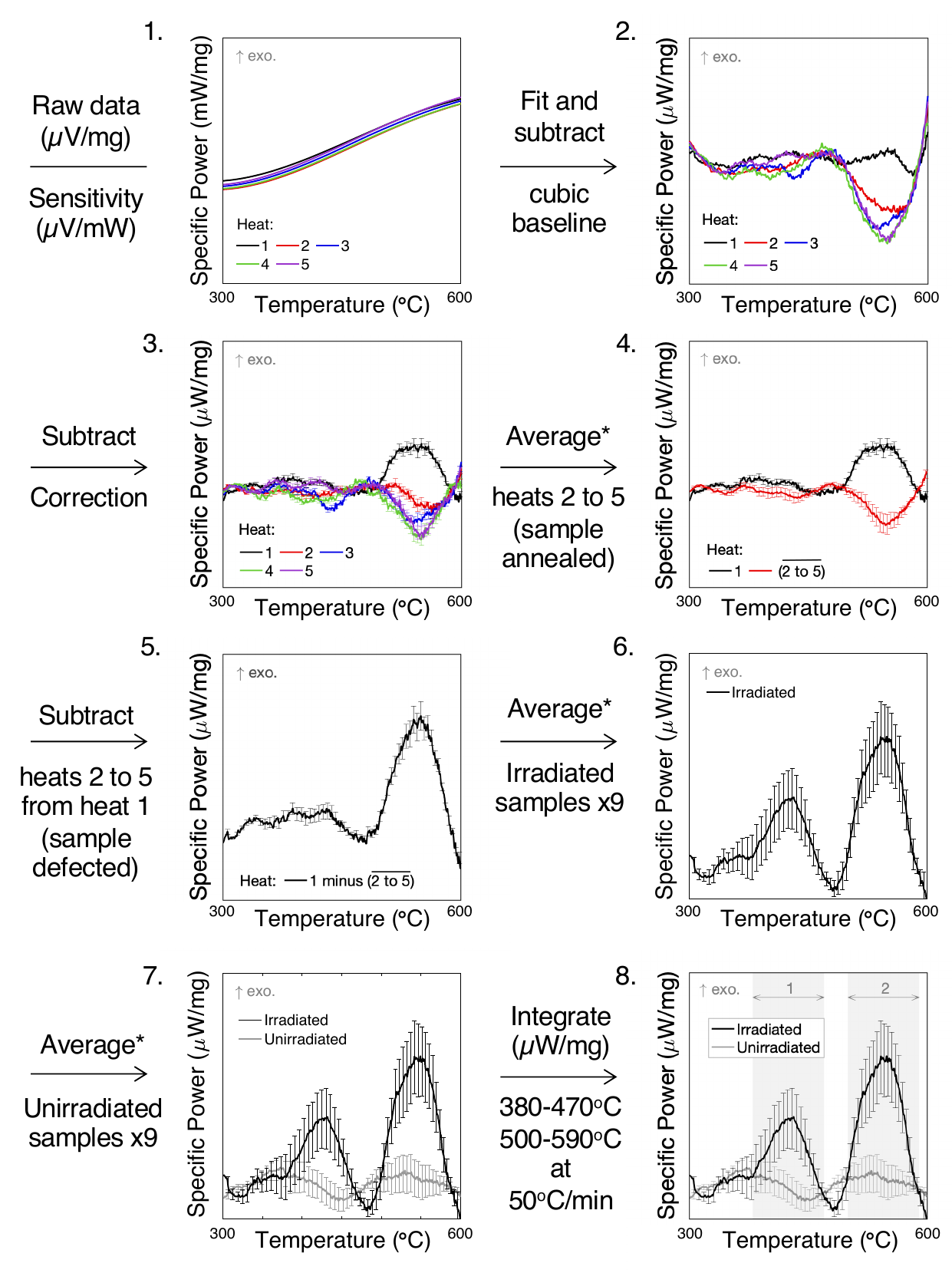}
\vspace{10pt}
\caption{\textbf{DSC data is analysed to convert from raw data in $(\frac{\text{\textmu} \text{V}}{\text{mg}})$ to an energy density in $(\frac{\text{J}}{\text{g}})$.} $^{*}$After averaging, the standard error is propagated through all subsequent steps, contributions are summed in quadrature.
\label{dsc_analysis_fig}}
\end{figure}

\noindent In order to evaluate the stored energy released during annealing, the raw data in $(\frac{\text{\textmu V}}{\text{mg}})$ must be converted to $(\frac{\text{mW}}{\text{mg}})$ and integrated to yield an energy density in $(\frac{\text{J}}{\text{g}})$. The full analysis process is depicted in figure \ref{dsc_analysis_fig}. The steps are as follows:\\

\begin{enumerate}
\addtocounter{enumi}{-1}
    \item The raw DSC data is measured in $(\frac{\text{\textmu} \text{V}}{\text{mg}})$. This is then divided by the sensitivity in $(\frac{\text{\textmu} \text{V}}{\text{mW}})$ to yield the commonly used DSC units of $(\frac{\text{mW}}{\text{mg}})$. For additional details on the sensitivity calibration see section S\ref{sensitivity}.\\
    \item Visually inspecting the DSC data in the temperature range of interest (300$\degree$C $< T <$ 600$\degree$C) shows there are no obvious enthalpic reactions. Note that the y-axis scale is in $(\frac{\text{mW}}{\text{mg}})$. Estimating the expected peak height, assuming a stored energy density of 0.1~J/g, an energy release peak width of 50$\degree$C, and a heating rate of 50$\degree$C/min gives an expected peak height of the order $(\frac{\text{\textmu} \text{W}}{\text{mg}})$. This is impossible to resolve at the current y-axis scale and therefore the data must have a baseline fitted and subtracted.\\
    \item For each individual heating run, a cubic polynomial baseline is fitted to the data between 300--350, 450--500, and 575--600$\degree$C and then subtracted from all the data between 300--600$\degree$C. This allows comparison between subsequent runs, note that now the y-range is now in $(\frac{\text{\textmu} \text{W}}{\text{mg}})$. For additional details regarding the baseline determination see section S\ref{baseline}.\\
    \item At this stage the data still represents the signal from the crucibles, the samples, and any enthalpic effects within the sample. In order to correct for the effect of the crucibles, the signal measured from empty crucibles (subject to the same heating profile) is subtracted from the data. The error bars are $\pm$ standard error of the signal from the crucibles, which is the average of 20 heating runs. For more details about the correction procedure refer to section S\ref{correction}.\\
    \item To generate a physical baseline from the annealed sample, heating runs 2--5 are averaged. The error bars are now the summation in quadrature of a) the standard error from the averaging of the crucible corrections, and b) the standard error from the averaging of heating runs 2--5.\\
    \item The annealed baseline is then subtracted from the first heating run to determine the DSC signal arising from irreversible changes to the sample during heat 1. Exothermic peaks have become evident.\\
    \item In order to increase the signal-to-noise ratio and to confirm the reliability of the results, 9 identical samples were measured and their signals averaged. The error bars now show the summation in quadrature of the previous contributions plus the standard error arising from averaging the different samples.\\
    \item The same procedure was conducted for 9 unirradiated samples sectioned from an identical Ti nut. The unirradiated samples do not exhibit significant exothermic peaks.\\
    \item The specific power was integrated as a function of time over the temperature ranges 380--470 and 500--590$\degree$C. This yields values of the stored energy released during annealing in $(\frac{\text{J}}{\text{g}})$. The integral limits were chosen as the temperatures where the signals diverged for the irradiated and unirradiated samples.
\end{enumerate}

The uncertainty on the integral values of stored energy in figure \ref{DSC} is calculated by integrating the signal from the individual samples and calculating the standard error from the values of stored energy, integrating the signal from the correction runs and calculating the standard error of these values, and summing both these errors in quadrature.

\subsection{Sensitivity calibration} \label{sensitivity}
\begin{figure}[h!]
     \centering
     \begin{subfigure}[b]{\textwidth}
         \centering
         \includegraphics[scale=0.5]{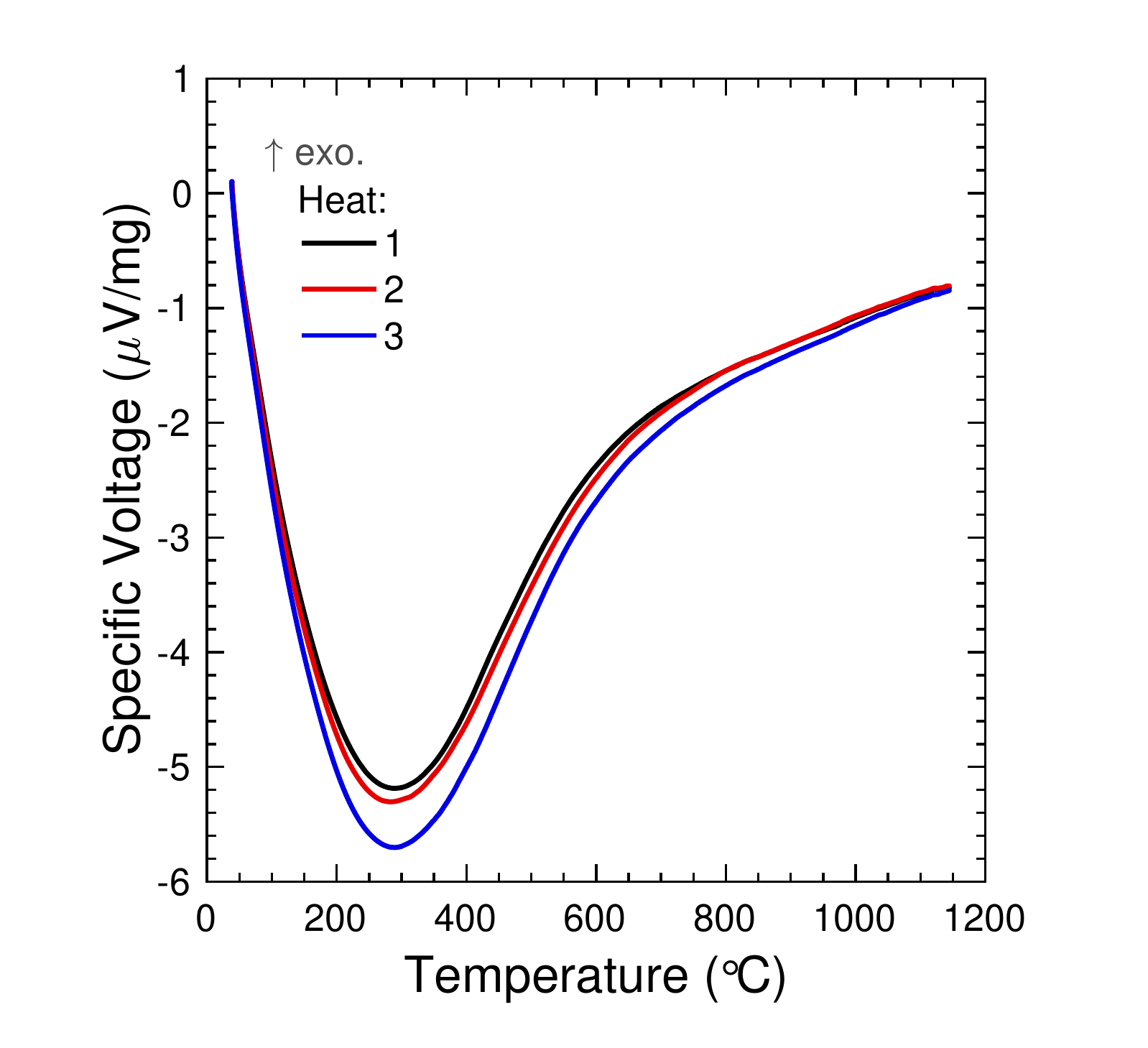}
         \caption{C$_{p}$ standard DSC measurement.}
         \label{sapph_fine}
     \end{subfigure}
     \begin{subfigure}[b]{\textwidth}
         \centering
         \includegraphics[scale=0.5]{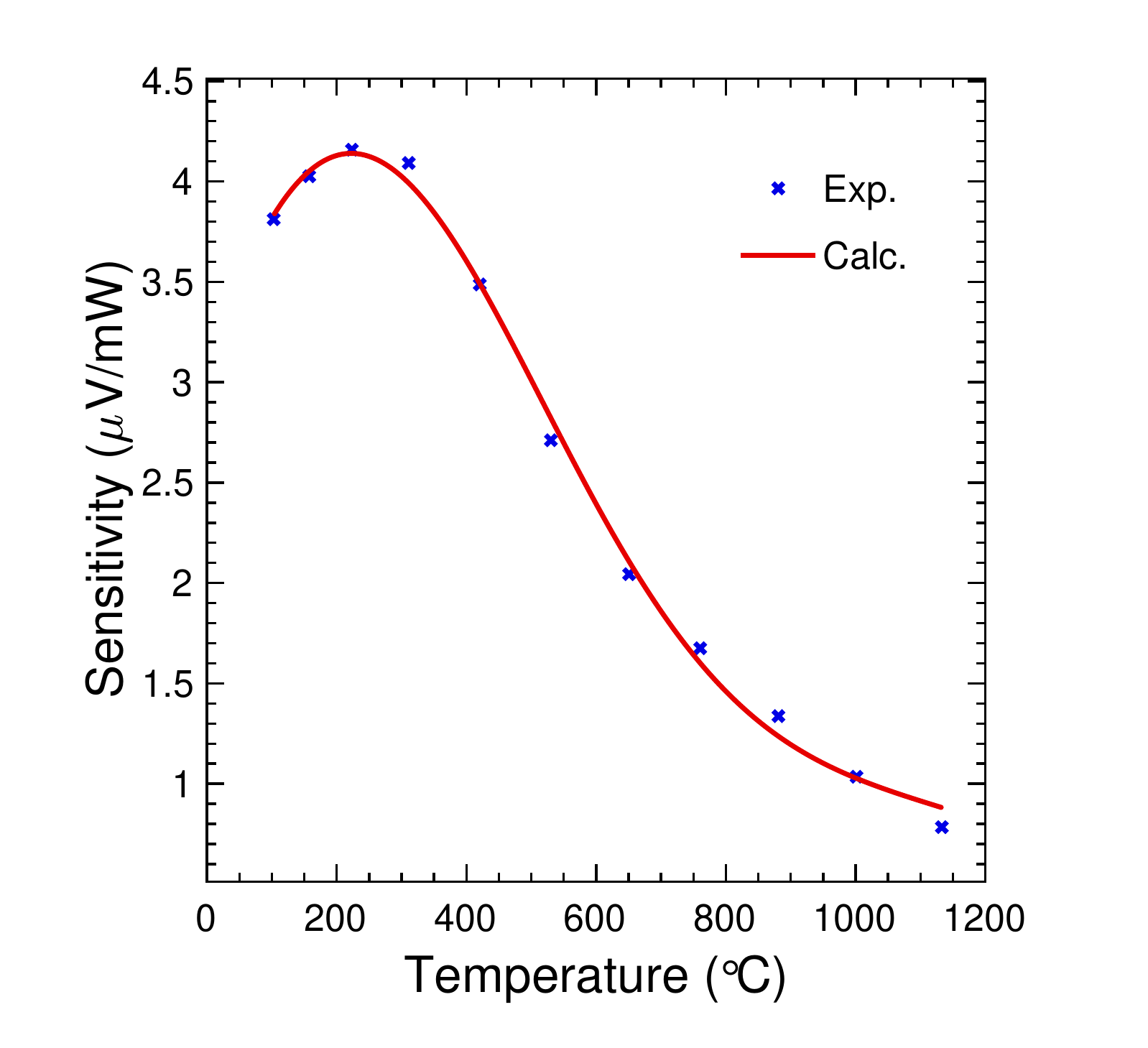}
         \caption{DSC sensitivity calibration function.}
         \label{sensitivity_fine}
     \end{subfigure}
        \caption{\textbf{Sensitivity calibration allows the conversion of measurements in $(\frac{\text{\textmu} \text{V}}{\text{mg}})$ to $(\frac{\text{mW}}{\text{mg}})$.}}
        \label{sensitivity_fig}
\end{figure}

\noindent To convert DSC data from $(\frac{\text{\textmu} \text{V}}{\text{mg}})$ to $(\frac{\text{mW}}{\text{mg}})$ the sensitivity of the DSC must be determined. This is done by heating a sapphire sample (1.0~mm thickness, 4~mm diameter) and comparing the measured signal to the reference values for heat capacity. Figure \ref{sapph_fine} shows the raw DSC data in $(\frac{\text{\textmu} \text{V}}{\text{mg}})$. At discrete points this is divided by the specific heat capacity to yield the `Exp.' sensitivity data in $(\frac{\text{\textmu} \text{V}}{\text{mW}})$ shown in figure \ref{sensitivity_fine}. In order to determine the sensitivity function across all temperatures, the data is fit to the following equations,

\begin{equation}
    \text{Sensitivity =}~f(T) = \sum_{i=1}^{4}P_{i+1}z^{i-1}e^{-z^{2}} 
\end{equation}
\begin{equation}
       z = (T-P_{0})/P_{1}
\end{equation}
\\where \textit{T} is the temperature and $P_{0-5}$ are coefficients, shown in table \ref{sensitivity_constants}. This function is plotted as the `Calc.' line in figure \ref{sensitivity_fine}. Note these coefficients are the mean of values from three sensitivity calibrations.

\begin{table}[h]
\centering
\begin{tabular}{|cccccc|}
\hline
P$_{0}$& P$_{1}$ & P$_{2}$ & P$_{3}$ & P$_{4}$ & P$_{5}$\\ \hline
223.1  & 683.4 & 4.315 & 0.174 & -5.895 & 4.706 \\ \hline
\end{tabular}
\caption{\textbf{Sensitivity calibration coefficients.}\label{sensitivity_constants}}
\end{table}

\subsection{Baseline determination} \label{baseline}
\noindent The inherent range of the DSC data in $(\frac{\text{mW}}{\text{mg}})$ is much greater than the expected peak height. To resolve these small peaks a baseline was fitted to the DSC data and subtracted. A cubic polynomial was chosen as the lowest order function to capture the macroscopic characteristics of the data and has also previously been used to model the heat capacity of Ti~\cite{Kaschnitz2001,Slezak2019}.

\begin{figure}[ht!]
     \centering
     \begin{subfigure}[b]{0.4\textwidth}
         \centering
         \includegraphics[scale=0.4]{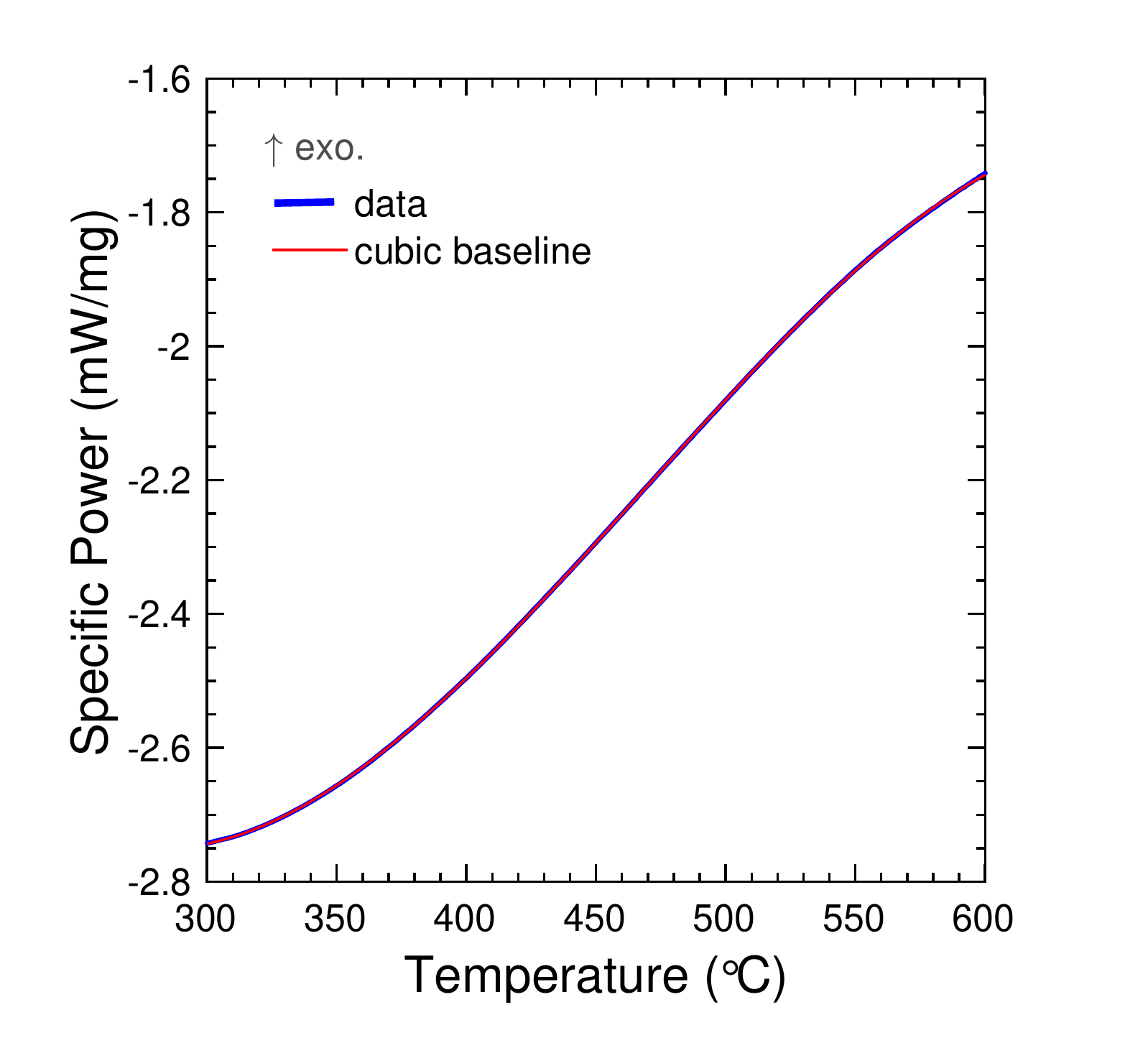}
         \caption{A baseline is fitted to all the data between 300--600$\degree$C. \vspace{8pt}}
         \label{baseline_all}
     \end{subfigure}
     \begin{subfigure}[b]{0.4\textwidth}
         \centering
         \includegraphics[scale=0.4]{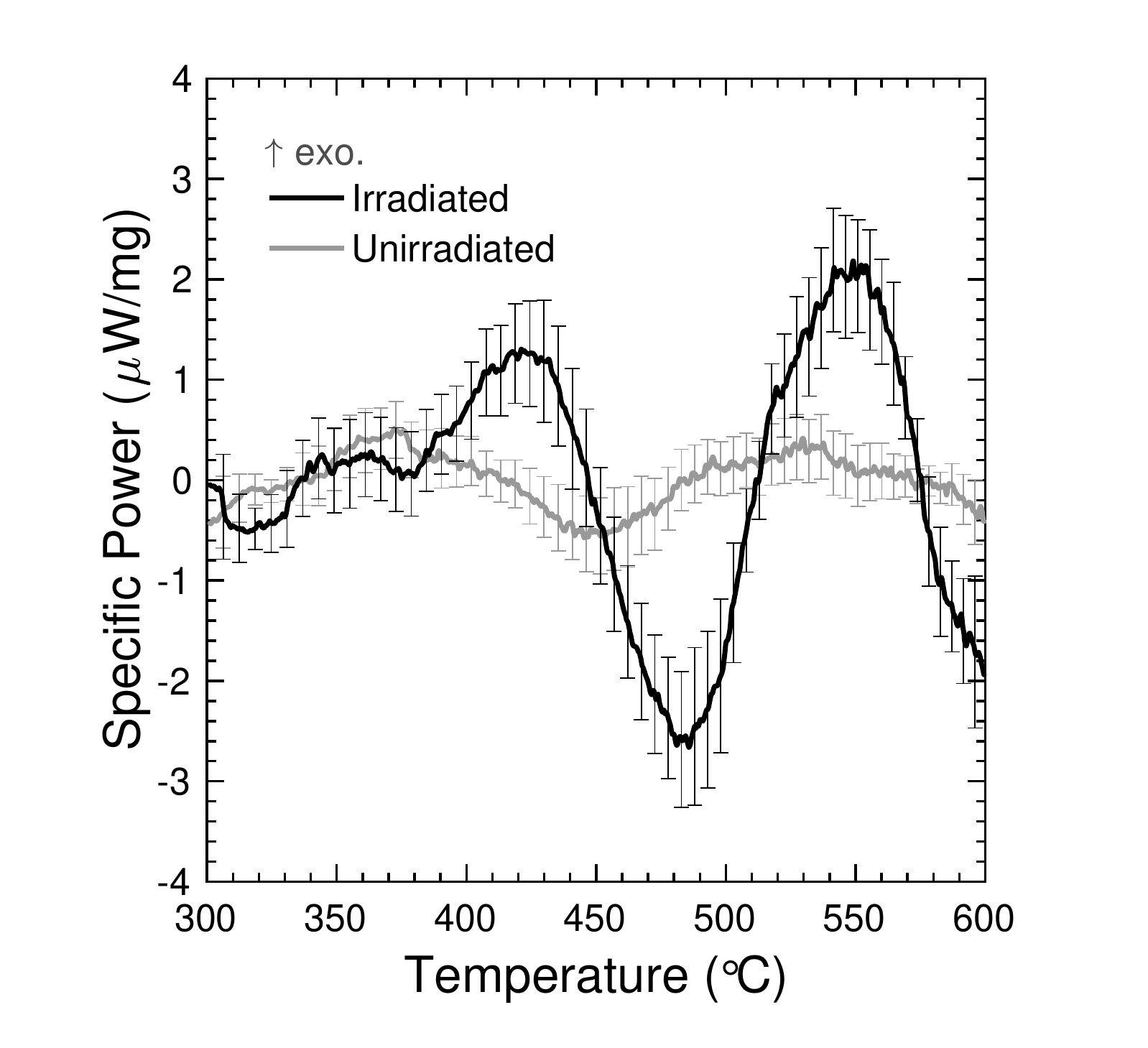}
         \caption{Fitting a baseline to data containing possible exo/endothermic peaks results in curves which are shifted in the y-direction.}
         \label{result_all}
     \end{subfigure}
     \begin{subfigure}[b]{0.4\textwidth}
         \centering
         \includegraphics[scale=0.4]{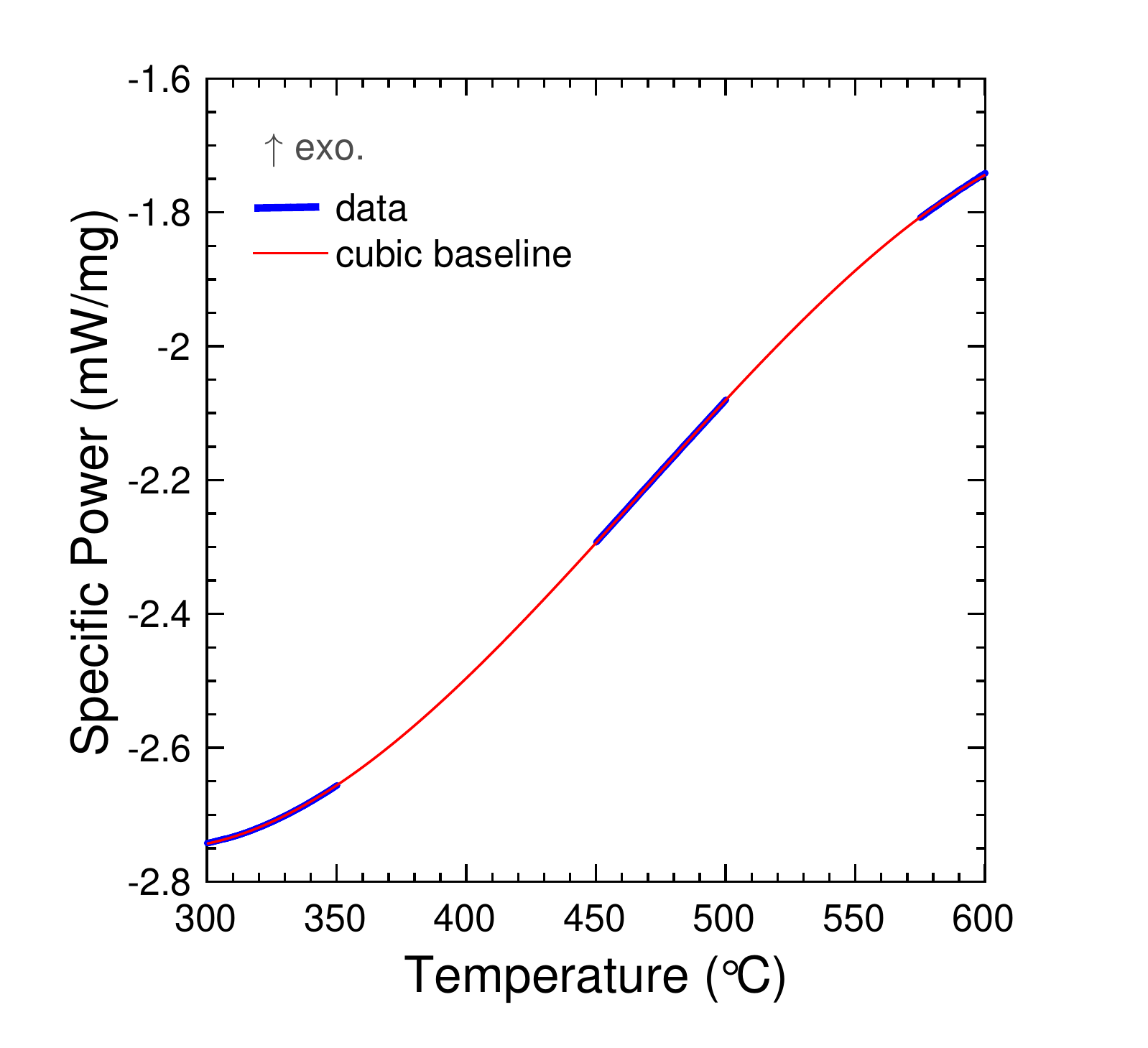}
         \caption{A baseline is fitted to the curve outside the regions of interest (ROIs), data between 300--350, 450--500, 575--600$\degree$C.}
         \label{baseline_subset}
     \end{subfigure}
     \begin{subfigure}[b]{0.4\textwidth}
         \centering
         \includegraphics[scale=0.4]{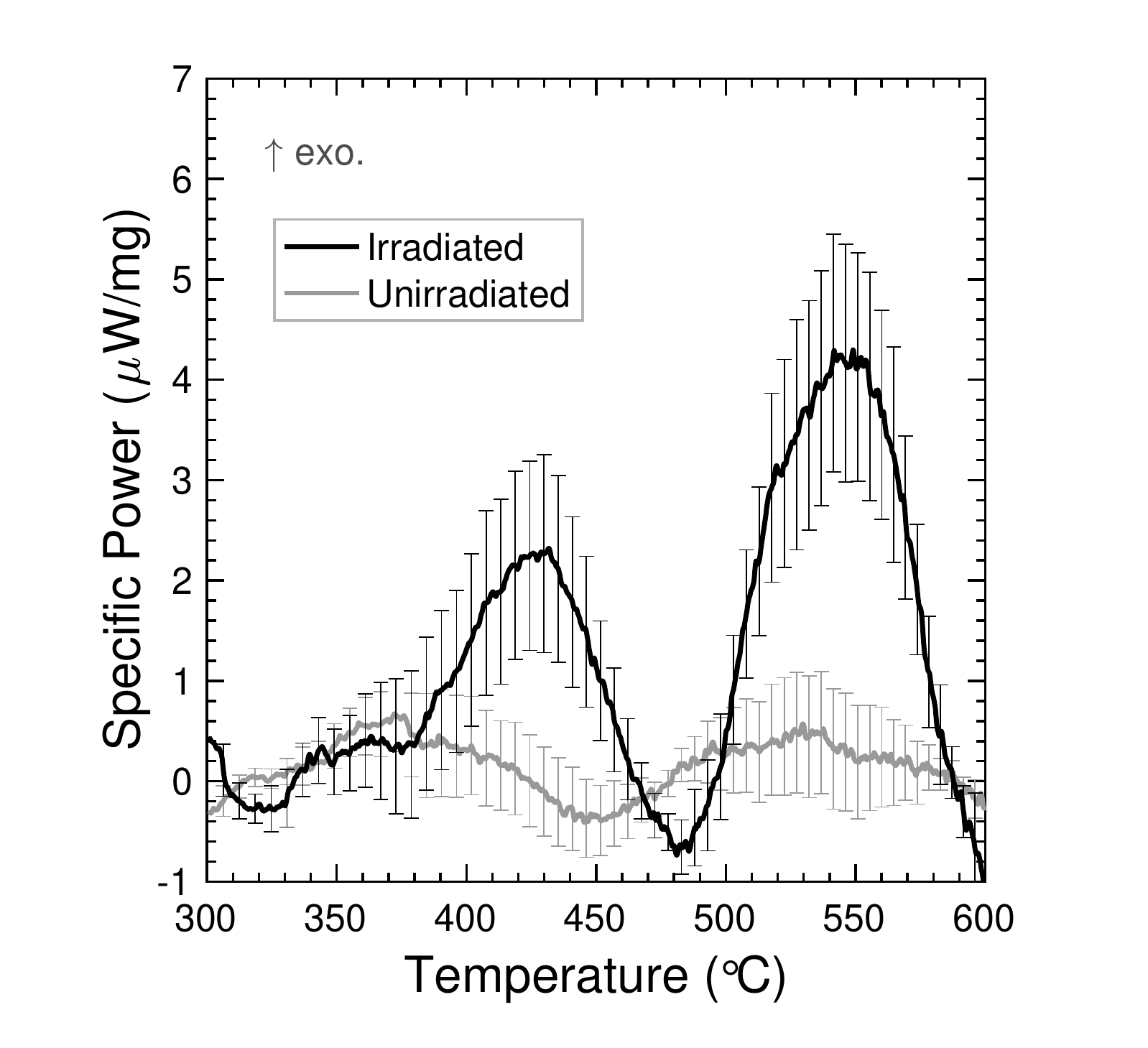}
         \caption{Excluding the ROIs from the baseline fit results in a curve where I$\approx$U$\approx$0 outside of ROIs.}
         \label{result_subset}
     \end{subfigure}
     \vspace{10pt}
        \caption{\textbf{The baseline is fit to data outside the ROIs to avoid artificially shifting the curves in the y-direction.}}
        \label{baseline_fig}
\end{figure}

Initially the baseline was fitted to all the data between 300--600$\degree$C, see figure \ref{baseline_all}. This resulted in an offset of the irradiated data, seen in figure \ref{result_all}. Considering the fitting of a polynomial function to a curve containing enthalpic peaks, the presence of peaks would result in a baseline that is displaced from the true datum. In the case of exothermic peaks, the baseline would be displaced in the positive direction which, when subtracted from the function, would lead to resultant data that is artificially displaced downwards. In order to determine the temperatures at which enthalpic reactions may occur, prior literature was consulted. PAS studies annealing neutron-irradiated Ti show a decrease in the vacancy-type defect density between 350--450$\degree$C~\cite{Hasegawa1982}. Additionally, previous DSC studies of cold-work recovery in Ti report dislocation recovery occurring at 530$\degree$C~\cite{Prabha2013}. Therefore to obtain a true baseline for the DSC curves, a cubic polynomial was fit to the data between 300--350, 450--500, 575--600$\degree$C, see figure \ref{baseline_subset}. This results in processed data in which the irradiated and unirradiated curves match each other outside the regions of interest, and also approximately equal zero specific power, see figure \ref{result_subset}.

\subsection{Crucible correction} \label{correction}
\begin{figure}[ht!]
     \centering
     \begin{subfigure}[b]{\textwidth}
         \centering
         \includegraphics[scale=0.4]{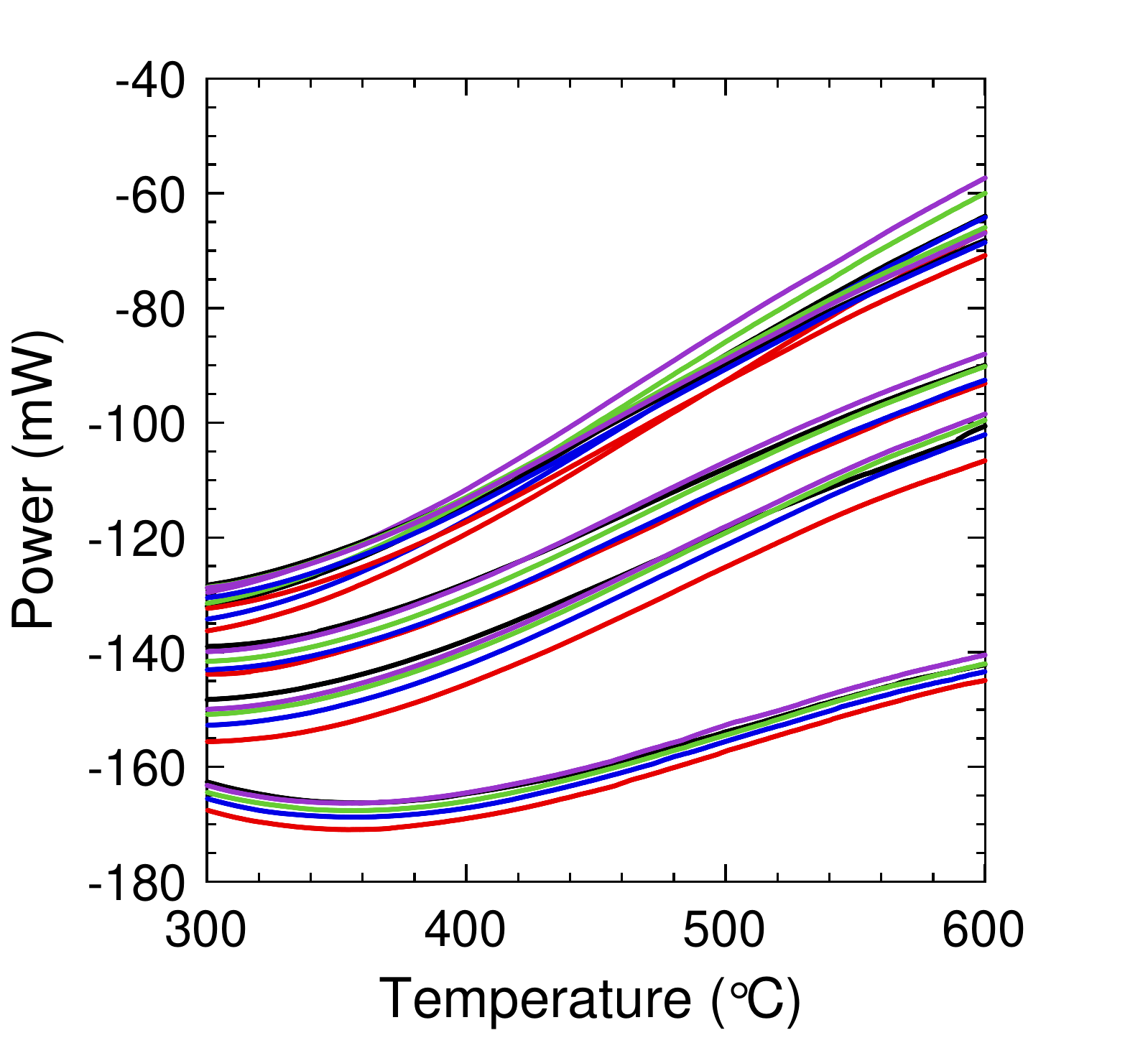}
         \caption{Empty crucibles are subject to the same heating profile as the samples.}
         \label{corr_raw}
     \end{subfigure}
     \begin{subfigure}[b]{\textwidth}
         \centering
         \includegraphics[scale=0.4]{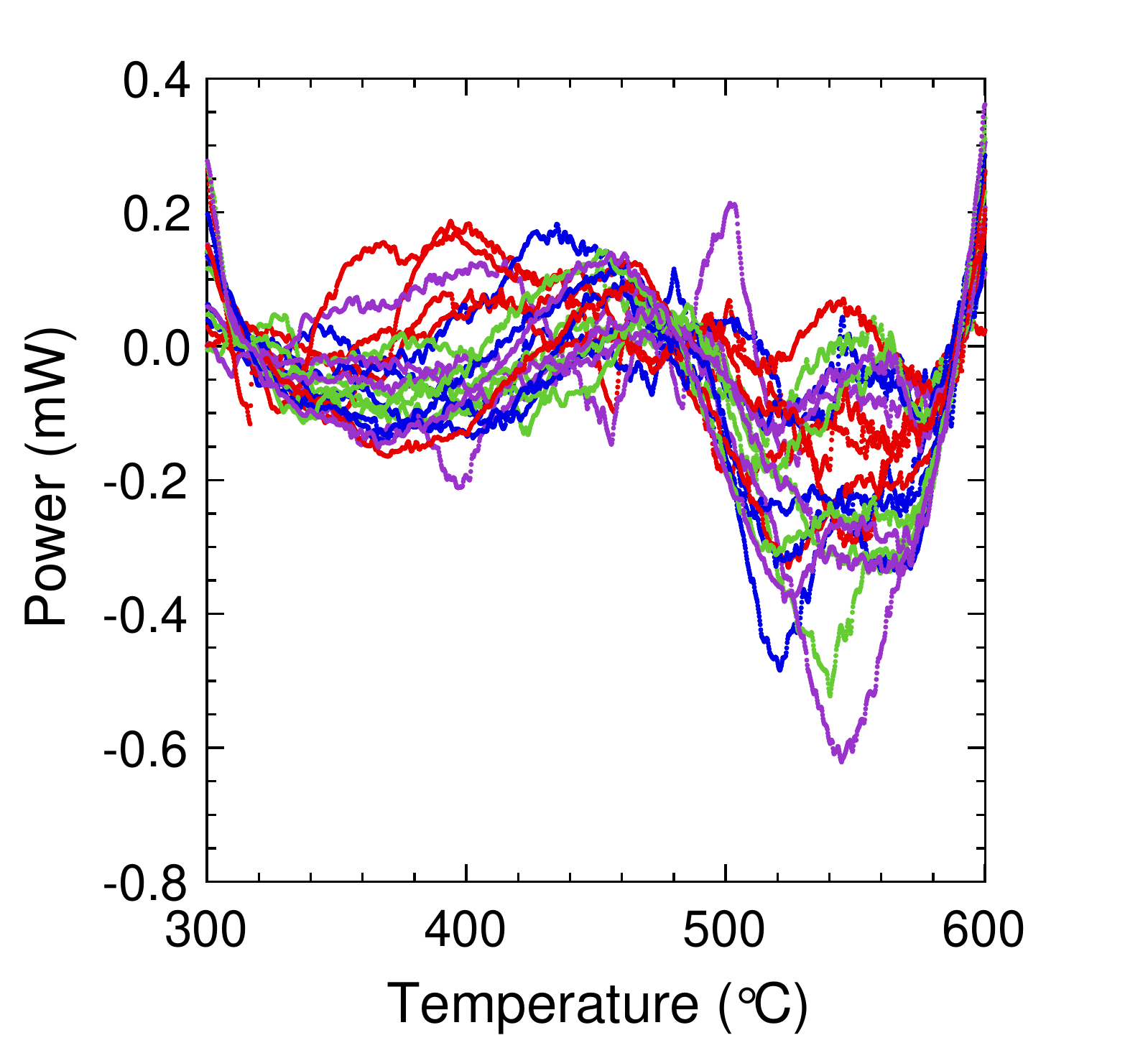}
         \caption{The data is then analysed in the same way as the sample runs.}
         \label{corr_sub}
     \end{subfigure}
     \begin{subfigure}[b]{\textwidth}
         \centering
         \includegraphics[scale=0.4]{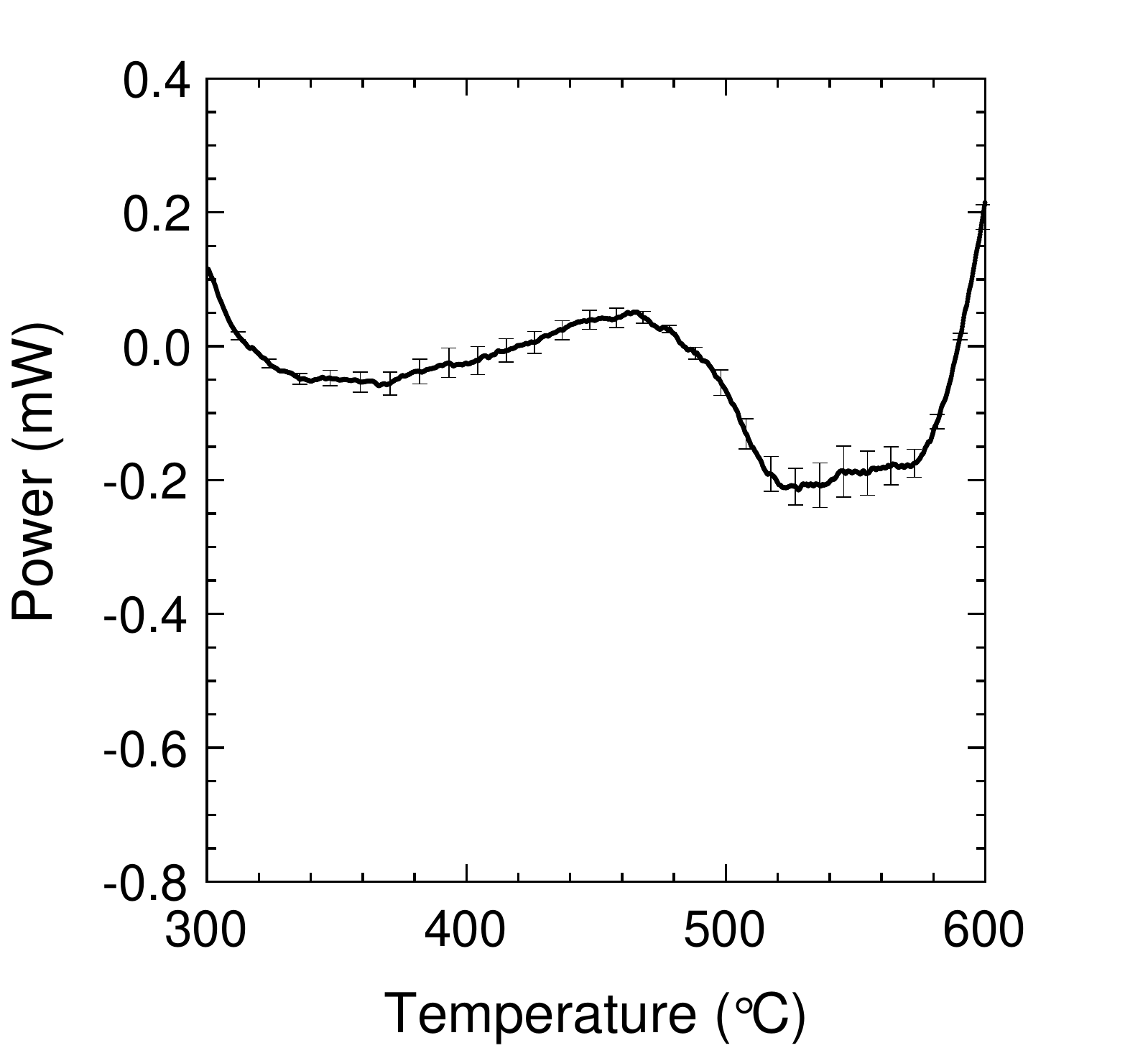}
         \caption{The signal arising from the crucibles can then be subtracted from the samples.}
         \label{corr_avg}
     \end{subfigure}
     \vspace{10pt}
        \caption{\textbf{Correction runs allow the effect of the crucibles and instrument to be subtracted from the data.}}
        \label{correction_fig}
\end{figure}

\noindent To correct for the effect of the crucibles and instrument, empty crucibles were heated through the same temperature profile as the samples (so-called `correction' runs). The signal was subject to the same baseline process, and then was subtracted from the sample data. In DSC artefacts are common; these can arise from a slight shifting of the crucibles which changes the thermal contact between the crucibles and the sensor. Therefore, in order to ensure that such artefacts were not introduced into the sample data, 20 corrections runs were conducted and then averaged before subtraction. This is shown in figure \ref{correction_fig}.

\subsection{$\alpha/\beta$ calibration} \label{alpha/beta}
\begin{figure}[ht!]
     \centering
     \begin{subfigure}[b]{0.4\textwidth}
         \centering
         \includegraphics[scale=0.4]{heatprofile.pdf}
         \caption{DSC heating profile. Heats 5--8 undergo $\alpha/\beta$ transition. \\ T$_{ReX}$ = recrystallisation temperature.}
         \label{supp_heatprofile}
     \end{subfigure}
     \begin{subfigure}[b]{0.4\textwidth}
         \centering
         \includegraphics[scale=0.37]{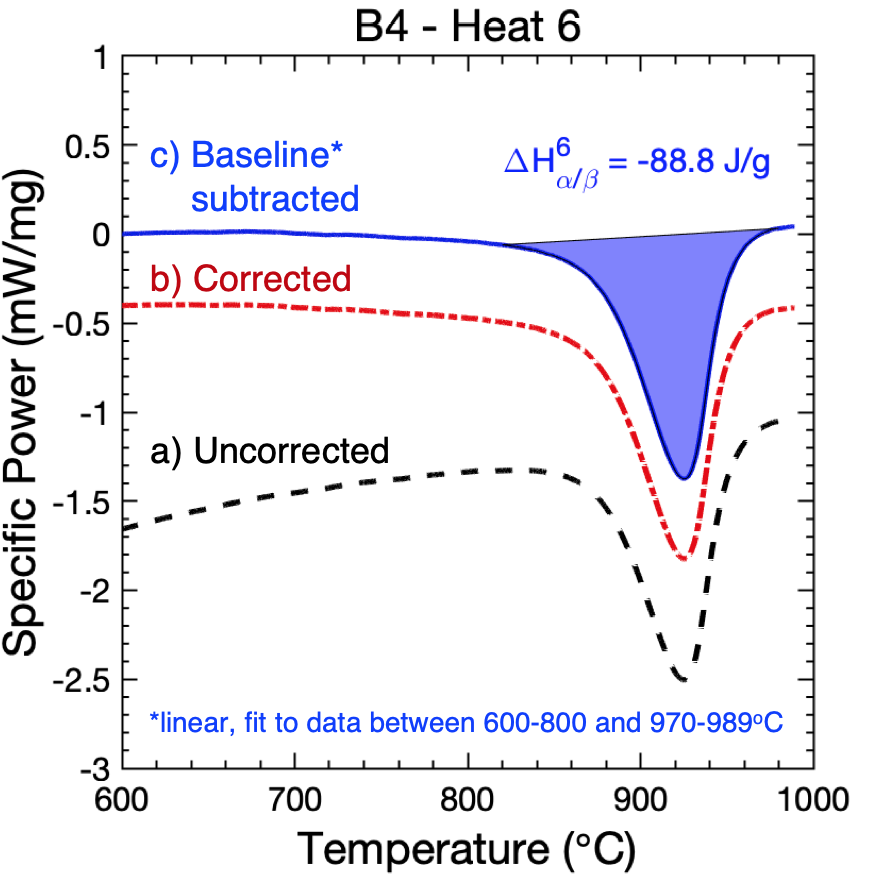}
         \caption{Data analysis process for $\Delta H_{\alpha/\beta}$.\vspace{9pt}}
         \label{B4_6}
     \end{subfigure}
     \begin{subfigure}[b]{0.4\textwidth}
         \centering
         \includegraphics[scale=0.4]{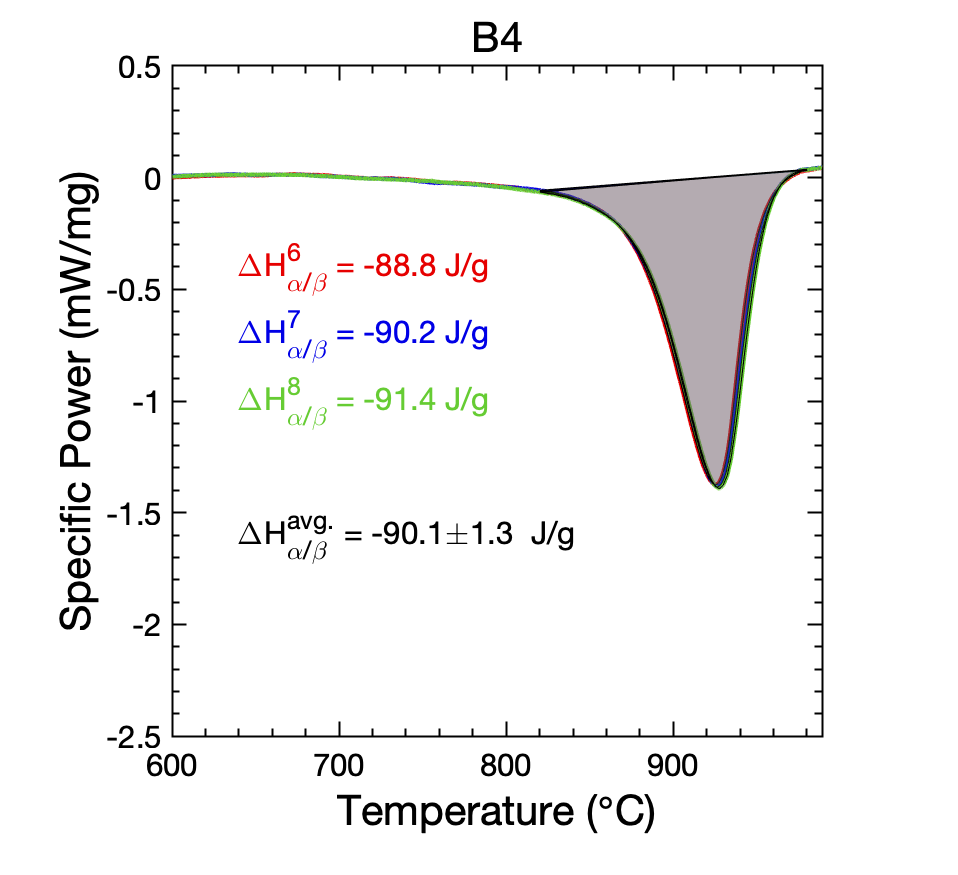}
         \caption{Heats 6--8 are averaged to determine $\Delta H^{avg.}_{\alpha/\beta}$. \\ Instrument reliability heat-heat is confirmed.}
         \label{B4_678}
     \end{subfigure}
     \begin{subfigure}[b]{0.4\textwidth}
         \centering
         \includegraphics[scale=0.4]{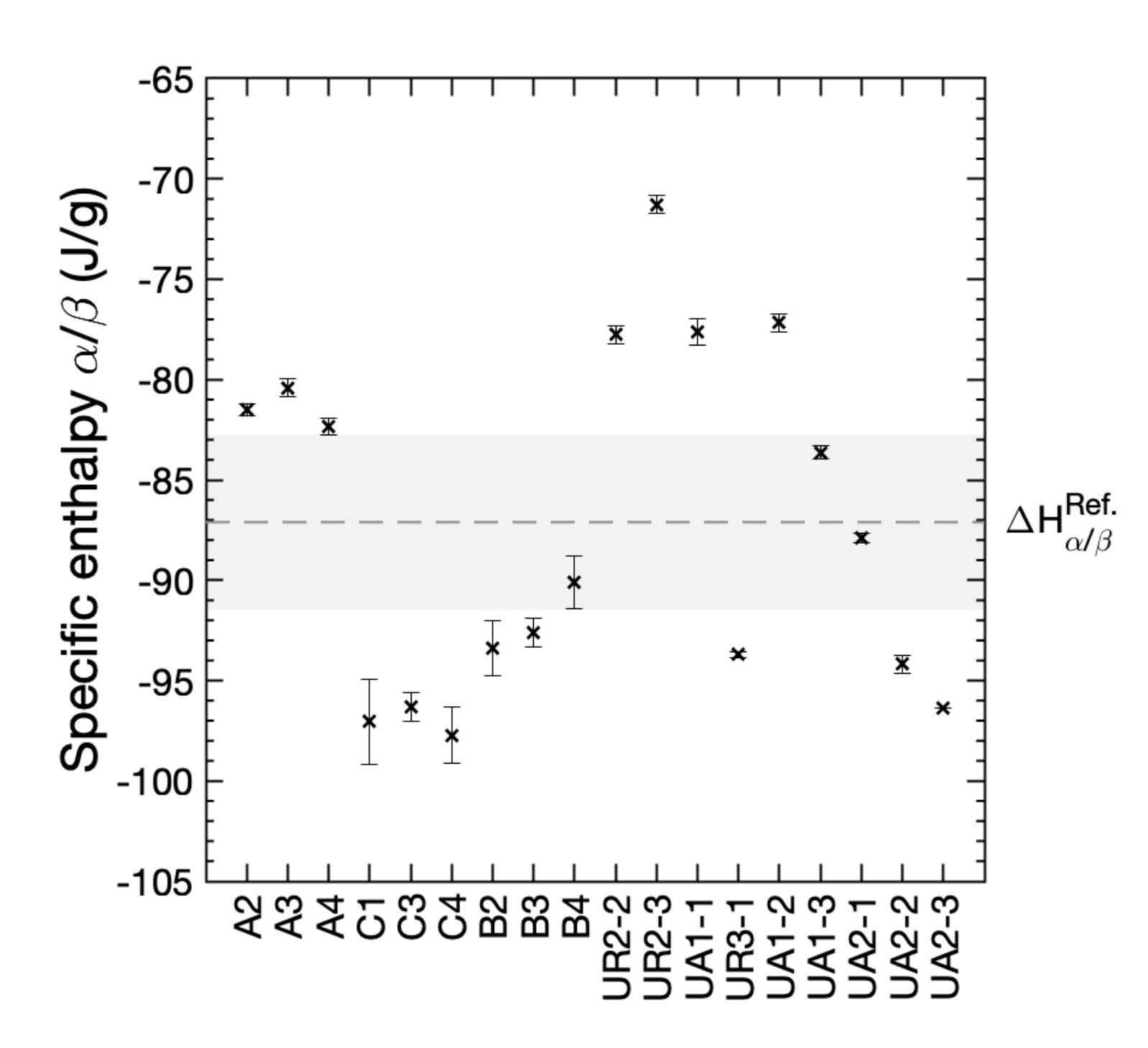}
         \caption{$\Delta H^{avg.}_{\alpha/\beta}$ for all samples in chronological order. Shows that instrument calibration does not drift.}
         \label{avg_aB}
     \end{subfigure}
     \vspace{15pt}
        \caption{\textbf{Enthalpy of $\alpha/\beta$ phase transformation is used to confirm DSC sensitivity calibration.}}
        \label{alpha/beta_fig}
\end{figure}

\noindent Sensitivity calibration of the DSC is carried out periodically. Therefore, to ensure confidence in the DSC results, experiments were designed to utilise the built-in enthalpy calibration. Ti exhibits a phase transformation from $\alpha$-BCC to $\beta$-HCP at 893$\degree$C with a specific enthalpy of $-$87.1~$\pm$~4.4~J/g~\cite{Cezairliyan1977}. After annealing the radiation damage the Ti samples were heated to 1000$\degree$C 4 times, shown in figure \ref{supp_heatprofile}. 

Heat 5 was not used for determination of the $\alpha/\beta$ enthalpy as there may be retained $\beta$-phase from processing. Heats 6 to 8 were each corrected for the effect of the crucible, and a linear baseline was fit to the data between 600--800$\degree$C and 970--990$\degree$C and subtracted. The data was then integrated between 820--980$\degree$C to determine the enthalpy of phase transformation. This process can be seen in figure \ref{B4_6}. The enthalpies from heats 6 to 8 were then averaged to improve the accuracy of the measurement. The uncertainty stated is the standard deviation of the averaged enthalpy and the small magnitude of this value confirms the reliability of the measurement. This can be seen in figure \ref{B4_678}.

The average specific enthalpy of the $\alpha/\beta$ transformation is plotted for all samples in figure \ref{avg_aB}. The nomenclature is as follows: samples A-, B-, and C- are irradiated, and samples U- are unirradiated. The samples are plotted in chronological order and it can be seen that there is no consistent drift of the values over time. This confirms that the sensitivity calibration is valid for all samples.

\subsection{Uncertainty evaluation} \label{uncertainty}
\begin{figure}[ht!]
     \centering
     \begin{subfigure}[b]{\textwidth}
         \centering
         \includegraphics[scale=0.4]{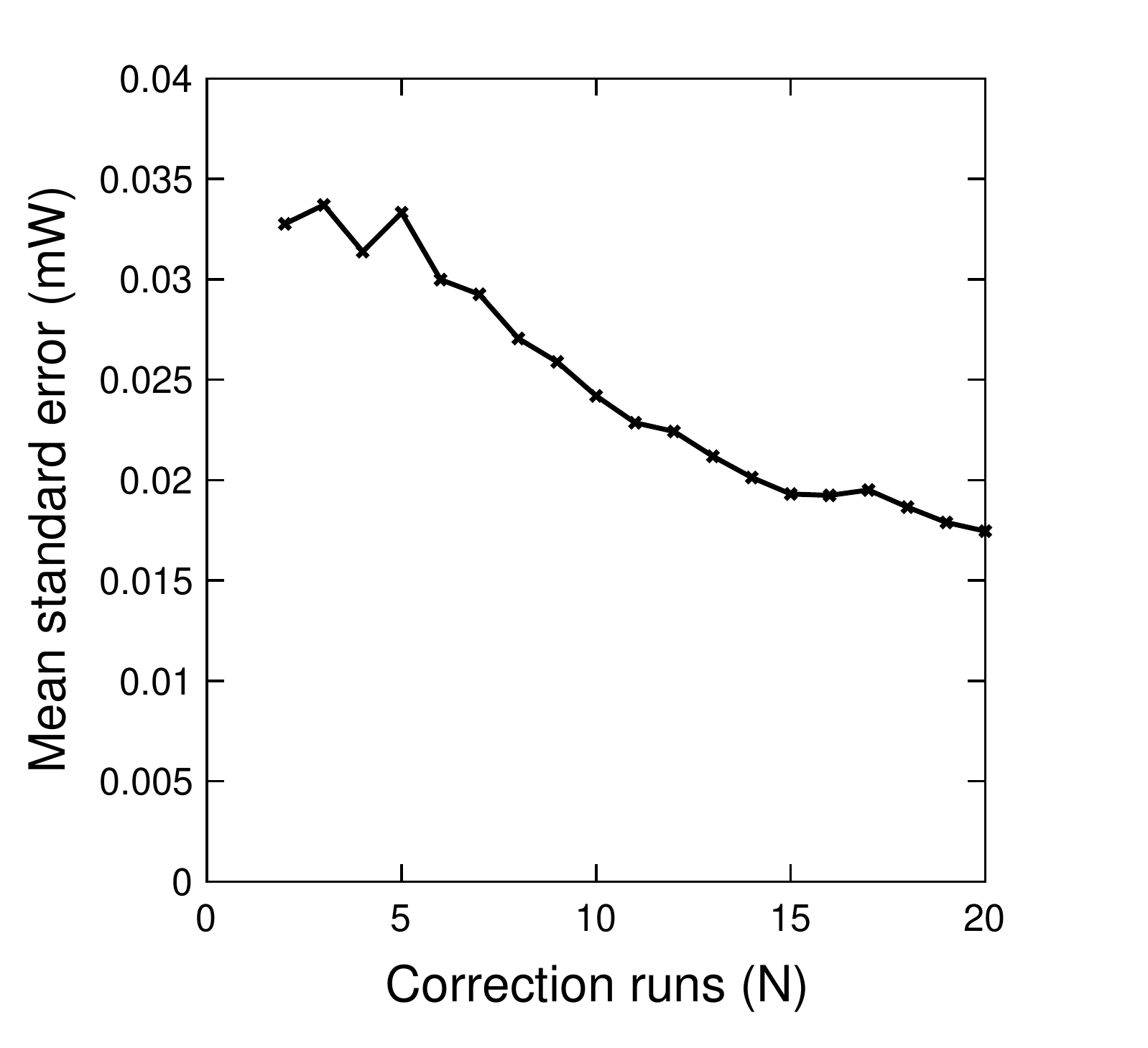}
         \caption{20 correction runs with empty crucibles were averaged.}
         \label{ste_corr}
     \end{subfigure}
     \begin{subfigure}[b]{\textwidth}
         \centering
         \includegraphics[scale=0.4]{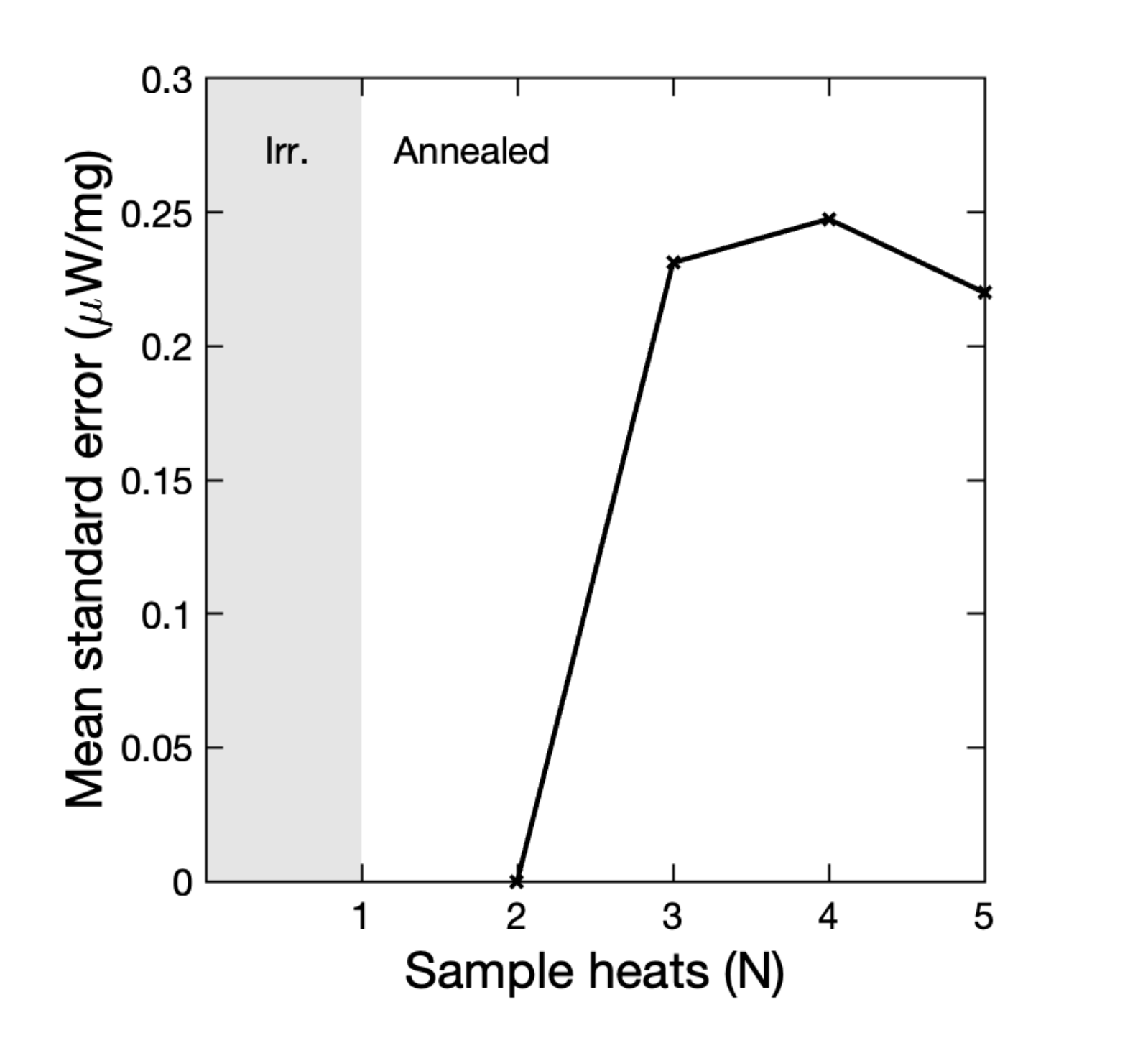}
         \caption{Heats 2--5 are averaged to form the annealed sample baseline.}
         \label{ste_heats}
     \end{subfigure}
     \begin{subfigure}[b]{\textwidth}
         \centering
         \includegraphics[scale=0.4]{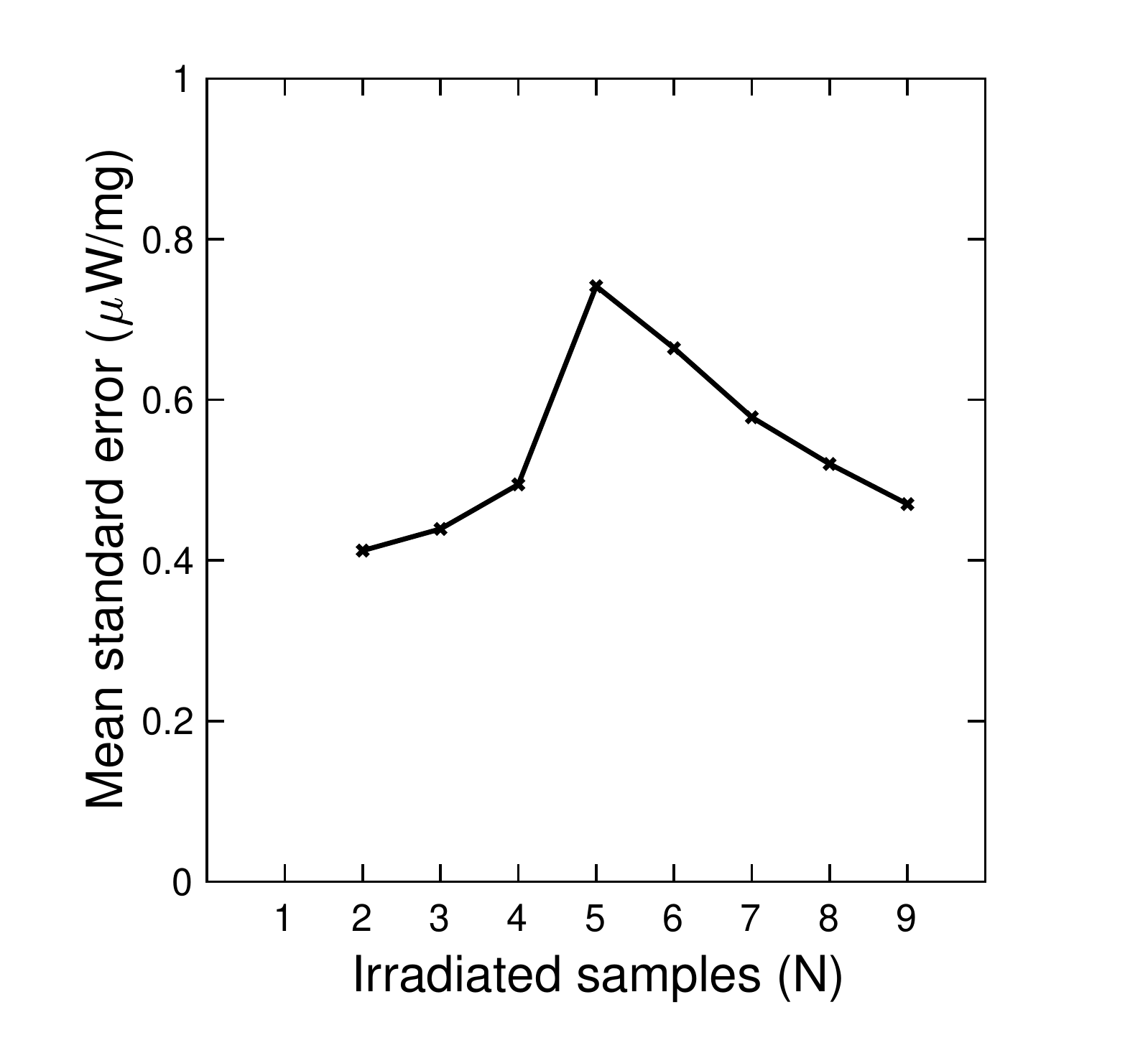}
         \caption{9 irradiated samples were averaged to increase the signal to noise ratio.}
         \label{ste_I}
     \end{subfigure}
     \vspace{10pt}
        \caption{\textbf{DSC uncertainty is reduced by averaging multiple repeats.} For all plots the standard error of the DSC signal is calculated at each temperature, as a function of the number of runs, then averaged over the full temperature range.}
        \label{repeats_fig}
\end{figure}

\noindent For the three instances where multiple runs are averaged (corrections, heats 2--5, 9 samples) the uncertainty can be evaluated as a function of iteration to determine if the number of repeats is sufficient. At each temperature, the standard error of the DSC signal is calculated as a function of the number of runs/samples. This is then averaged across the whole temperature range to give one value of standard error as a function of the population size. This is shown in figure \ref{repeats_fig}.

For the correction runs, the mean standard error starts to a plateau after 15 corrections runs. This indicates that 20 corrections is sufficient to minimise the uncertainty from this contribution. Next, analysing the mean standard error as a function of heats, the uncertainty increases from the second to third sample heat (first to second annealed heat), as expected due to the inherent variation between runs, but then is approximately constant with increasing number of sample heats. This shows that the baseline does not strongly depend on the number of heats and would not be improved with further iteration. Finally, looking at the uncertainty with respect to the number of samples averaged, initially there is an increase for 2--5 samples, followed by a decrease from 5--9 samples. The notable increase between samples 4 and 5 may be due to the 5th sample being dissimilar to previous ones, however this effect is mitigated when averaging over 9 samples. The mean standard error is still decreasing at 9 samples, but the uncertainty is approaching the magnitude of the lowest values observed. Constraints on the availability of neutron-irradiated samples present an upper limit on the number of samples that can be measured, which ultimately determined the minimum uncertainty that can be achieved.

\begin{figure}[ht!]
     \centering
     \begin{subfigure}[b]{\textwidth}
         \centering
         \includegraphics[scale=0.4]{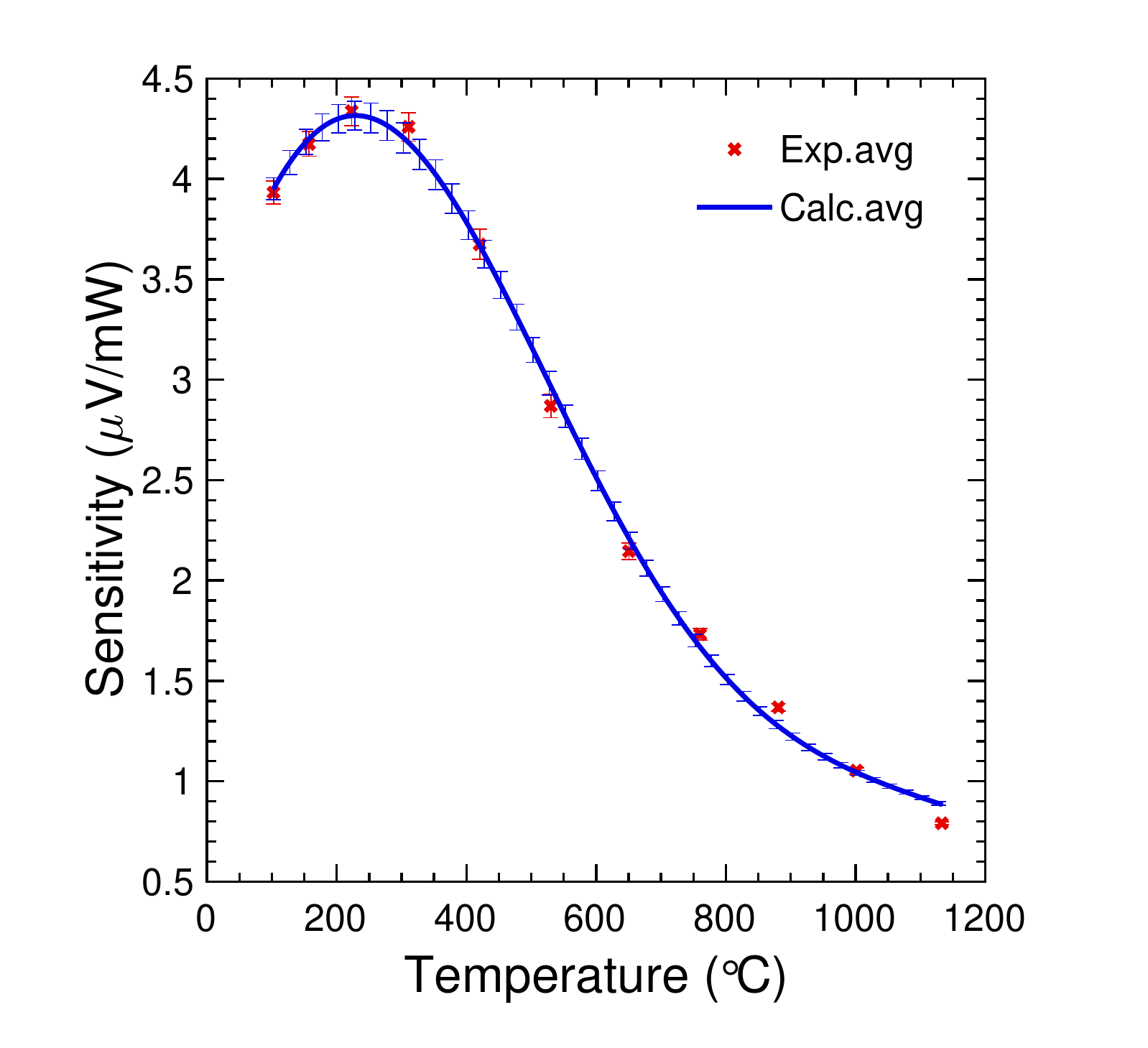}
         \caption{The sensitivity calibration is formed from the average of 3 heats. Error bars show $\pm$ the standard error.}
         \label{sensitivity_errorbars}
     \end{subfigure}
     \begin{subfigure}[b]{\textwidth}
         \centering
         \includegraphics[scale=0.4]{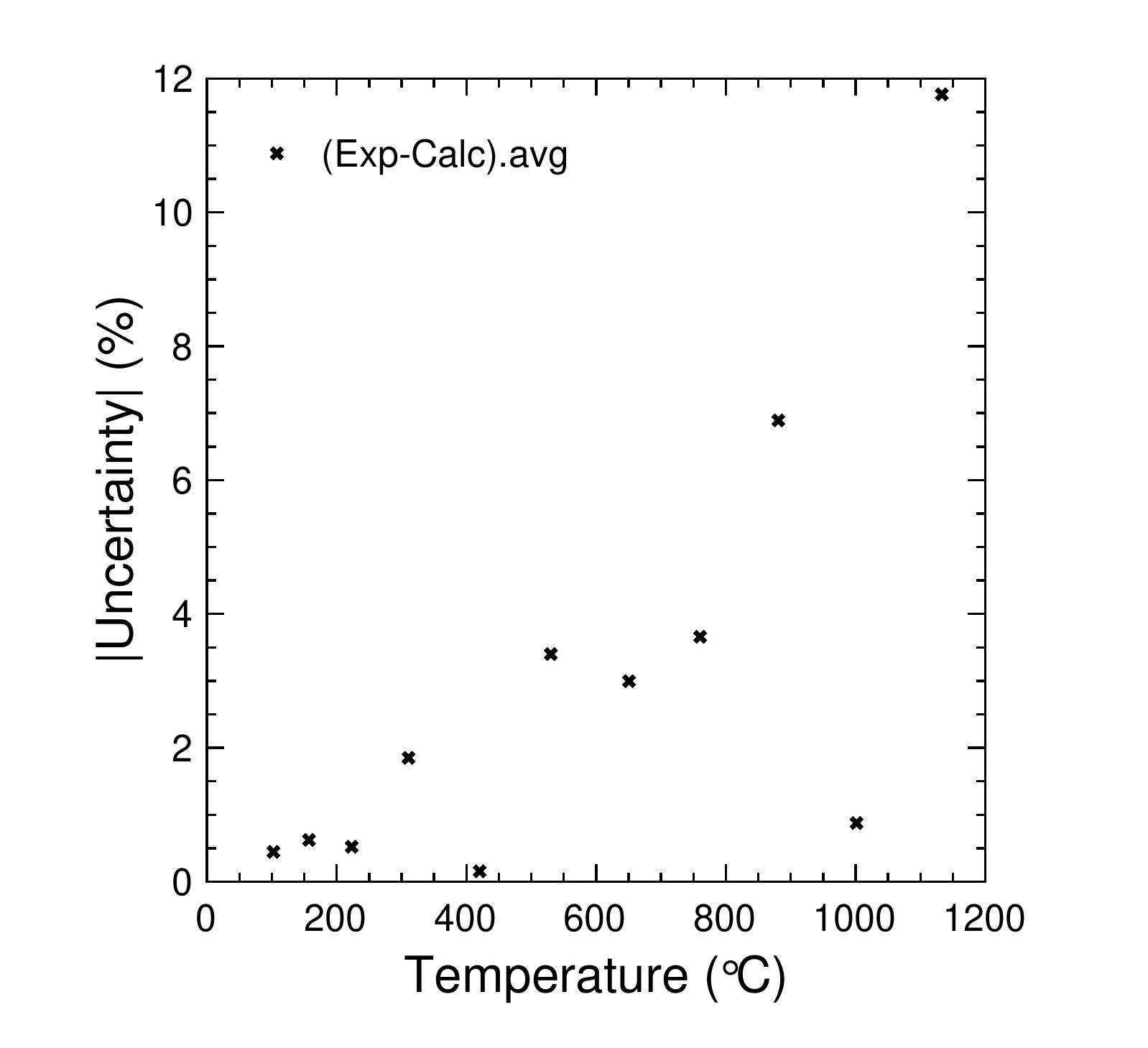}
         \caption{The maximum uncertainty of the sensitivity calibration in the temperature range of experiments is $<$8\%.}
         \label{sensitivity_uncertainty}
     \end{subfigure}
     \begin{subfigure}[b]{\textwidth}
         \centering
         \includegraphics[scale=0.4]{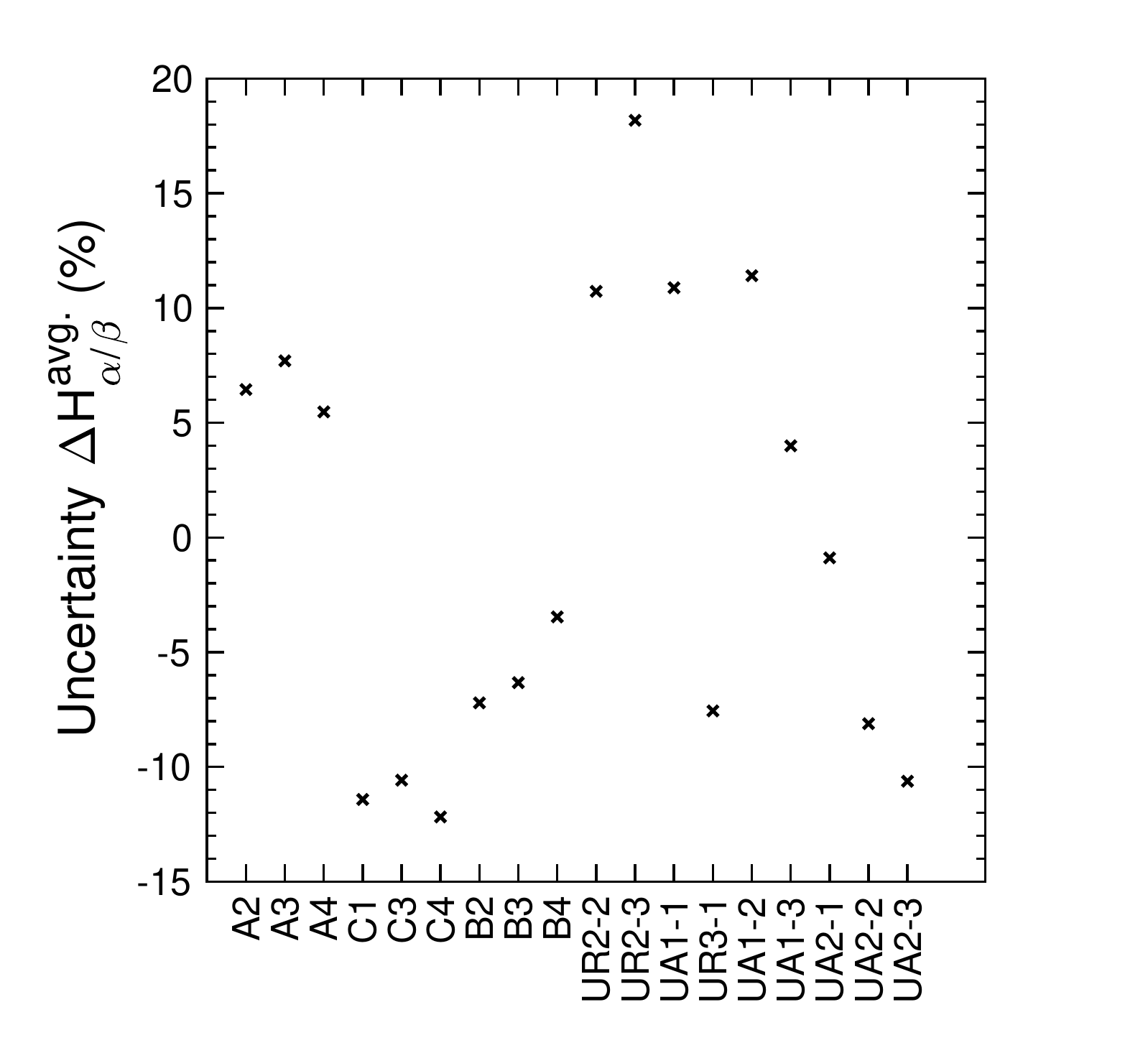}
         \caption{The $\alpha/\beta$ calibration allows the sensitivity uncertainty for \textit{each} sample to be determined. The maximum uncertainty is 19\%.}
         \label{uncertainty_aB}
     \end{subfigure}
     \vspace{10pt}
        \caption{\textbf{DSC uncertainty in sensitivity is less than 19\%.}}
        \label{calibration_uncertainty_fig}
\end{figure}

The sensitivity calibration used was the average of sensitivity functions derived from 3 consecutive heats of the sapphire standard. The mean values and variation are shown in figure \ref{sensitivity_errorbars}. The difference between the experimental data and calculated function gives a measure of the uncertainty, shown in figure \ref{sensitivity_uncertainty}. While the largest relative difference is $\sim$12\%, which is likely due to the decreasing value of sensitivity with temperature, in the temperature range of the experiments (40--1000$\degree$C) the greatest uncertainty is $\sim$8\%. 

One other evaluation of the uncertainty is the difference between the measured enthalpy of transformation $\Delta H^{avg.}_{\alpha/\beta}$ and the reference value. These are shown in figure \ref{uncertainty_aB} where it can be seen that the maximum uncertainty is 19\% (for sample `UR2-3').

\section{TEM methods \& analyses} \label{extraTEM}
\subsection{Additional images \& loop size distribution}
\begin{figure}[ht!]
     \centering
     \begin{subfigure}[b]{\textwidth}
         \centering
         \includegraphics[scale=0.6]{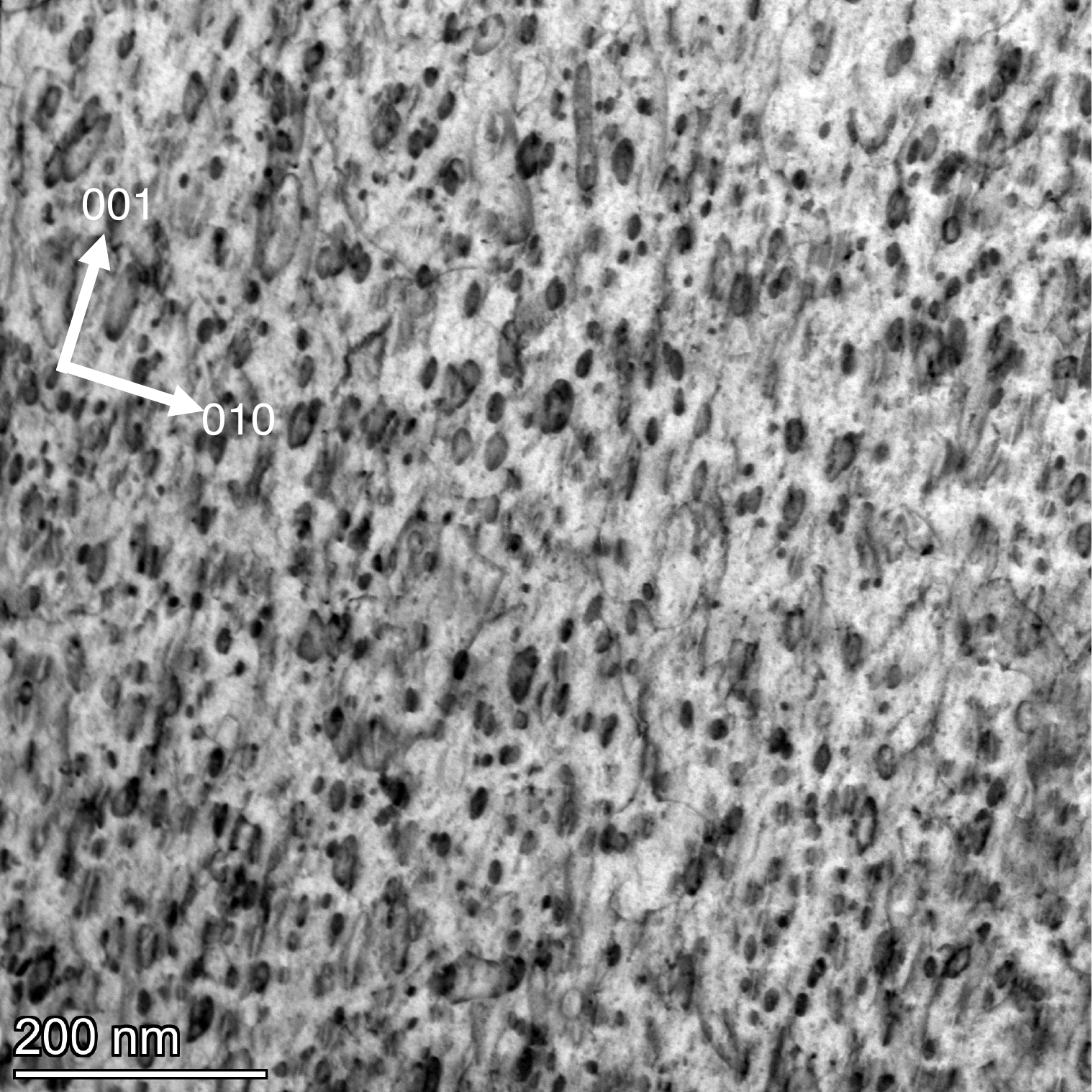}
         \caption{STEM} \vspace{10pt}
     \end{subfigure}
     \begin{subfigure}[b]{\textwidth}
         \centering
         \includegraphics[scale=0.93]{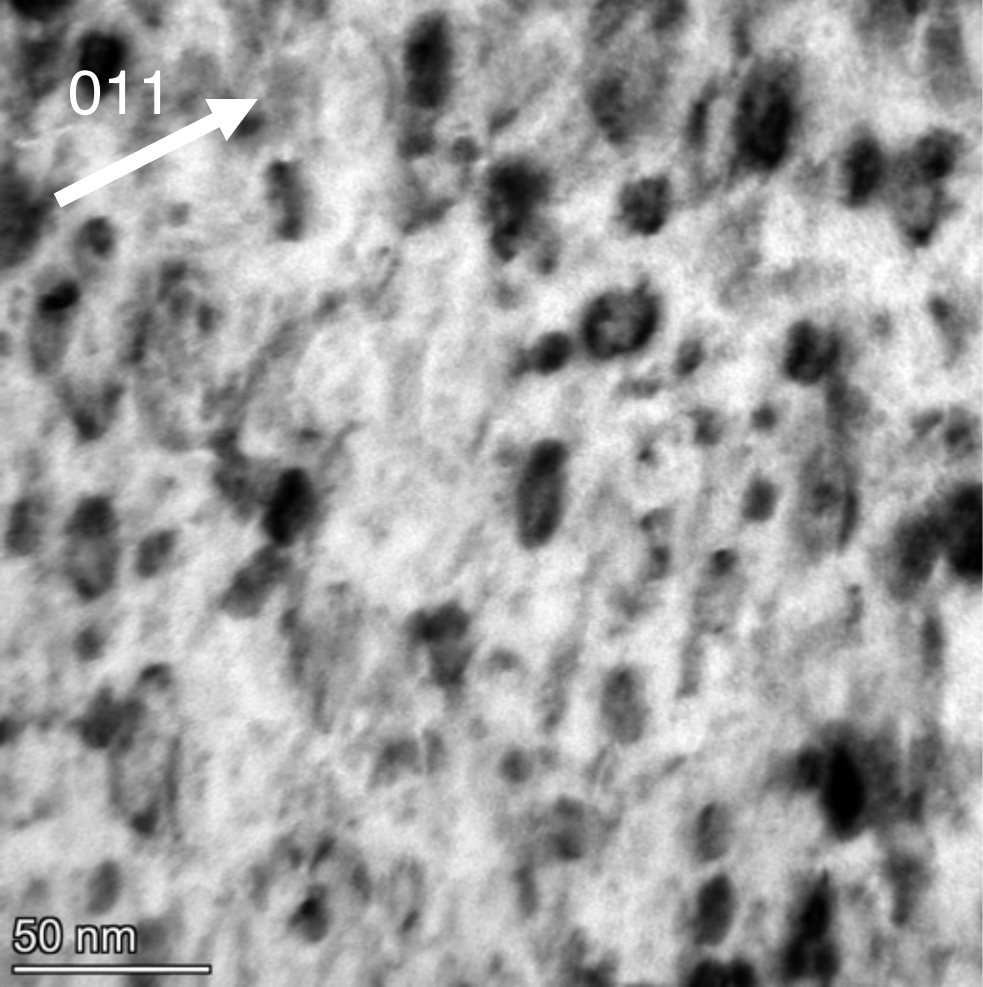}
         \caption{TEM}
     \end{subfigure}
        \caption{\textbf{Full-scale, indexed, and additional images of the as-irradiated sample.} The diffraction conditions for both images were the two-beam condition with g = 011 and zone axis = [100]. The indexed image shows that the elliptical dislocation loops have a major axis $\parallel$ to \textbf{c} and minor axis $\parallel$ to the \textbf{a} direction, in agreement with prior literature. \label{extraTEM_A}}
\end{figure}

\begin{figure}[ht] 
     \centering
     \begin{subfigure}[b]{\textwidth}
         \centering
         \includegraphics[scale=0.93]{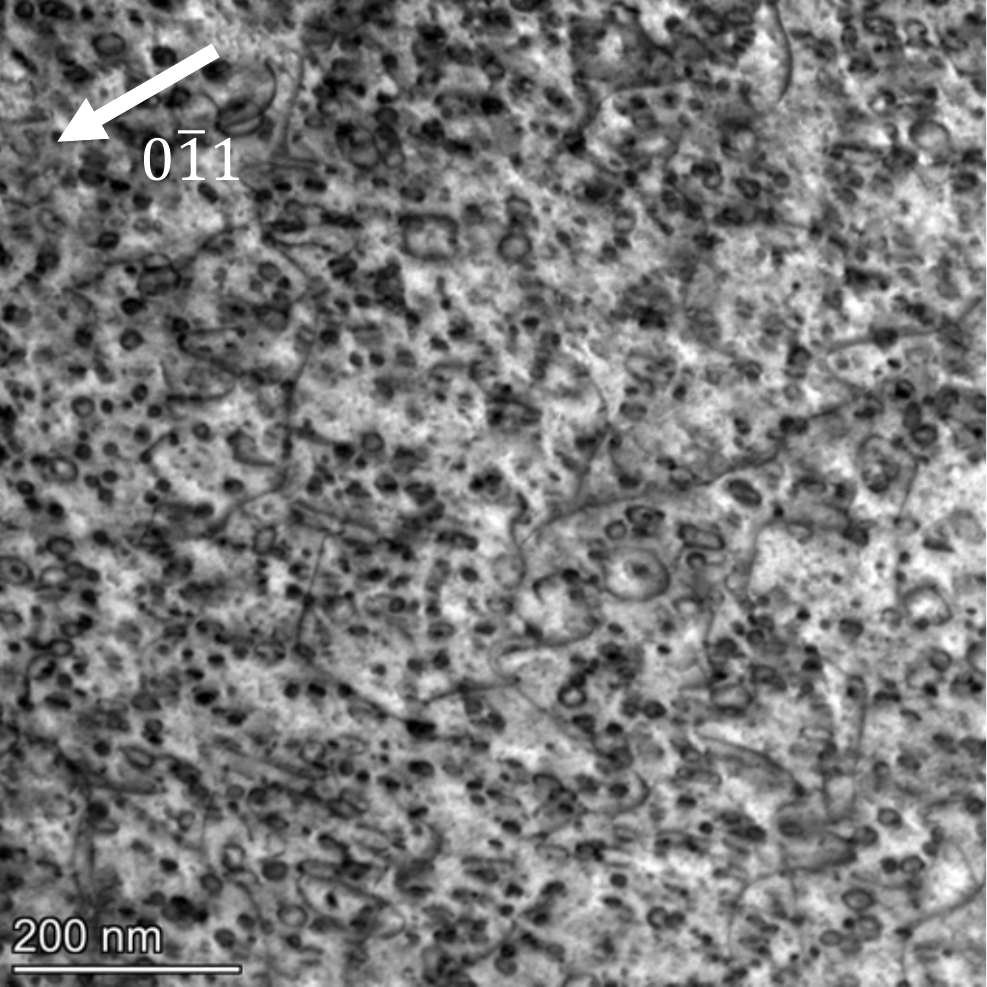}
         \caption{STEM}  \vspace{10pt}
     \end{subfigure}
     \begin{subfigure}[b]{\textwidth}
         \centering
         \includegraphics[scale=0.93]{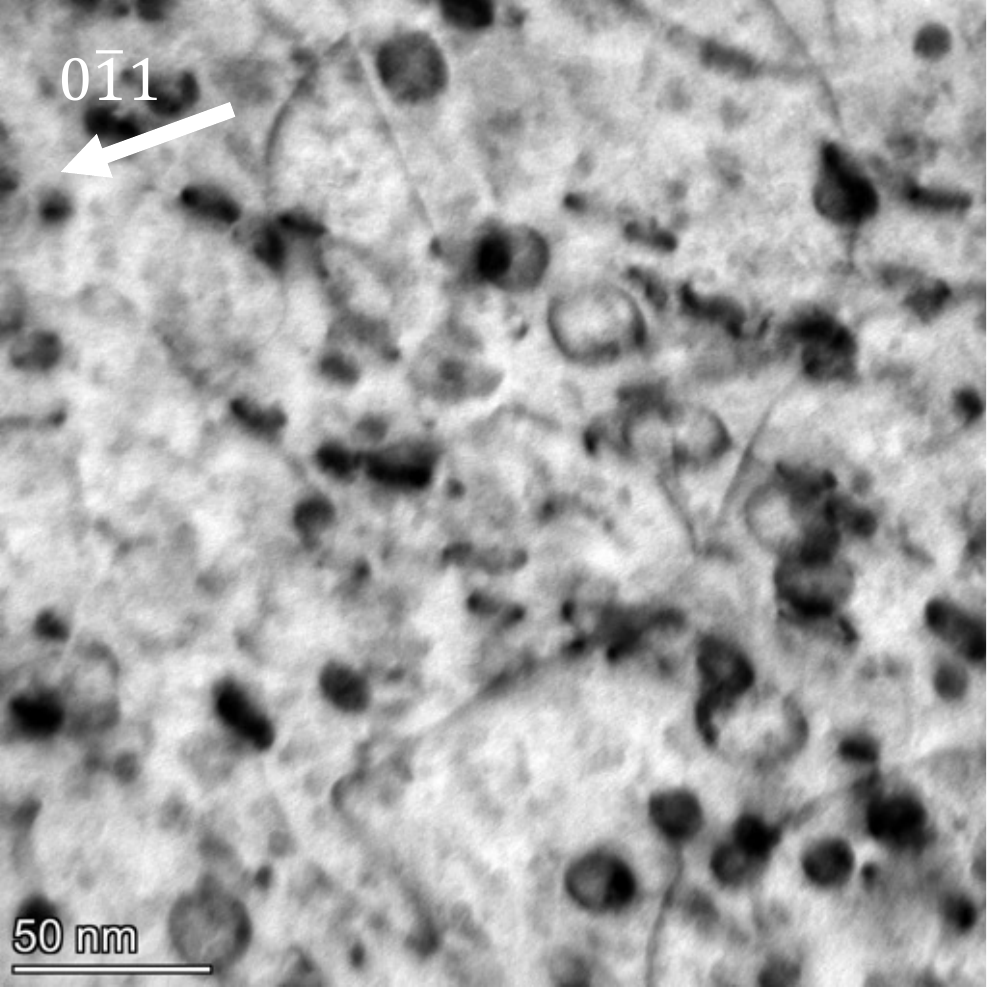}
         \caption{TEM}
     \end{subfigure}
        \caption{\textbf{Full-scale and additional images of the sample irradiated and annealed to 480$\degree$C.} The diffraction conditions for both images were the two-beam condition with g = 0$\bar{1}$1 and zone axis = [311].} \label{extraTEM_B}
\end{figure}

\begin{figure}[ht]
     \centering
     \begin{subfigure}[b]{\textwidth}
         \centering
         \includegraphics[scale=0.93]{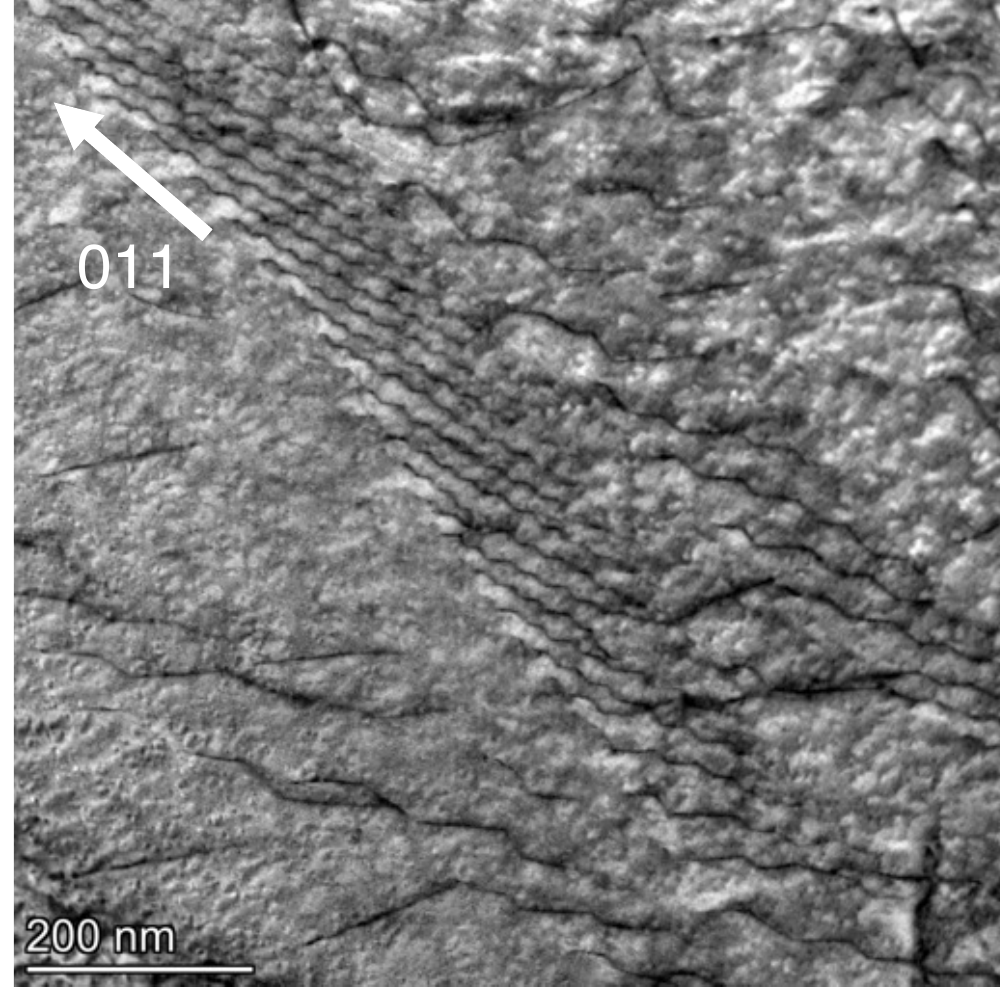}
         \caption{STEM}   \vspace{10pt}
     \end{subfigure}
     \begin{subfigure}[b]{\textwidth}
         \centering
         \includegraphics[scale=0.93]{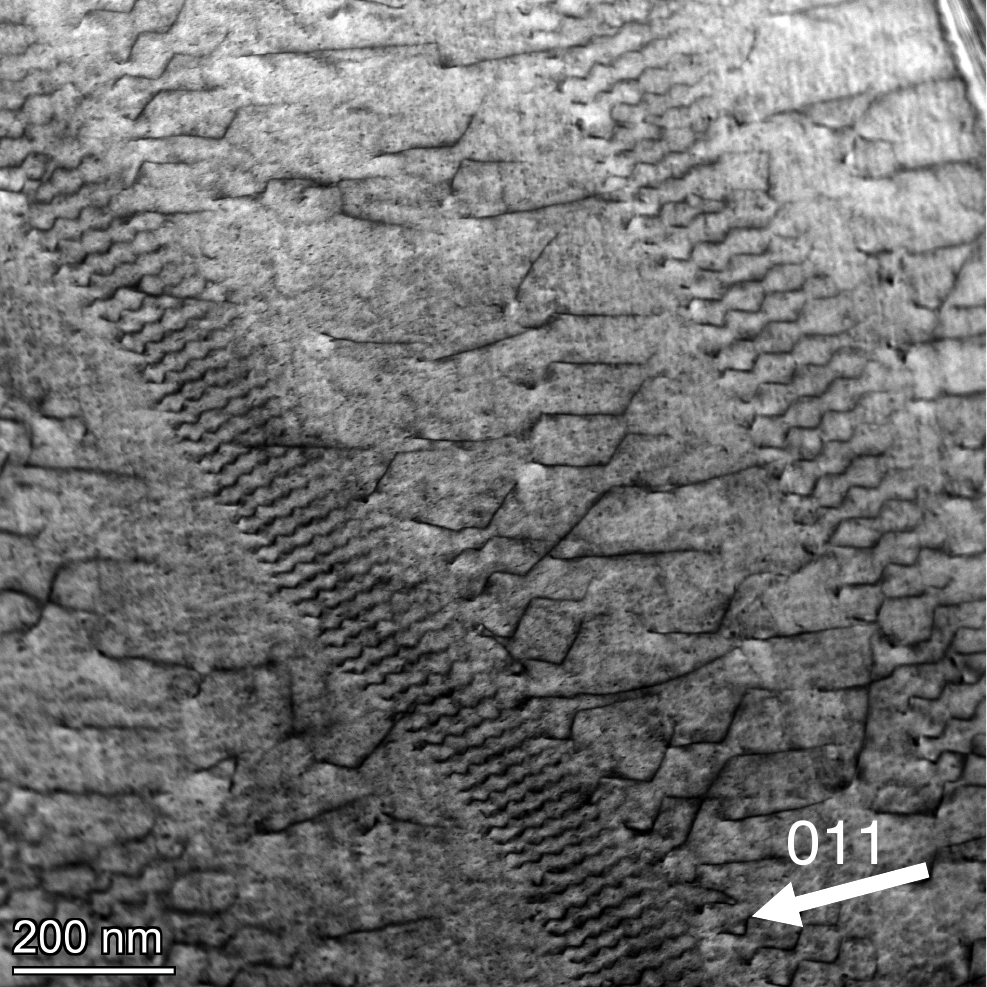}
         \caption{STEM}
     \end{subfigure}
        \caption{\textbf{Full-scale and additional STEM images of the sample irradiated and annealed to 600$\degree$C.} The diffraction conditions for both images were the two-beam condition with g = 01$\bar{1}$ and zone axis = [111].} \label{extraTEM_C}
\end{figure}

\begin{figure}[ht!]
     \centering
     \begin{subfigure}[b]{\textwidth}
         \centering
         \includegraphics[scale=0.93]{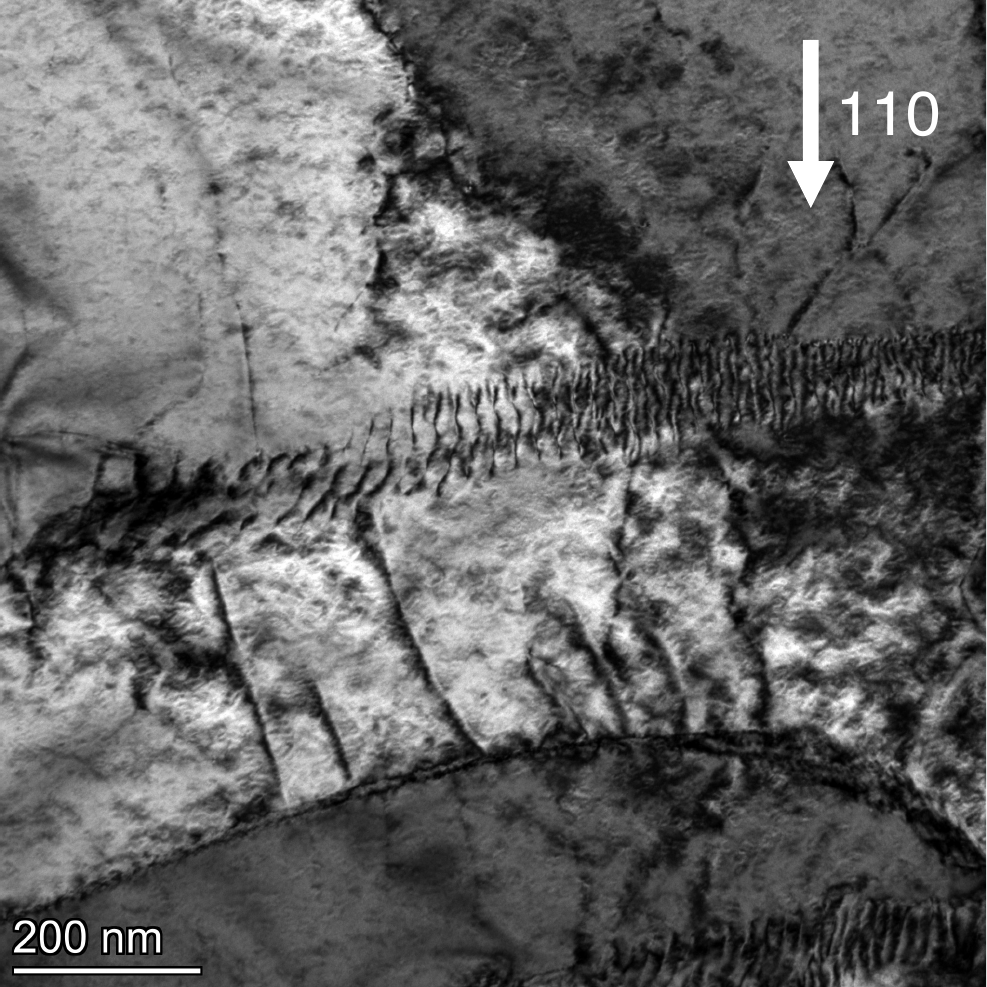}
         \caption{TEM}   \vspace{10pt}
     \end{subfigure}
     \begin{subfigure}[b]{\textwidth}
         \centering
         \includegraphics[scale=1.25]{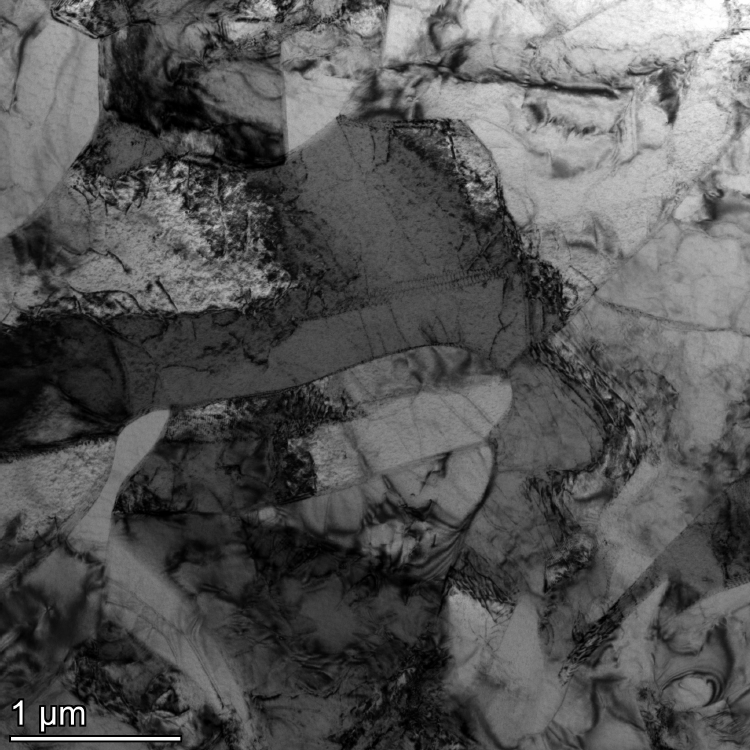}
         \caption{TEM}
     \end{subfigure}
        \caption{\textbf{Additional TEM images of the unirradiated sample.} Image a) was taken in the two-beam condition with g = 1$\bar{2}$0 and zone axis = [210]. Image b) contains several grains and thus does not have defined diffraction conditions. \label{extraTEM_U}}
\end{figure}

\begin{figure}[ht!]
    \centering
    \includegraphics[scale=0.5]{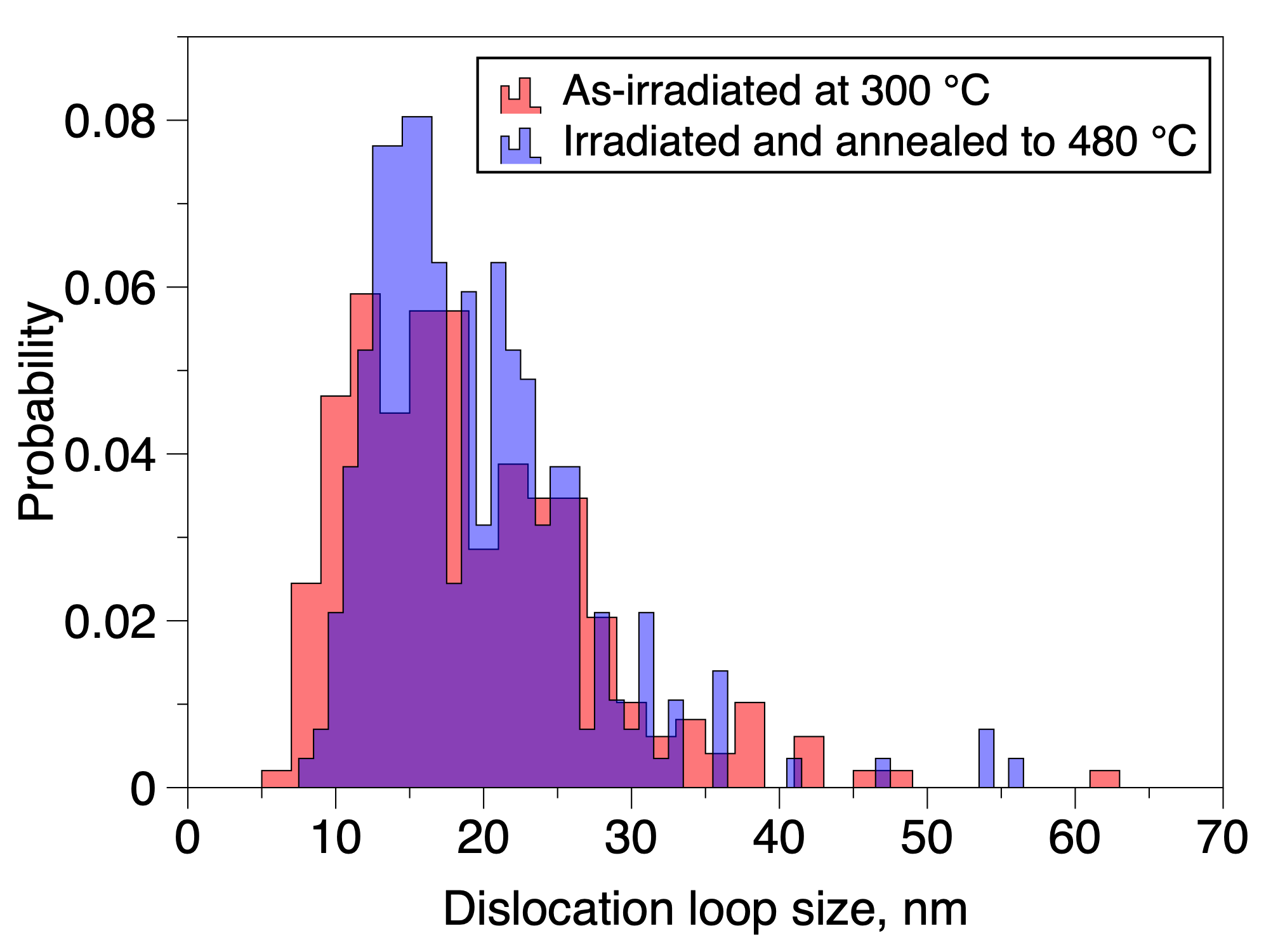}
    \caption{\textbf{The size distribution of dislocation loops in the as-irradiated sample is similar to that for the sample annealed to 480$\degree$C.} Dislocation loop size is defined as the Feret diameter measured using ImageJ, and 245 and 287 loops were measured for each sample, respectively. The mean loop diameter is 19~nm for both samples.} \label{loopSize}
\end{figure}

\noindent Additional scanning transmission electron microscopy (STEM) and TEM images can be seen in figures \ref{extraTEM_A}, \ref{extraTEM_B}, \ref{extraTEM_C}, and \ref{extraTEM_U} for the irradiated, annealed, and unirradiated samples. The as-irradiated sample and the irradiated \& annealed to 480$\degree$C sample have similar microstructures with $<$\textbf{a}$>$-type dislocation loops and linear dislocations, as shown in figures \ref{extraTEM_A} and \ref{extraTEM_B}. Furthermore, the size distribution of the dislocation loops between these samples are similar, as shown in figure \ref{loopSize}. In contrast, the irradiated \& annealed to 600$\degree$C sample exhibits a typical recovered microstructure with dislocation substructures, sporadic distribution of line dislocations and no dislocation loops. This is similar to the microstructure of unirradiated sample (see figures \ref{extraTEM_C} and \ref{extraTEM_U}). Such microstructure with dislocation substructures and linear dislocations has previously been observed in dynamically recovered hexagonal closed packed (HCP) alloys~\cite{Kombaiah2015}.

The dislocation loops have been identified as $<$\textbf{a}$>$-type by indexing the TEM images. Figure \ref{extraTEM_A} shows that the dislocation loops are elliptical with major axis $\parallel$ to \textbf{c} and minor axis $\parallel$ to the \textbf{a} axis. This is in agreement with multiple studies from prior literature~\cite{Brimhall1971,Jostsons1980}.
Regarding the presence of $<$\textbf{c}$>$-type dislocation loops, previous literature reporting on neutron irradiation of Ti under similar conditions (our work: T$_{irr}$=300$\degree$C and fluence ($>$~0.1~MeV) = 6.3$\times$10$^{20}$~cm$^{-2}$, Brimhall et al.~\cite{Brimhall1971}: T$_{irr}$=450$\degree$C and fluence ($>$~0.1~MeV) = 3$\times$10$^{21}$~cm$^{-2}$, Jostsons et al.~\cite{Jostsons1980}: T$_{irr}$=400$\degree$C and fluence ($>$~1~MeV) = 3.9$\times$10$^{19}$~cm$^{-2}$, and Griffiths et al.~\cite{Griffiths1983}: T$_{irr}$=316$\degree$C and fluence ($>$~0.1~MeV) = 1.7$\times$10$^{21}$~cm$^{-2}$) has found that radiation damage consists only of $<$\textbf{a}$>$-type dislocation loops. Thus $<$\textbf{c}$>$-type dislocation loops are not expected to be produced in our samples. Furthermore, if $<$\textbf{c}$>$-type dislocation loops (with \textbf{b}~=~$\frac{1}{6}\langle20\bar{2}3\rangle$) were present, the imaging conditions used in our study would result in them being visible as the product of $\textbf{g}\cdot$\textbf{b} is non-zero.

\subsection{Defect density quantification} 
\begin{figure}[ht!]
     \centering
     \begin{subfigure}[b]{\textwidth}
         \centering
         \includegraphics[scale=0.93]{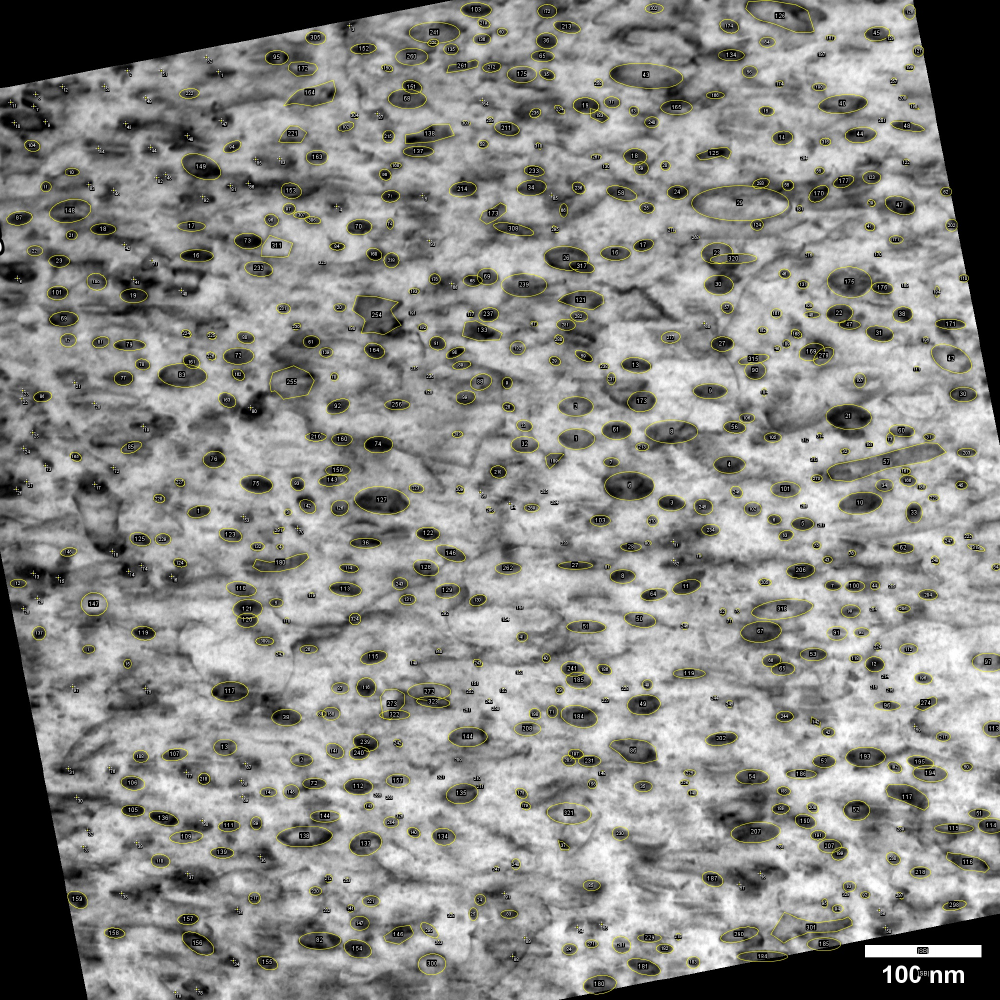}
         \caption{Image analysis of the sample as-irradiated at 300$\degree$C.} \vspace{10pt}
     \end{subfigure}
     \begin{subfigure}[b]{\textwidth}
         \centering
         \includegraphics[scale=0.93]{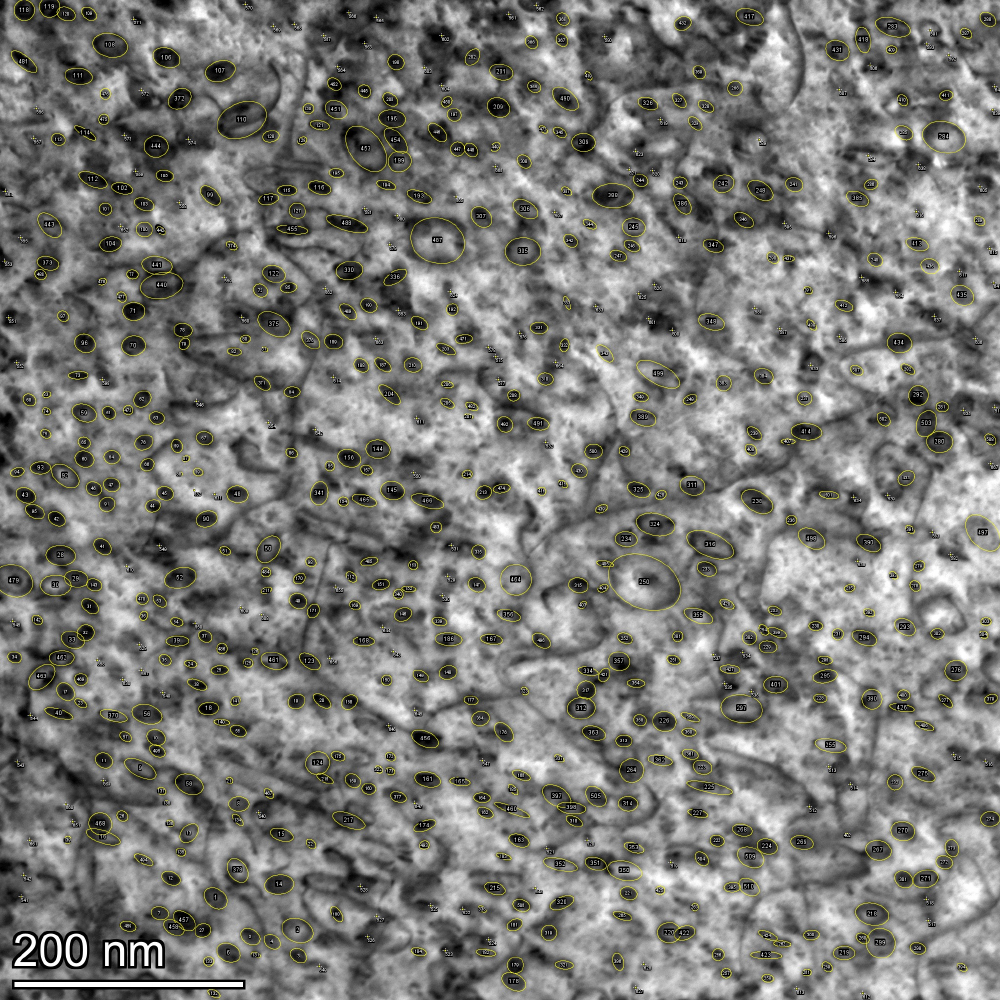}
         \caption{Image analysis of the sample irradiated and annealed to 480$\degree$C.}
     \end{subfigure}
        \caption{\textbf{ImageJ image analysis software was used to measure the dislocation loop size and number density}. The number density was determined by counting the number of loops within the image and dividing by the sample volume. The as-irradiated sample contains a dislocation loop number density of (4.0$\pm$0.7)$\times$10$^{21}$~m$^{-3}$ and the irradiated and annealed to 480$\degree$C sample contains (4.0$\pm$0.8)$\times$10$^{21}$~m$^{-3}$.} \label{defect_measure}
\end{figure}

\begin{figure}[ht!]
     \centering
     \begin{subfigure}[b]{\textwidth}
         \centering
         \includegraphics[scale=0.13]{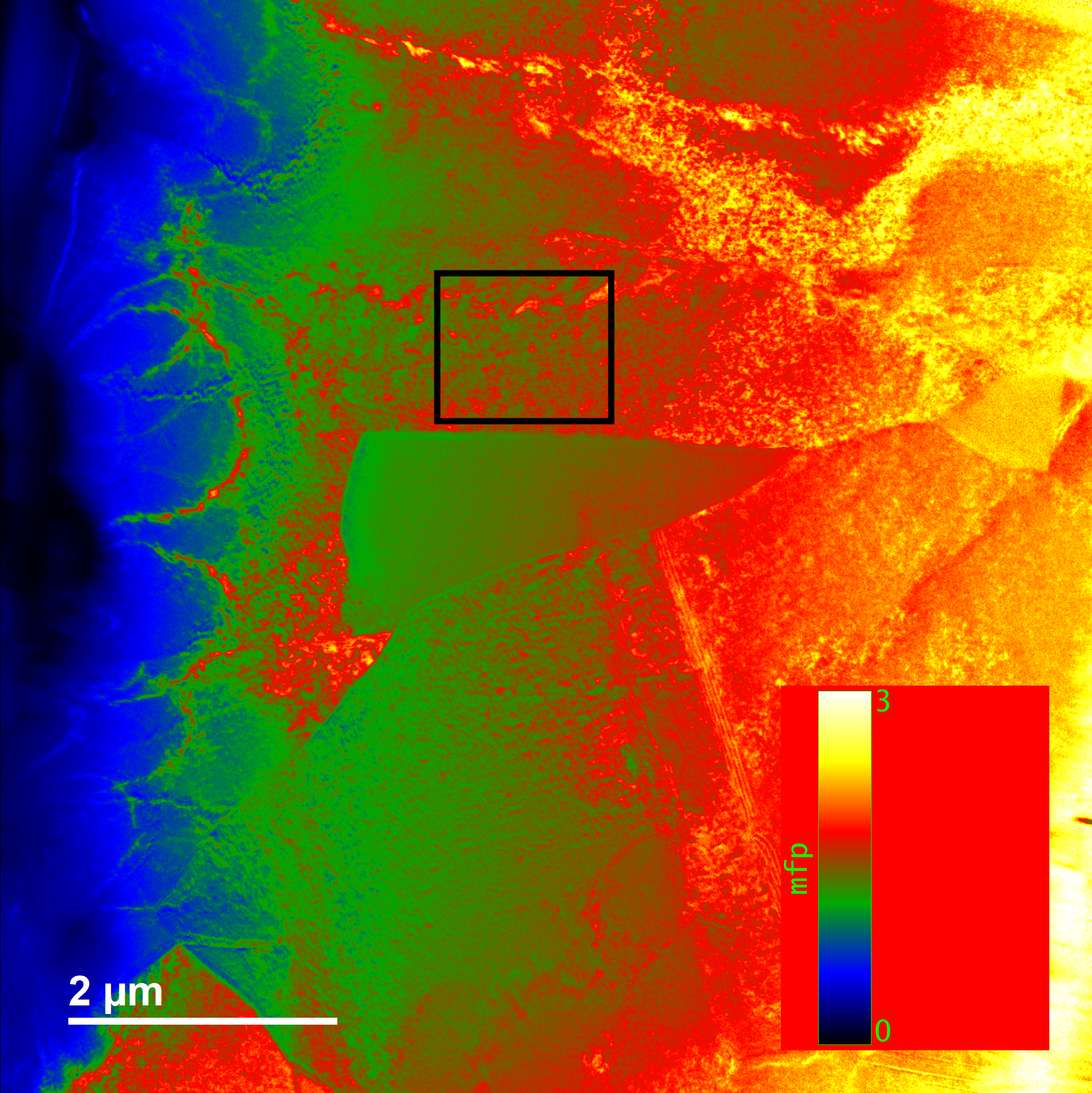}
         \caption{EFTEM of sample as-irradiated at 300$\degree$C.} \vspace{10pt}
     \end{subfigure}
     \begin{subfigure}[b]{\textwidth}
         \centering
         \includegraphics[scale=0.26]{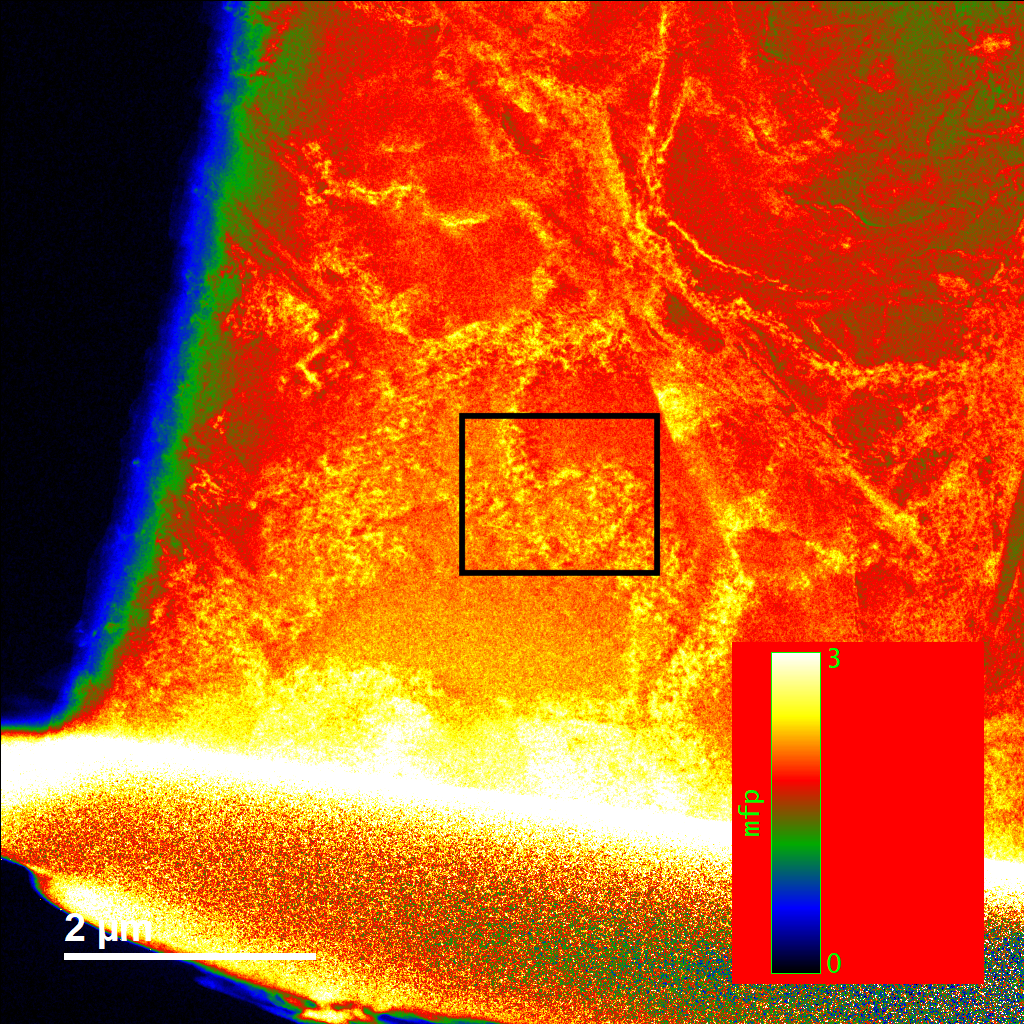}
         \caption{EFTEM of sample irradiated and annealed to 480$\degree$C.}
     \end{subfigure}
        \caption{\textbf{The thickness of the lamellae was determined using energy-filtered TEM (EFTEM).} The black boxes show the locations of the images in figure \ref{defect_measure}, and their dimensions are (1029~nm)$^2$ and (858~nm)$^2$, respectively. The color scale indicates the inelastic mean free path of 200~keV electrons.} \label{eftem}
\end{figure}

\vspace{5pt} \label{defect}
\noindent The number density and size distribution of dislocation loops (from TEM images of the as-irradiated sample) were used to calculate the energy stored in dislocation loops. ImageJ image analysis software was used to count the number ($N$) of dislocation loops and obtain their size using tracing method as shown in figure \ref{defect_measure}. The Feret diameter of the dislocation loops was used as a measure of the dislocation loop size. The thickness of the TEM lamellae, at the locations of the images, was determined using energy-filtered TEM (EFTEM) maps shown in figure \ref{eftem}. The EFTEM maps provide the ratio between the TEM specimen thickness ($t$) and the inelastic mean free path ($\lambda$) of electrons at 200~keV. The $\lambda$ value for Ti at 200~keV and the uncertainty in its estimation are 106~nm and 19\%, respectively, obtained from literature~\cite{Shinotsuka2015}. The thickness of the TEM lamellae was used to determine the volume imaged ($V_{T}$), and thus calculate the dislocation loop number density ($D_{loop}$) using equation \ref{ND}. The error in the dislocation loop number density calculation was estimated from the uncertainty in the TEM thickness measurement.

\begin{equation}\label{ND}
    D_{loop} = N / V_{T}
\end{equation} \vspace{5pt}

Comparing our calculated number density of loops to previous literature shows that our values match closely. Brimhall et al.~\cite{Brimhall1971}, Jostsons et al.~\cite{Jostsons1980}, and Griffiths et al.~\cite{Griffiths1983} report loop number densities of 2.5$\times$10$^{21}$~m$^{-3}$, (7$\pm$1)$\times$10$^{20}$~m$^{-3}$, and 7$\times$10$^{20}$~m$^{-3}$, respectively, compared to our value of (4.0$\pm$0.7)$\times$10$^{21}$~m$^{-3}$. The match between these values establishes confidence that our measurements are accurate.

\begin{figure}[ht!]
         \centering
         \begin{subfigure}[b]{\textwidth}
             \centering
              \includegraphics[scale=0.9]{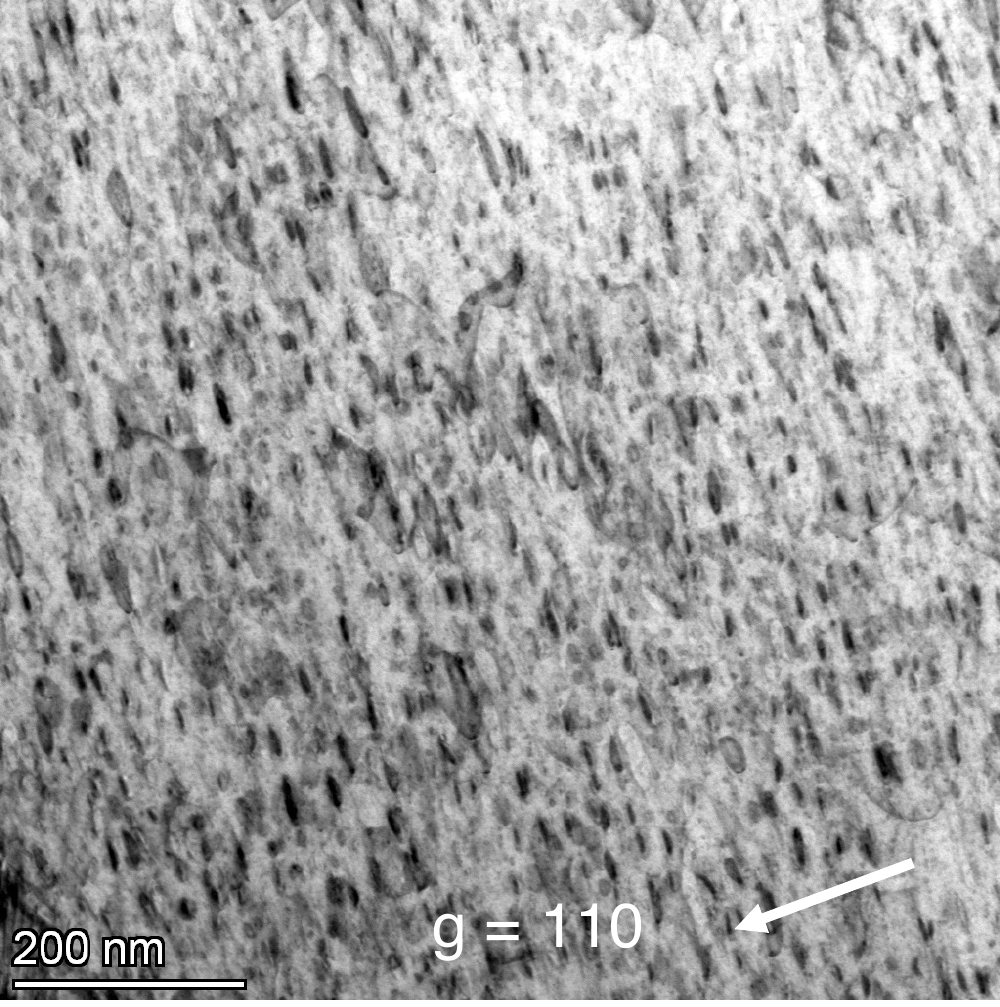}
             \caption{}
             \label{A1_other_TEM}
         \end{subfigure}
         \begin{subfigure}[b]{\textwidth}
             \centering
             \includegraphics[scale=0.9]{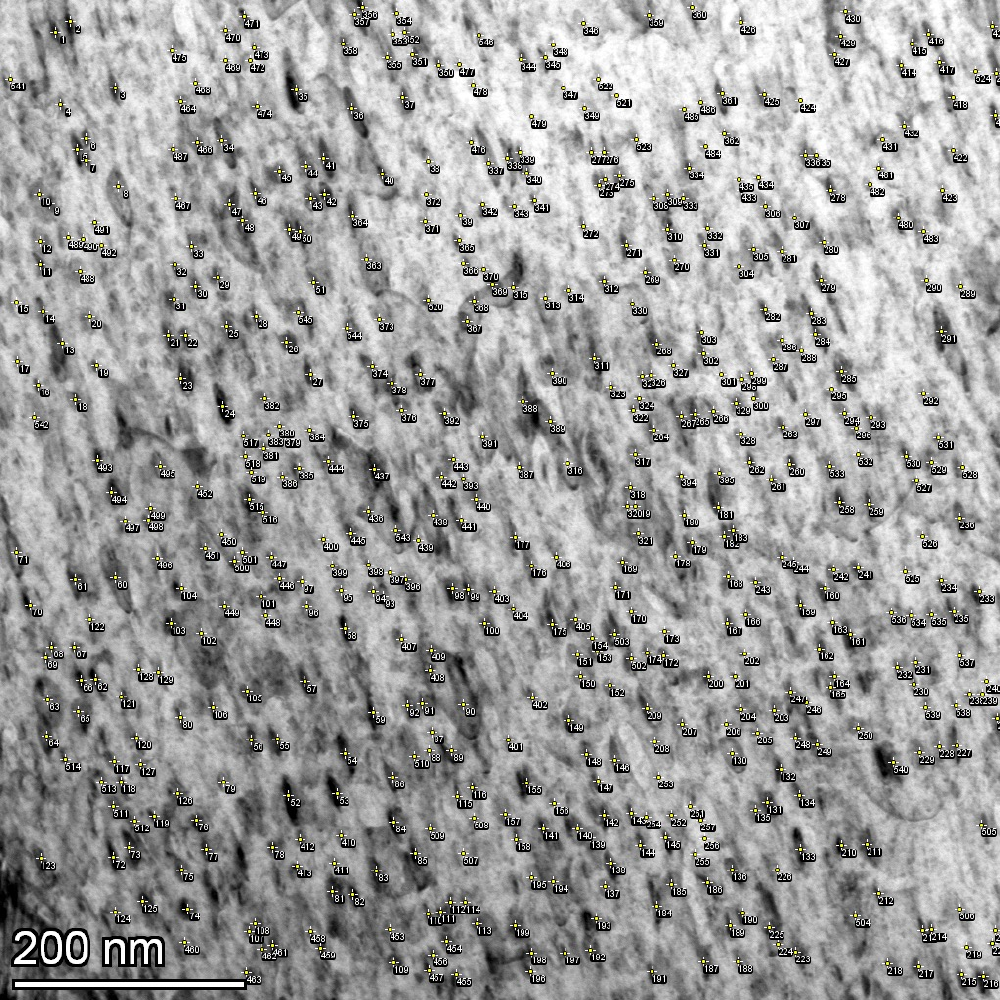}
             \caption{}
             \label{A1_other_count}
         \end{subfigure}
         \caption{\textbf{Imaging with different conditions returns the same number density of loops.} (a) STEM image taken with zone axis = [$\bar{1}$10] and \textbf{g}~=~110 in the two beam condition. (b)  The same image with annotations showing the 546 loops that were counted. This results in a number density of (4.4$\pm$0.8)$\times$10$^{21}$~m$^{-3}$ which is identical (within uncertainty) to the number density obtained earlier under different imaging conditions.}
         \label{A1_other}
         \end{figure}

To further confirm the accuracy of our TEM experiments we conducted analysis of an image taken from the same grain under different imaging conditions. In figure \ref{A1_other}, the [$\bar{1}$10] zone axis and \textbf{g}~=~110 were used. It is noted that, should there be any, $<$\textbf{c}$>$-loops would not show contrast under these imaging conditions. Despite this, the number density of dislocation loops calculated from the image is (within uncertainty) the same as previously determined, namely (4.4$\pm$0.8)$\times$10$^{21}$~m$^{-3}$. Additionally, analysis of an image from a different grain taken under these imaging conditions yields a very similar number density of (4.1$\pm$0.8)$\times$10$^{21}$~m$^{-3}$. That the dislocation loop density is consistent under different imaging conditions and at different locations provides confidence in the accuracy of our results.

\begin{figure}[ht!]
         \centering
         \begin{subfigure}[b]{\textwidth}
            \centering
            \includegraphics[scale=0.6]{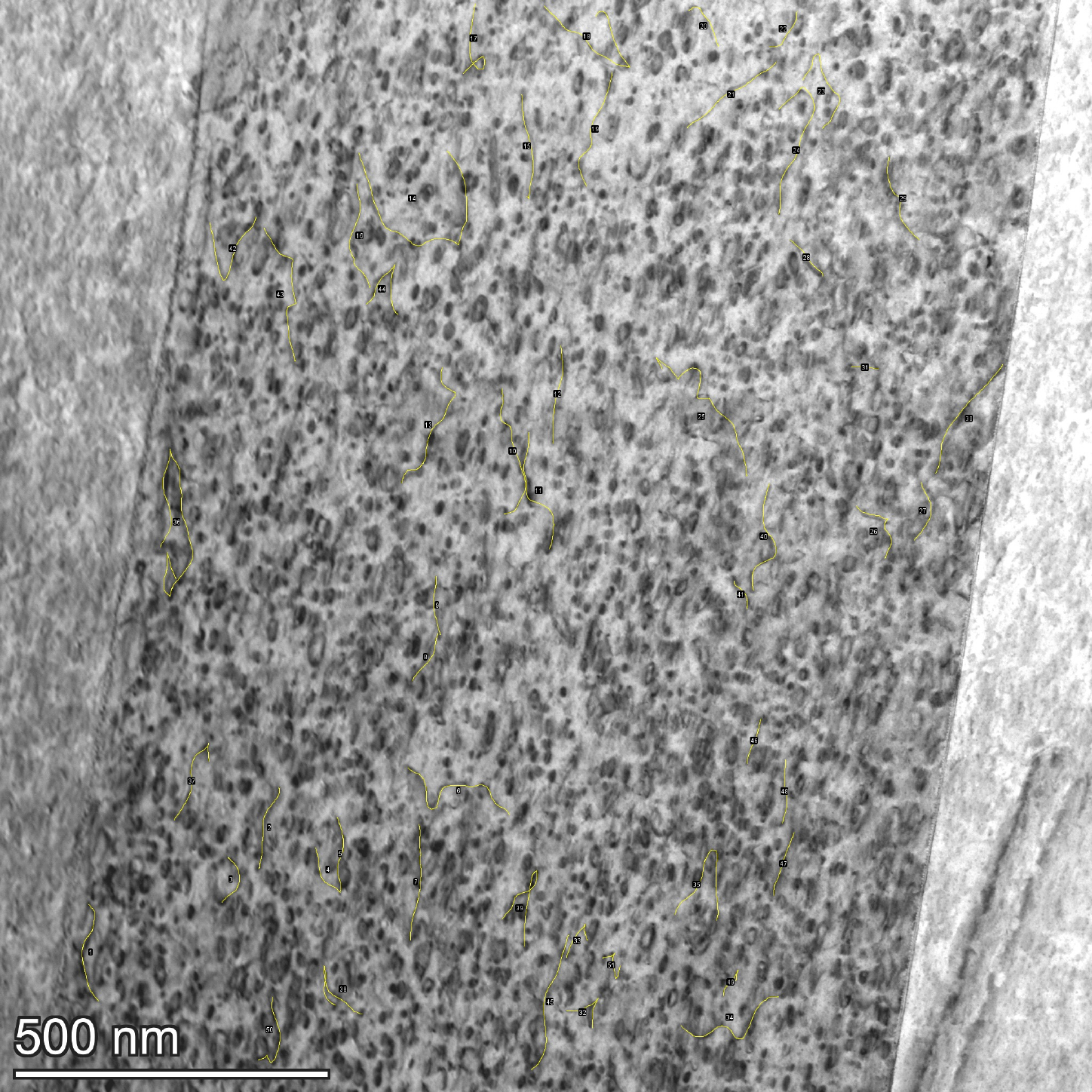}
            \caption{\textbf{As-irradiated sample.} STEM image taken with a zone axis of [100] and \textbf{g} = 011.} \vspace{10pt}
         \end{subfigure}
          \begin{subfigure}[b]{\textwidth}
         \centering
            \includegraphics[scale=0.9]{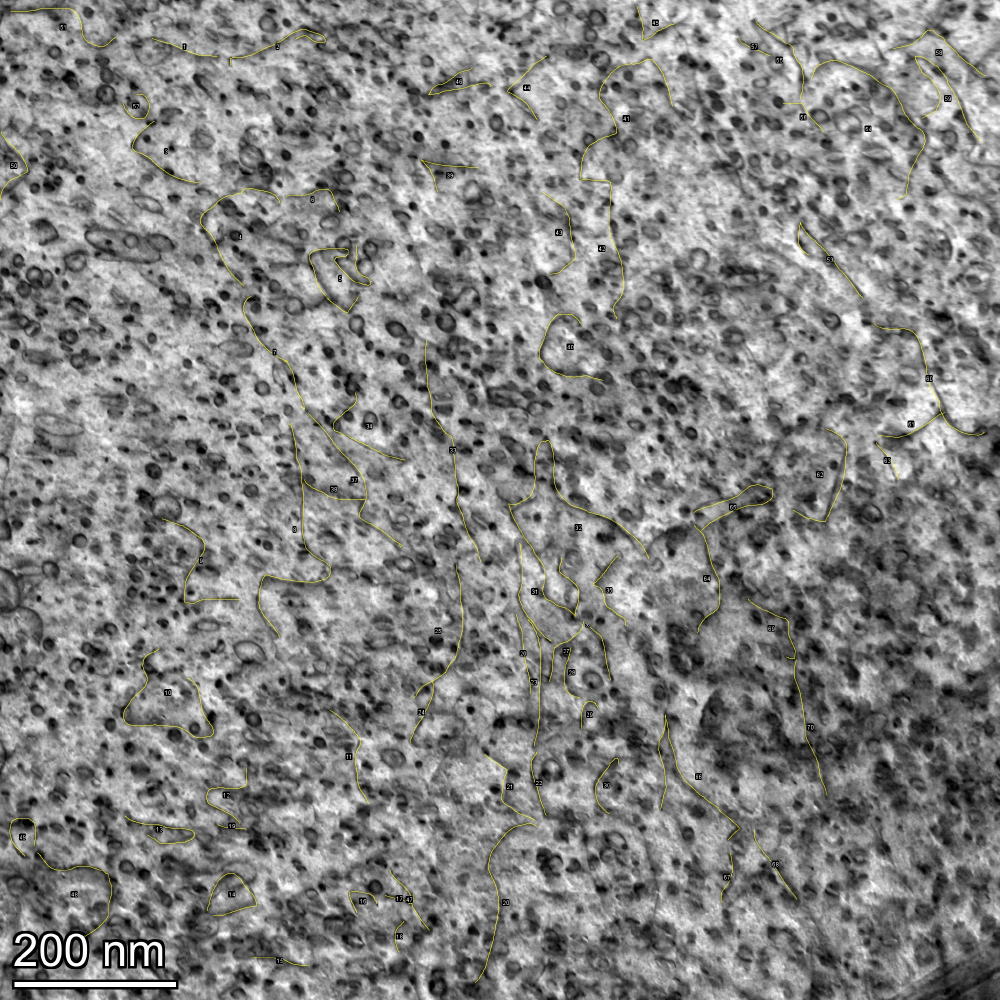}
            \caption{\textbf{Sample irradiated and annealed to 480$\degree$C.}} \vspace{10pt}
         \end{subfigure}
         \caption{\textbf{Linear dislocations have an order of magnitude lower areal density than loops.} Network dislocation density in the as-irradiated sample is (2.4$\pm$0.5)$\times$10$^{13}$~m$^{-2}$ and (3.2$\pm$0.6)$\times$10$^{13}$~m$^{-2}$ in the sample irradiated and annealed to 480$\degree$C.}
         \label{network_disns}
        \end{figure}

The areal number density of linear dislocations in the as-irradiated sample and the sample irradiated and annealed to 480$\degree$C was determined as (2.4$\pm$0.5)$\times$10$^{13}$~m$^{-2}$ and (3.2$\pm$0.6)$\times$10$^{13}$~m$^{-2}$, respectively. This contributes a stored energy density of less than 9\% when compared to the dislocation loops. Annotated images are shown in figure \ref{network_disns}. Additionally, so-called `black spot defects' were determined to have a number density of $(5.0\pm0.9)\times10^{20}$~m$^{-3}$ and average size of 3~nm. These contribute a stored energy density of less than 2\% when compared to the dislocation loops.

\section{Dislocation loop energy calculations} \label{energy_calc}
\noindent To make a comparison between DSC and TEM measurements the energy per defect must be determined. Inspired by recent work by Liu et al.~\cite{Liu2020}, elasticity theory was used to calculate the energy for an elliptical $<$\textbf{a}$>$-type dislocation loop on the prismatic plane in Ti.
In order to accurately calculate the dislocation loop energy, the effects of temperature and anisotropy were accounted for. Temperature-dependent elastic constants for Ti were obtained from Fisher et al.~\cite{Fisher1964}. The temperature of most relevance to our experiments is 550$\degree$C which corresponds to the peak temperature of the second annealing peak observed in the DSC. The following equations and the parameters detailed in table \ref{table} were used to calculate the energy per loop as

\begin{equation} \label{eq_full}
    E_{loop} = 2\pi R \cdot K \cdot \frac{b^2}{4\pi} \left( \ln \left[ \frac{4R}{(b/2\alpha)\cdot \exp(\gamma)}\right]-1\right)
\end{equation}

\begin{table}[h]
\centering
\begin{tabular}{llll}
\textbf{Variable}\vspace{5pt}          & \textbf{Description}                                                          & \textbf{Value}                      & \textbf{Ref.} \\
R                          & Radius of dislocation loop                                                    & 9.5~nm                                &  this work             \\
b                          & Burgers vector of prismatic loop                                              & 2.95~\AA          &   \cite{Boyer1994}            \\
$\alpha$      & Dislocation core parameter                                                    & 1.25                                &  \cite{Hirth1982}             \\
exp($\gamma$) & exp($\gamma$) = \text{exp}(1-2$\nu$)/4(1-$\nu$) & 1.17                                &   \cite{Hirth1982}            \\
\end{tabular}
\caption{\textbf{Parameters used for dislocation loop energy calculation.}} \label{table}
\end{table}

\subsection{Anisotropy dependence}

\noindent Parameter $K$ is the energy factor which contains the temperature-dependent elastic constants and orientation dependence of the dislocation line. From Savin et al.~\cite{Savin1976}, the energy factor for an edge dislocation on the prismatic plane, aligned along the \textbf{c} axis with \textbf{b}=$\frac{1}{3}\langle1\bar{2}10\rangle$, is given by

        \begin{equation}
            K_\textbf{c} = \frac{C_{11}^{2}-C_{12}^{2}}{2C_{11}}
        \end{equation}
        
\noindent The energy factor for an edge dislocation on the basal plane, with \textbf{b}=$\frac{1}{3}\langle1\bar{2}10\rangle$, is given by

		\begin{equation}
            K_\textbf{a} = (\lambda^{2}C_{33} + C_{13})\times
            \left[ \frac{C_{44}(\lambda^{2}C_{33}-C_{13})}
            {C_{33}(\lambda^{2}C_{33}+C_{13}+2C_{44})} \right]^{\frac{1}{2}}
        \end{equation}
where
        \begin{equation}
            \lambda^{2} = \left( \frac{C_{11}}{C_{33}} \right)^{\frac{1}{2}}
        \end{equation}

\noindent Evaluating these expressions at 550$\degree$C gives energy factors of K$_\textbf{c}$= 31.0~GPa and K$_\textbf{a}$= 53.2~GPa and a ratio of K$_\textbf{c}$/K$_\textbf{a}$ = 0.58. These calculations are supported by Brimhall et al.~\cite{Brimhall1971} who determine the same ratio to be K$_\textbf{I}$/K$_\textbf{III}$ = 0.57, demonstrating an excellent match between our calculations and prior literature.
From our TEM experiments, the major ($\parallel$ to \textbf{c}) and minor ($\parallel$ to \textbf{a}) axes of the 244 observed elliptical dislocation loops in the as-irradiated sample are averaged to give a mean ratio of 1.70~$\pm$~0.03 ($\pm$ standard error). This results in a weighted average energy factor of
        \begin{equation}
            K_{eff} = \frac{1.7}{1.7+1} \times K_\textbf{c} + \frac{1}{1.7+1} \times K_\textbf{a} = 39.2~GPa
        \end{equation}

\noindent Combining this value with equation \ref{eq_full} resulted in an energy per length of 7.8~$\frac{\text{eV}}{\text{nm}}$ and a total energy per dislocation loop of 467~eV.

\section{MD methods \& analyses}
\subsection{PKA and annealing input files}
\noindent Sample input files for the PKA simulations and instructions how to run them are provided in the data repository: \url{https://github.com/shortlab/2021-Ti-Wigner}. Also included are input files for the annealing simulations and all the atomic configurations before and after annealing at 300, 480, and 600$\degree$C. The `before' configurations are 10 unique supercells, each containing 492,800 atoms, which have seen 8000 PKAs corresponding to a dose of 0.6~dpa.

\subsection{Videos of annealing simulations}
\noindent Videos showing the annealing of radiation damage at 300, 480, and 600$\degree$C are included in the data repository: \url{https://github.com/shortlab/2021-Ti-Wigner} 

\newpage
\section*{FIGURES - SUPPLEMENTARY}

\newpage

\clearpage

\clearpage

\clearpage

\clearpage

\clearpage

\clearpage

\clearpage

\clearpage

\clearpage

\clearpage

\clearpage

\clearpage

\clearpage

\clearpage

\end{document}